\documentclass[aps,twocolumn,prc,superscriptaddress,showpacs,nofootinbib,floatfix,amssymb,amsfonts,amsmath]{revtex4-1}
\usepackage{graphicx}
\usepackage{dcolumn}
\usepackage{bm}
\usepackage{amsmath}    
\usepackage{amsfonts}   
\usepackage{amssymb}
\usepackage{graphicx}   

\begin{document}

\title{The propagation of statistical errors in covariant density 
functional theory: ground state observables and single-particle
properties.}

\author{S.\ E.\ Agbemava}
\affiliation{Department of Physics and Astronomy, Mississippi
State University, MS 39762}

\author{A.\ V.\ Afanasjev}
\affiliation{Department of Physics and Astronomy, Mississippi
State University, MS 39762}

\author{A.\ Taninah}
\affiliation{Department of Physics and Astronomy, Mississippi
State University, MS 39762}

\date{\today}

\begin{abstract}
  Statistical errors in ground state observables and single-particle 
properties of spherical even-even nuclei and their propagation to the 
limits of nuclear landscape have been investigated in covariant density 
functional theory (CDFT) for the first time. In this study we consider 
only covariant energy density functionals with non-linear density 
dependency. Statistical errors for binding energies and neutron 
skins significantly increase on approaching two-neutron drip line.
On the contrary, such a trend does not exist for statistical errors 
in charge radii and two-neutron separation energies. The absolute and 
relative energies of the single-particle states in the vicinity of 
the Fermi level are characterized by low statistical errors 
($\sigma(e_i)\sim 0.1$ MeV). Statistical errors in the predictions of 
spin-orbit splittings are rather small. Statistical errors in physical 
observables are substantially smaller than related systematic 
uncertainties. Thus, at the present level of the development of theory, 
theoretical uncertainties at nuclear limits are dominated by systematic 
ones. Statistical errors in the description of physical observables 
related to the ground state and single-particle degrees of freedom are 
typically substantially lower in CDFT as compared with Skyrme density 
functional theory. The correlations between the model parameters are 
studied in detail. The parametric correlations are especially pronounced 
for the $g_2$ and $g_3$ parameters which are responsible for the density 
dependence of the model. The accounting of this fact potentially allows 
to reduce the number of free parameters of non-linear meson coupling 
model from six to five.
\end{abstract}


\maketitle

\section{INTRODUCTION}

 Although significant progress has been achieved over the years in
the development of theoretical tools for the description of low-energy
nuclear phenomena, some simplifications and approximations are still
necessary because of the complexity of nuclear many-body problem and
the impossibility of its exact solution for the systems with large number
of particles. In addition, fine details of nuclear force and its
dependence on density are still not fully resolved. As a result,
it becomes necessary to estimate theoretical uncertainties in the
description of physical observables \cite{RN.10,DNR.14,AARR.14}. This 
is especially important when one deals with the extrapolations beyond 
the known regions, as, for example, in particle number or deformation, 
since experimental data which acts as a substitute of exact solution 
are not available there. Such  estimates are also required for the evaluation 
of predictive power of the models and the robustness of their predictions. 
The need for such estimates has been clearly recognized by nuclear theory 
community as illustrated by a substantial number of the studies aiming at
the quantification of theoretical uncertainties in nuclear structure, nuclear 
reactions and nuclear astrophysics (see Refs.\ 
\cite{Eet.12,AARR.13,GDKTT.13,AARR.14,MSHSWN.15,MANO.16,MSMA.16,LNSW.17} 
and references quoted  therein).

  There are two types of theoretical uncertainties: systematic and statistical
ones \cite{Stat-an,DNR.14}. {\it Systematic} theoretical uncertainties emerge from 
underlying theoretical approximations. In the framework of density functional
theory (DFT), there are two  major sources of these approximations, namely, the 
range of interaction and the form of the density dependence of the effective 
interaction \cite{BHP.03,BB.77}. In the non-relativistic case one has zero range 
Skyrme and finite range Gogny forces and different density dependencies 
\cite{BHP.03,PM.14}. A similar situation exists also in the relativistic case: point 
coupling and meson exchange models have an interaction of zero and of finite range, 
respectively \cite{VALR.05,DD-ME2,NL3*,DD-PC1}. The density dependence is introduced 
either through an explicit dependence of the coupling constants \cite{TW.99,DD-ME2,DD-PC1} 
or via non-linear meson couplings \cite{BB.77,NL3*}. This ambiguity in the definition 
of the range of the interaction and its density dependence leads to several major 
classes of the covariant energy density functionals (CEDF) which were discussed in 
Ref.\ \cite{AARR.14}.

  It is necessary to recognize that precise quantification of {\it systematic errors} 
in the regions of nuclear chart for which experimental data is not available is not 
possible due to a number of reasons \cite{DNR.14,AARR.14}. Thus, we prefer to use 
the notation {\it systematic uncertainties} (instead of {\it systematic errors} 
introduced in Ref.\ \cite{DNR.14}) which has more narrow meaning since they are 
defined with respect of selected set of the functionals (see 
introduction of Ref.\ \cite{AARR.14}).

  An additional source of theoretical uncertainties is related to the 
details of the fitting protocol such as the choice of experimental data 
and the selection of adopted errors. It applies only to a given 
functional and the related theoretical uncertainties are called 
{\it statistical errors} \cite{Stat-an,DNR.14}. Note that the selection 
of adopted errors is to a degree subjective, in particular, if one deals 
with quantities of different dimensions.

  The covariant density functional theory (CDFT) represents a relativistic 
extension of the DFT framework to the nuclear many-body problem \cite{VALR.05}. 
It  exploits basic properties of QCD at low energies, in particular symmetries 
and the separation of scales \cite{LNP.641}. It provides a consistent treatment 
of the spin degrees of freedom and spin-orbit splittings \cite{LA.11} and 
includes the complicated interplay between the large Lorentz scalar and
vector self-energies induced on the QCD level by the in-medium changes of the scalar
and vector quark condensates \cite{CFG.92}. In addition, CEDFs include {\it nuclear 
magnetism} \cite{KR.89}, i.e. a consistent description of currents and time-odd mean 
fields \cite{AA.10}, for which no new adjustable parameters are required because
of Lorentz invariance. The CDFT has been successfully applied to the
description of a large variety of nuclear phenomena (see reviews in Refs.\ 
\cite{VALR.05,MTZZLG.06,NVR.11,RDFNS.16}). 

   However, only recently systematic efforts have been undertaken to quantify
theoretical uncertainties in the description of physical observables within 
the CDFT framework. Systematic uncertainties, their sources and their
propagation to the extremes of neutron number have been studied globally for 
the ground state masses, deformations, two-particle separation energies,
charge radii and neutrons skins of even-even nuclei, as well as for
the  positions of drip lines in Refs.\ 
\cite{AARR.13,AARR.14,AARR.15,AANR.15,AAR.16,AA.16}. In Ref.\ \cite{AANR.15}
systematic uncertainties in the predictions of the ground state properties 
of superheavy nuclei have been investigated. Above mentioned investigations 
were restricted to even-even nuclei which are either spherical or have
only even-multipole deformations in the ground state. This restriction has
been removed in Refs.\ \cite{AAR.16,AA.17-oct} in which the global studies of 
octupole deformed nuclei and related systematic uncertainties in their description 
have been performed. The investigations of systematic uncertainties have 
also been carried out for excited states such as inner fission barriers in 
superheavy nuclei (Ref.\ \cite{AARR.17}), deformed one-quasiparticle states
in odd-mass actinides \cite{AS.11,DABRS.15} and rotational states in even-even
and odd-mass actinides (Ref.\ \cite{AO.13}). 

  Although impressive amount of data on systematic uncertainties in the 
description of physical observables in the CDFT has been collected within
last few years, very little is known about related statistical errors. So 
far, they have been investigated only for potential energy curves in a single 
superheavy nucleus for two CEDFs in Ref.\ \cite{AARR.17}. To fill this gap 
in our knowledge of the performance of CEDFs, the present manuscript aims 
on a systematic study of statistical errors in the description of the ground 
state and single-particle properties of spherical nuclei. 

  A second goal of the present investigation is to establish the correlations 
between the CEDF parameters and to see whether these parameters are independent. 
Such correlations have not been studied systematically so far in the CDFT 
framework. Their presence affects statistical errors in the description of 
physical observables \cite{DD.17}. The removal of parametric correlations leads 
to the reduction of the dimensionality of the parameter hyperspace and to the 
decrease of statistical errors. The latter was illustrated in Ref.\ \cite{DD.17} 
on the example of the study of statistical errors in the single-particle energies
of spherical nuclei performed with Woods-Saxon potential.

 So far, mostly the covariance analysis has been used in the studies of 
statistical errors for physical observables in the DFT framework (see, for 
example, Refs.\ \cite{GDKTT.13,DNR.14,NPRV.15,HK.17}. However, in the
calculations of the covariance matrix a linearized least-square system 
in the vicinity of the minimum of objective function $\chi^2_{norm}$ is 
usually assumed \cite{Stat-an}.
  This means that the covariance 
analysis assumes that full information about statistical errors is contained 
in the derivatives taken at the optimum parametrization and ignores potential 
non-linear dependence of the observables on the coupling constants in the 
parts of the parameter hyperspace away from the optimum parametrization. However, 
there is no guarantee that linear approximation is valid for non-linear problems in
the region of the parameter hyperspace away from the optimum parametrization
(see Ref.\ \cite{Stat-an}).  If that is a case the covariance matrix loses 
its validity \cite{Stat-an}. A priori such non-linearities, which are present
in the CDFT models, cannot be disregarded.
For example, the analysis of the correlations between the $\alpha_S$ and 
$\alpha_V$ parameters in point coupling models performed in Ref.\ \cite{BMR.04} 
clearly indicates their presence.

  Thus, we use alternative approach based on the Monte-Carlo method in which 
randomly generated functionals are accepted/rejected based on the condition 
of Eq.\ (\ref{cond}) given in Sec.\ \ref{sect-form} below.
 As a result, the set of reasonable functionals is
generated which is used for the calculations of statistical errors in the 
physical observables of interest (see Sec.\ \ref{sect-form}). The advantage 
of this method is that its outcome is defined by full parameter hyperspace 
used in the analysis (and not only by the neighborhood of optimum
parametrization as in covariance analysis). This allows to take existing 
non-linearities between the parameters fully into account. Thus, the Monte-Carlo 
approach is free from above mentioned deficiencies of the covariance 
analysis. However, it is much more numerically time consuming than relatively 
simple covariance analysis and requires significant computational power. As 
a result, so far it was applied only to the analysis of statistical errors in 
the single-particle energies of phenomenological Woods-Saxon potential (see 
Ref.\ \cite{DD.17}). Because of this reason we focus in the present exploratory 
study on the non-linear (NL) CEDFs which are characterized by the minimum number 
of the  parameters (and, as a consequence, by minimal dimensionality of the parameter 
hyperspace) among the state-of-the-art CEDFs. Their six parameters are the mass 
$m_{\sigma}$ and coupling constant $g_{\sigma}$ of the $\sigma$-meson, the coupling constant 
$g_{\omega}$ of the $\omega$-meson, the coupling constant $g_{\rho}$ 
of the $\rho$ meson which is responsible for the isovector channel 
of the functional and the coupling constants $g_2$ and $g_3$ which 
define the density dependence of the functional.

  The paper is organized as follows. Section \ref{sect-form} describes 
the details of the analysis of statistical errors. Fitting protocols and
related theoretical uncertainties are discussed in Sec.\ \ref{fitting-prot}. 
The ranges of the parameters and parametric correlations in the functionals
are discussed in Sec.\ \ref{sect-param}. Statistical errors in the description 
of the ground state observables of spherical even-even nuclei and their 
propagation towards neutron drip line are investigated in Sec.\ \ref{sect-GS}. 
Sec.\ \ref{sect-sp} is devoted to the discussion of statistical errors in 
the description of single-particle energies and their consequences for the 
predictions at the extremes of neutron number and charge. Note that statistical 
errors discussed in Secs.\ \ref{sect-GS} and \ref{sect-sp} are compared 
with available systematic uncertainties estimated previously in Refs.\ 
\cite{AARR.13,AARR.14,AANR.15,AA.16}. Finally, Sec.\ \ref{concl} summarizes 
the results of our work.

\section{Statistical errors and the
         details of the calculations}
\label{sect-form}

 The normalized objective function is defined for model having $N_{par}$ adjustable 
parameters ${\bf p}=(p_1, p_2, ..., p_{N_{par}})$ as
\begin{eqnarray}
\chi^2_{norm}({\bf p})=\frac{1}{s}\sum_{i=1}^{N_{type}} \sum_{j=1}^{n_i} \left( \frac{O_{i,j}({\bf p})-O_{i,j}^{exp}}
{\Delta O_{i,j}} \right)^2
\label{Ksi}
\end{eqnarray}
where 
\begin{eqnarray}
s=\frac{\chi^2({\bf p}_0)}{N_{data}-N_{par}}
\end{eqnarray}
is global scale factor (Birge factor \cite{Birge.32})  defined at the minimum of the 
penalty function (optimum parametrization 
${\bf p}_0$\footnote{Because of the experimental errors and incompleteness of the 
physical modelling optimum parametrizations of the models are known 
only up to their uncertainty probability distributions \cite{DD.17}.})
which leads to the average $\chi^2 ({\bf p}_0)$ per degree of freedom 
equal to one \cite{DNR.14} and
\begin{eqnarray}
N_{data}= \sum_{i=1}^{N_{type}}n_i
\end{eqnarray}
is the total number of data points of different types. 
Here,  
$N_{type}$ stands  for the number of different data types. 
The calculated and experimental/empirical values of physical 
observable $j$ of the $i-$th type are represented by $O_{i,j}({\bf p})$ 
and $O^{exp}_{i,j}$, respectively. $\Delta O_{i,j}$ is adopted error 
for physical observable $O_{i,j}$. These quantities for the functionals 
under study are summarized in Table \ref{table-inp}. 

  The acceptable functionals are defined from the condition
\cite{DNR.14}
\begin{eqnarray}
\chi^2_{norm} ({\bf p}) \leq \chi^2_{norm}({\bf p}_0) + 1 .
\label{cond}
\end{eqnarray} 
This condition specifies the 'physically reasonable' domain around 
${\bf p}_0$ in which the parametrization ${\bf p}$ provides 
a reasonable fit and thus can be considered as acceptable. This domain 
is the $N$-dimensional parameter hyperspace 
${\bf P_{space}}=[p_{1_{min}} - p_{1_{max}}, p_{2_{min}} - p_{2_{max}}, ..., p_{N_{min}} - p_{N_{max}}]$,
where $p_{i_{min}}$ and $p_{i_{max}}$ represent the lower and upper boundaries
for the variation of the $i-th$ parameter. These boundaries are defined in
such a way that their further increase (for $p_{i_{max}}$) or decrease (for 
$p_{i_{min}}$) does not lead to additional points in parameter hyperspace which 
satisfy Eq.\ (\ref{cond}).

  The numerical calculations are performed in the following way. New 
parametrizations ${\bf p}$ are randomly generated in the $N$-dimensional 
parameter hyperspace  and they are accepted if the condition  
(\ref{cond}) is satisfied. Using the set 
  $[{\bf p}_1, {\bf p}_2, ..., {\bf p}_M ]$ 
of $M$ accepted functional variations the calculations are performed for spherical 
nuclei in the Ca, Ni, Sn and Pb isotope chains from proton to neutron drip 
lines. For each nucleus the mean values of physical observables 
\begin{eqnarray}
\bar{O}_{i,j}=\frac{1}{M} \sum_{k=1}^{M} O_{i,j}({\bf p}_k)
\label{mean}
\end{eqnarray}
and their standard deviations
\begin{eqnarray}
\sigma_{i,j} = \sqrt{ \frac{1}{M}  \sum_{k=1}^{M} [O_{i,j}({\bf p}_k) - \bar{O}_{i,j}]^2}
\label{st-dev}
\end{eqnarray}
are calculated. The latter serves as a measure of statistical error.

 As mentioned in the introduction we consider here only non-linear meson
coupling models which are characterized by the minimal set of the parameters
amongst different classes of the CDFT models. In the meson-exchange models 
\cite{SW.86,NL3*}, the nucleus is described as a system of Dirac nucleons 
interacting via the exchange of mesons with finite masses leading to finite-range 
interactions. The starting point is a standard Lagrangian density 
\cite{GRT.90}
\begin{align}
\label{lagrangian}%
\mathcal{L}  &  =\bar{\psi}\left[%
\gamma\cdot(i\partial-g_{\omega}\omega-g_{\rho
}\vec{\rho}\,\vec{\tau}-eA)-m-g_{\sigma}\sigma
\right]%
\psi\nonumber\\
&+\frac{1}{2}(\partial\sigma)^{2}
-U(\sigma)
-\frac{1}{4}\Omega_{\mu\nu}\Omega^{\mu\nu}+\frac{1}{2}m_{\omega}^{2}\omega^{2}
\nonumber \\
&-\frac{1}{4}{\vec{R}}_{\mu\nu}{\vec{R}}^{\mu\nu}+\frac{1}{2}m_{\rho}^{2}\vec{\rho}^{\,2}
-\frac{1}{4}F_{\mu\nu}F^{\mu\nu}
\end{align}
which contains nucleons described by the Dirac spinors $\psi$ with the mass 
$m$ and several effective mesons characterized by the quantum numbers of 
spin, parity, and isospin. They create effective fields in a Dirac equation, 
which corresponds to the Kohn-Sham equation~\cite{KS.65} of non-relativistic 
density functional theory. The density dependence is introduced into model 
via a non-linear meson coupling \cite{BB.77}
\begin{equation}
U(\sigma)~=~\frac{1}{2}m_{\sigma}^{2}\sigma^{2}+\frac{1}{3}g_{2}\sigma
^{3}+\frac{1}{4}g_{3}\sigma^{4}.
\label{g2g3-eq}
\end{equation}
  In simplest ansatz, the Lagrangian (\ref{lagrangian}) contains as 
parameters the mass $m_{\sigma}$ of the $\sigma$ meson, the coupling 
constants $g_{\sigma}$, $g_{\omega}$, and $g_{\rho}$ as well as density 
dependent parameters $g_2$ and $g_3$. The masses $m$, $m_\omega$ and 
$m_\rho$ are typically fixed in non-linear meson coupling models.
$e$ is the charge of the protons and it vanishes for neutrons.

  The calculations have been performed using the spherical RHB code. The 
truncation of the basis is  performed in such a way that all states 
belonging to the shells up to $N_F = 20$ fermionic shells and $N_B = 20$ 
bosonic shells are taken into account. In order to avoid the uncertainties 
connected with the definition of the size of the pairing window we use 
the separable form of the finite range Gogny pairing interaction introduced 
by Tian et al \cite{TMR.09} with the strength of pairing defined in Ref.\ 
\cite{AARR.14}. This is also done for the consistency with previous global 
studies   of systematic uncertainties in the description of physical 
observables within the CDFT framework 
\cite{AARR.13,AARR.14,AARR.15,AANR.15,AAR.16,AA.16}.

\begin{table*}[ht]
\begin{center}
\caption{Input data for fitting protocol of the NL5() CEDFs. The number 
$n_i$ of experimental (empirical) data points and adopted errors 
$\Delta O_{i,j}$ are presented for each type of data. The binding
energies of the $^{16}$O, $^{40}$Ca,  $^{48}$Ca, $^{72}$Ni, $^{90}$Zr,
 $^{116}$Sn, $^{124}$Sn, $^{132}$Sn, $^{204}$Pb, $^{208}$Pb, 
$^{214}$Pb and $^{210}$Po nuclei, the charge radii of  $^{16}$O, 
$^{40}$Ca,  $^{48}$Ca, $^{90}$Zr, $^{116}$Sn, $^{124}$Sn, $^{204}$Pb, 
$^{208}$Pb and $^{214}$Pb nuclei as well as neutron skins of the 
$^{90}$Zr, $^{116}$Sn, $^{124}$Sn\ and $^{208}$Pb nuclei are used
in the fitting protocol. In addition, employed empirical values 
$O_{i,j}$ for the properties of symmetric nuclear matter at 
saturation are provided; these are the density $\rho_0$, the 
energy per particle $E/A$, the incompressibility $K_0$ and 
the symmetry energy $J$. The columns 3-6 show only the changes
with respect of the values provided in column 2.  
\label{table-inp}
}
\begin{tabular}{|c|c|c|c|c|c|} \hline 
                         &    NL5(A)      &  NL5(B)     & NL5(C)           & NL5(D)        &  NL5(E)  \\ \hline         
 1 & 2 & 3 & 4 & 5 & 6 \\ \hline
        \multicolumn{6}{|c|}{1. Masses $E$ (MeV)}  \\ \hline    
  $n_1$                  &    12          &             &                  &               &          \\          
$\Delta E$ [MeV]         &  $0.001 E$     &             &                  &               &          \\ \hline         
        \multicolumn{6}{|c|}{2. Charge radii $r_{ch}$ (fm)}  \\ \hline    
  $n_2$                  &    9           &             &                  &               &          \\            
$\Delta r_{ch}$ [fm]      & $0.002\,r_{ch}$ &            &                  &               &          \\ \hline         
        \multicolumn{6}{|c|}{3. Neutron skin $r_{skin}$ (fm)} \\ \hline    
  $n_3$                  &    4            &            & 3 [no $^{90}$Zr]  &   0          &           \\      
$\Delta r_{skin}$ [fm]    & $0.05\,r_{skin}$ &            &                  &              &   see text        \\ \hline         
        \multicolumn{6}{|c|}{4. Nuclear matter properties}  \\ \hline    
  $n_4$                  &    4            &            &                  &              &            \\          
$E/A$ [MeV]              &  -16.0          &            &                  &              &            \\          
$\Delta E/A$ [MeV]       & $0.05 E/A$      &            &                  &              &            \\          
$\rho$ [fm$^{-3}$]         &  0.153          &            &                  &              &            \\          
$\Delta \rho$ [fm$^{-3}$]  & $0.1 \rho$      &            &                  &              &            \\         
$K_0$ [MeV]              &  250.0          &            &                  &              &             \\         
$\Delta K_0$ [MeV]       & $0.1 K_0$       & $0.025 K_0$ &   $0.025 K_0$   &   $0.025 K_0$ & $0.025 K_0$ \\          
$J$  [MeV]               &  33.0           &            &                  &              &             \\          
$\Delta J$ [MeV]         & $0.1 J$         &            &                  &              &             \\ \hline         
        \multicolumn{6}{|c|}{Parameters of Eq. (\ref{Ksi})}  \\ \hline 
$N_{data}$                &   29            &            &        28        &  25          &             \\ 
$N_{par}$                 &    6            &            &                  &              &             \\ 
$N_{type}$                &    4            &            &                  &  3           &             \\ \hline
\end{tabular}
\end{center}
\end{table*}

\section{Fitting protocols: an example of the origin
of the uncertainties}
\label{fitting-prot}

 Previous fits of the non-linear CEDFs have been performed in the
RMF+BCS framework with simple pairing (see Refs.\ \cite{NL1,NLZ,NLSH,NL3,NL3*} 
for details). Since in the present work the RHB framework with separable 
pairing of finite range is used in fitting protocol for the first time,
the investigation of the dependence of the optimum parametrization on 
the details of the fitting protocol is performed.

\begin{table*}[ht]
\begin{center}
\caption{Different functionals obtained in the present work. The 
details of the fitting protocols are discussed in the text. First 
part of the table shows the parameters of the functionals. Note
that the masses of nucleon, $\omega$ and $\rho$ mesons are fixed 
at $m_N=939.0$ MeV, $m_{\omega}=782.6$ MeV and $m_{\rho}=763.0$ MeV, 
respectively. Related nuclear matter properties are 
displayed in the second part of the table. In addition to those quoted 
in Table \ref{table-inp}, they also include the slope of the 
symmetry energy $L_0$. Total penalty function  $\chi _{total}^{2}$ 
and the contributions to it coming from nuclear matter properties
($\chi _{NM}^{2}$), binding energies ($\chi _{E}^{2}$), charge
radii ($\chi _{Rch}^{2}$) and neutron skins ($\chi _{Nskin}^{2}$)
[and, in particular, the contribution of neutron skin of
$^{90}$Zr ($\chi _{Nskin(Zr)}^{2}$)] are presented in the last part 
of the table. Note that these contributions are given both in
absolute values and in percentages [in parantheses] with respect 
of $\chi _{total}^{2}$. In addition, the penalty function per degree
of freedom $\chi _{total}^{2}$/degree (Birge factor) is provided.
For comparison, the NL1 \cite{NL1} (which is historically the 
first succesful CEDF) and NL3* \cite{NL3*} CEDFs and their 
nuclear matter properties are included in columns 2 and 3.
The global performance of the NL3*, NL5(C), NL5(D) and
NL5(E) functionals in the description of ground state properties
of even-even nuclei is discussed in Appendix \ref{Glob-perfom}. 
\label{table-forces}
}
\begin{tabular}{|c|c|c|c|c|c|c|c|} \hline
             &     NL1      &    NL3*      &    NL5(A)   &   NL5(B)     &  NL5(C)      &  NL5(D)    &   NL5(E)    \\ \hline\hline
\multicolumn{8}{|c|}{1. Parameters}  \\ \hline
 $m_{\sigma}$ &   492.250    & 502.574200   & 516.993054  &  503.253177   & 502.481217 & 503.122989 & 503.298890 \\ \hline
 $g_{\sigma}$ &   10.1377    & 10.094400    &  10.165747  &    9.896631   &  9.900244  & 10.187753  &  10.263955 \\ \hline
 $g_{\omega}$ &   13.2846    & 12.806500    &  12.658290  &  12.457831    & 12.489590  & 12.940276  & 13.052487  \\ \hline 
 $g_{\rho}$   &   4.9757     &  4.574800    & 4.277136    &   4.202553    &  4.318575  & 4.589814   &  4.582673   \\ \hline
 $g_{2}$     &  -12.1742    & -10.809300   & -8.350509   &  -10.925997   & -10.821667 & -10.858440 & -10.976703  \\ \hline   
 $g_{3}$     &  -36.2646    & -30.148600   & -19.260373  &  -28.502727   & -28.27378  & -30.993091 & -32.006687 \\ \hline
\multicolumn{8}{|c|}{2. Nuclear matter properties}  \\ \hline
 $E/A$      &    -16.42    &   -16.31     & -16.25      &  -16.20      & -16.24      &  -16.29     & -16.27 \\ \hline
 $\rho_{0}$  &    0.152     &   0.150      & 0.146       &  0.150       & 0.150      &   0.150     &  0.150  \\ \hline
 $K_{0}  $   &    211.11   &   258.27     & 318.42       &  259.22      & 260.673    &   256.50    & 252.96  \\ \hline
 $J$        &     43.46    &   38.68      & 34.92        &  34.92       & 35.925     &   38.87     &  38.93  \\ \hline
 $L_0$      &    140.07     &   122.68     & 108.85      &  108.33      & 112.31     &   123.98    &  124.96  \\ \hline
\multicolumn{8}{|c|}{2. Penalty function contributions}  \\ \hline
 $\chi _{total}^{2}$        &     &         &   343.901    & 367.822      &  273.014   &    74.973    &   85.049  \\ \hline
 $\chi _{total}^{2}$/degree &     &         &   14.95      & 15.99        &  12.41     &    3.9459    &   3.698   \\ \hline
 $\chi _{NM}^{2}$ ($\%$)  &  &   &  8.120 ( 2.3 $\%$) &  2.626 (0.5 $\%$)   &  3.842 (1.1 $\%$)   & 4.434 (5.3 $\%$) &  3.625  (3.5 $\%$) \\ \hline 
 $\chi _{E}^{2}$ ($(\%$)  &  &       &128.318 (37.3 $\%$) &145.727 (39.6 $\%$)  &111.550 (40.9 $\%$)  &55.221 (73.7 $\%$) & 51.231 (60.2 $\%$)\\ \hline
 $\chi _{Rch}^{2}$ ($\%$) &  &       & 34.231 ( 9.9 $\%$) & 18.363 (5.0 $\%$)   & 16.124 (5.9 $\%$)   &15.318 (20.4 $\%$) & 16.802 (19.7 $\%$)\\ \hline
 $\chi _{Nskin}^{2}$ ( $\%$)  &  &    &173.235 (50.4 $\%$) &201.105 (54.6 $\%$)  &141.498 (51.8 $\%$)  &  0    (0.0 $\%$)  & 13.390 (15.7 $\%$)\\ \hline
 $\chi _{Nskin(Zr)}^{2}$($\%$) &  &    & 76.071 (22.1 $\%$) & 86.591 (23.5 $\%$)  &  0     (0.0 $\%$)   &  0     (0.0 $\%$)  & 1.178 (1.4 $\%$)\\ \hline
\end{tabular}
\end{center}
\end{table*}

  The starting point is the fitting protocol of the NL3* functional 
(see Ref.\ \cite{NL3*}). The types of the input data for this protocol 
and related adopted errors are summarized in column 2 of Table 
\ref{table-inp}. The minimization within this protocol leads to
optimum functional labeled NL5(A) (see Table \ref{table-forces}).
When considering the quality of the 
functional we take into account the ranges of the nuclear matter 
properties recommended for relativistic functionals in Ref.\ 
\cite{RMF-nm}. These are $\rho_0 \sim 0.15$ fm$^{-3}$, $E/A \sim -16$ MeV,
$K_0= 190-270$, $J=25-35$ MeV ($J=20-35$ MeV) and $L_0=25-115$ 
($L_0=30-80$) for the SET2a (SET2b) sets of the constraints on
the experimental/empirical ranges for the quantities of interest.
As compared with the CEDF NL3*, the NL5(A) functional has better 
$J$ and $L_0$ values but much worse $K_{0}$ value. The analysis of the 
contributions of physical observables of different classes 
shows that $\chi^2_{total}$ is dominated by the contributions from
4 data points on neutron skins with the contribution of the single
data point on the neutron skin of $^{90}$Zr providing 22.1\% of
$\chi^2_{total}$. These large contributions from neutron skin data
to $\chi^2_{total}$ clearly illustrate that these functionals 
cannot accurately describe presently adopted experimental values
of $r_{skin}$ obtained by means of hadronic probes\footnote{There 
is a significant contraversy in 
the adopted experimental values of neutron skins (see discussion
in Sect.\ \cite{AARR.14} and in Ref.\ \cite{PREX.12}). For example, 
the experiments based on hadronic probes provide neutron skin in 
$^{208}$Pb around 0.2 fm or slightly smaller. However, these 
experimental data are extracted in model-dependent ways. Alternatively, 
a measurement using an electroweak probe has been carried 
out in parity violating electron scattering on nuclei (PREX) and it 
brings $r_{skin}=0.33\pm 0.17$ \cite{PREX.12}. A central value of 0.33 
fm is particularly intriguing because it is around 0.13 fm higher 
than central values obtained in other experiments. The electroweak 
probe has the advantage over experiments using hadronic probes 
that it allows a nearly model-independent extraction of the 
neutron radius that is independent of most strong interaction 
uncertainties \cite{PREX-CREX}. Note that non-linear CEDFs typically 
give $r_{skin} \sim 0.3$ fm, so if the central value $r_{skin}\sim 0.33$
obtained in PREX experiment would be confirmed in future PREX-2 
experiment \cite{PREX-CREX}, this would lead to substantial 
reduction of $\chi^2_{skin}$.}.

 To compensate for too large value of $K_{0}$ and to force it
to more acceptable value, the adopted error for $K_{0}$ has been 
reduced from 10\% to 2.5\% in the fitting protocol of the NL5(B) 
functional (see column 3 of Table \ref{table-inp}). This functional 
and related nuclear matter properties as well as penalty function 
contributions are shown in Table \ref{table-forces}. Its $E/A$, 
$\rho_0$ and $K_{0}$ values are close to the NL3* ones, but it 
has better symmetry energy $J$ and the slope of symmetry energy 
$L_0$. However, similar to NL5(A) it suffers from too large 
contribution of neutron skins (and especially, the one
coming from $^{90}$Zr) to $\chi^2_{total}$.

  To reduce this problem, the neutron skin of $^{90}$Zr has been 
dropped from the fitting protocol of the NL5(C) functional\footnote{The
same procedure has been employed in the fitting protocol of the
DD-ME2 functional in Ref.\ \cite{DD-ME2}.} (see column 
4 in Table \ref{table-inp}.) This functional has $E/A$, $\rho_0$ and $K_{0}$ 
values similar to the ones of the NL3* and NL5(B) functionals
(see Table \ref{table-forces}). Its $J$ and $L_0$ values are better 
than those of the NL3* functional but slightly worse as compared with 
those of the NL5(B) one. However, the NL5(C) functional provides better 
description of binding energies and charge radii as compared with the 
NL5(A) and NL5(B) ones; these are physical observables which are precisely 
measured in experiment. On the other hand, it gives slightly worse description
of neutron skins in the $^{116}$Sn, $^{124}$Sn and $^{208}$Pb nuclei but 
as mentioned before this physical observable is characterized by 
substantial experimental uncertainties.

   One can consider removing the neutron skins (as least reliable
experimental data on finite nuclei) from the fitting protocol;
this leads to the NL5(D) functional (see Table \ref{table-forces}).
It is characterized by a substantial reduction (by a factor of 2 as 
compared with the CEDF NL5(C)) of the error in the reproduction of 
experimental data on binding energies. However, for this functional 
the $J$ and $L_0$ values deviate more from recommended values as 
compared with the NL5(C) one.

  Alternatively, one can use experimental errors (from Refs.\ 
\cite{PhysRevLett.82.3216,Nskin.90Zr,Nskin.208Pb}) 
as adopted errors for neutron skins which are substantially 
larger than adopted errors (5\%) of the fitting protocol of the
NL5(A-C) functionals (Table  \ref{table-inp}). This leads to 
the NL5(E) functional in which the impact of neutron skins on total 
$\chi^2_{total}$ is substantially reduced as compared with NL5(A-C) 
functionals. As a consequence, the nuclear matter properties of the 
NL5(D) and NL5(E) functionals are similar and they provide 
comparable description of binding energies and charge radii.

  The results presented here clearly show that the selection of 
the fitting protocol (physical observables and related adopted
errors) is to a degree subjective. Definitely the use of
more experimental data of different types is expected to reduce
this level of subjectivity but it cannot be completely 
eliminated. Since the Monte-Carlo analysis of statistical 
errors is  numerically extremely time-consuming, we restrict our
investigation of statistical errors of non-linear CEDFs to
NL5(C) and NL5(A) functionals.

\begin{figure*}
\centering
\includegraphics[width=16.0cm]{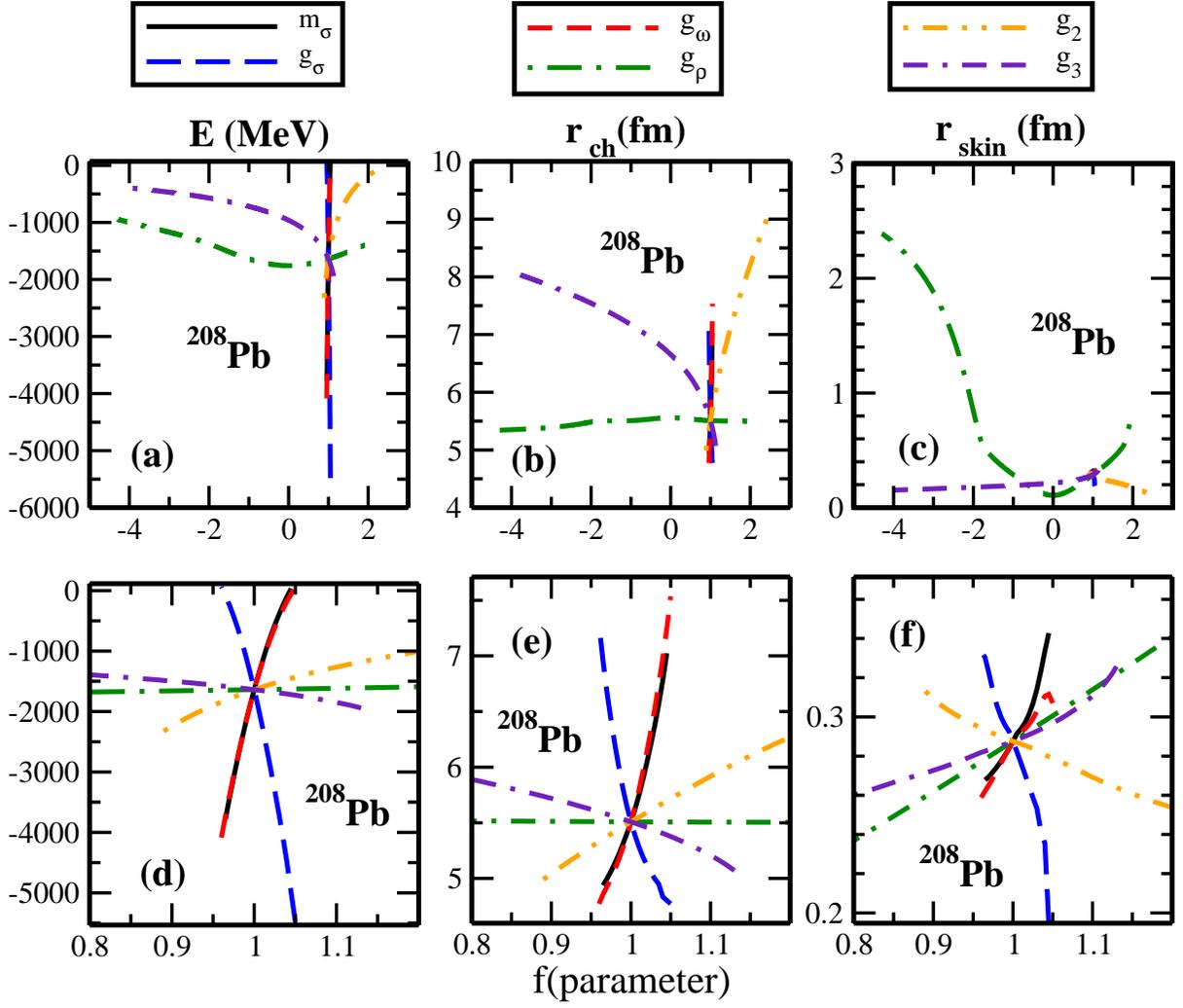}
\caption{(Color online) The range of the variations of the parameters 
of the NL5(C) CEDF and related changes of physical observables such as total 
binding energy $E$ (left columns), charge radius $r_{ch}$ (central columns) 
and neutron skin $r_{skin}$ (right columns). Upper row shows the full range 
of the parameter variations, while the bottom panels magnify the region in 
the vicinity of optimum parametrization ($f_i=1.0$). On the one end, the 
range of parameters is limited by the condition that the total energy of 
the nucleus is negative. On the other end, it is defined by the collapse 
of numerical solution due to underlying numerical instabilities. For each 
line, only the indicated parameter is changing while the remaining parameters 
are kept at the values corresponding to optimum NL5(C) functional.
}
\label{Paramet-range}
\end{figure*}

\section{The ranges of the parameters and parametric 
         correlations in the functionals}
\label{sect-param}

  In meson exchange models the general features of the nuclei 
are dominated by the properties of the $\sigma$ and $\omega$ mesons 
which are responsible for intermediate range attraction and short 
range repulsion. They lead to attractive scalar $S$ and repulsive 
vector $V$ potentials with magnitudes of $S\approx -400$ MeV/nucleon 
and $V\approx +350$ MeV/nucleon \cite{Rei.89,Ring1996_PPNP37-193}. 
Their sum defines the depth of nucleonic potential ($\sim -50$ 
MeV/nucleon).

   Because of these large magnitudes, very small variations of the 
masses and coupling constants of these two mesons lead to substantial 
changes in binding energies (see Figs.\ \ref{Paramet-range}a and d). 
Note that in this chapter instead of functional parameters $par_i$ we 
are using the ratio 
\begin{equation}
f_i=\frac{par_i}{par_i^{opt}}
\end{equation}
where $par_i^{opt}$ is the value of the parameter in the optimum 
functional and $i$ indicates the type of the parameter. This 
allows to better understand the range of the variations of the 
parameters and related parametric correlations in the functionals.

  Coming back to Figs.\ \ref{Paramet-range}a and d, one can see 
that $\pm 5$\% change in the values of $m_{\sigma}$, $g_{\sigma}$ and
$g_{\omega}$ leads to the changes of binding energies in the range 
of 2000-3000 MeV. Other physical observables used in the fitting 
protocols such as charge radii $r_{ch}$ also sensitevely depend on 
$f_i$ (see Figs.\ \ref{Paramet-range}b and e). However, some 
flexibility in acceptable ranges of these parameters is provided 
by the fact that binding energies and charge radii have different 
dependencies on $f(m_\sigma)$, $f(g_\sigma)$ and $f(g_{\omega})$
(see Fig.\ \ref{Paramet-range}).

\begin{figure*}
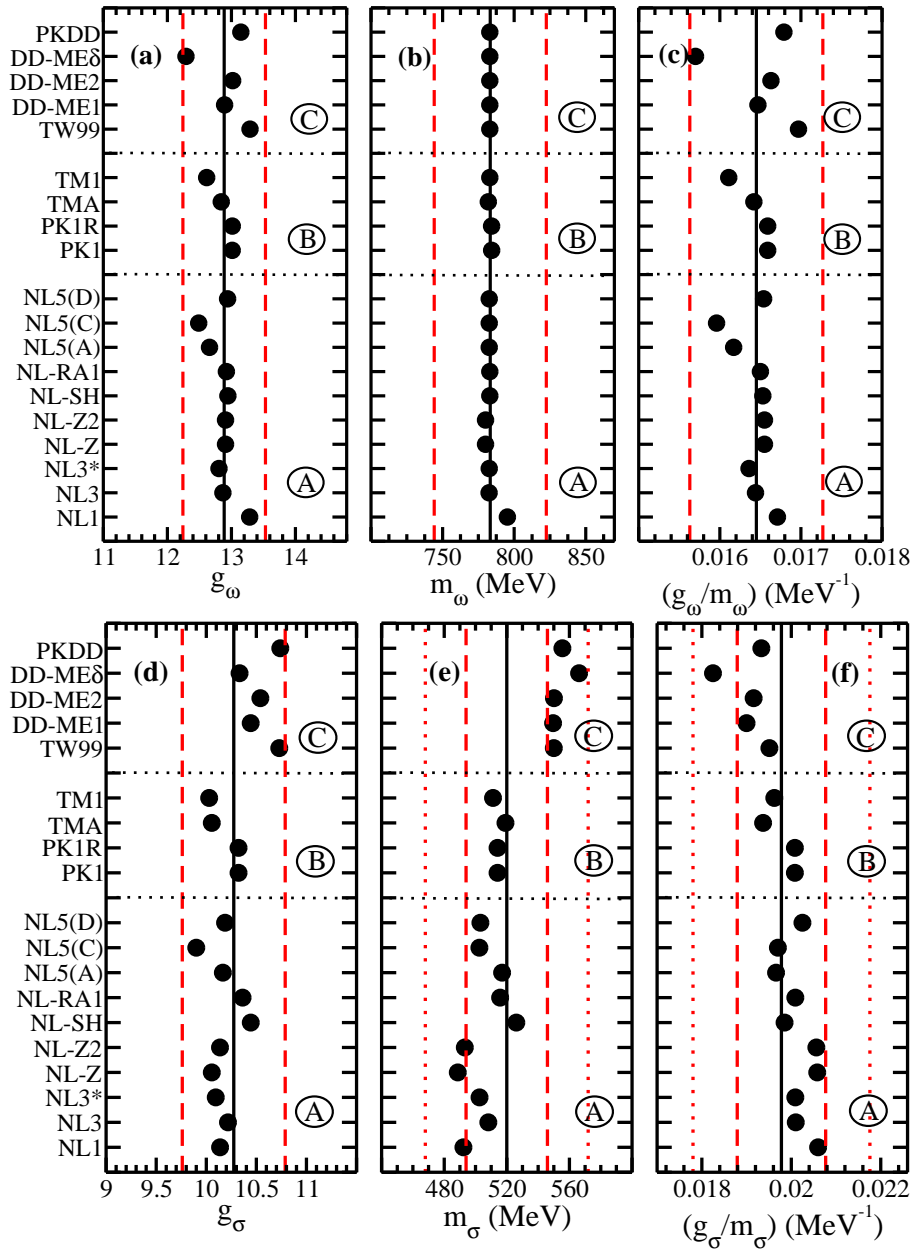

\centering
\includegraphics[width=12.0cm]{fig-2-a.eps}
\includegraphics[width=12.0cm]{fig-2-b.eps}
\caption{(Color online) The masses and coupling constants of the $\sigma$ and $\omega$ mesons 
in different CEDFs which contain meson exchange. They are combined into three 
groups dependent on how self- and mixed-couplings are introduced.  Group A represents 
the parametrizations which include non-linear self-couplings only for the $\sigma$-meson.  
Group B contains the parametrizations which include self-couplings for the $\sigma$- 
and $\omega$-mesons (and $\rho-$mesons in the case of PK1R). Group C represents the 
parametrizations which include density-dependent meson-nucleon couplings for the 
$\sigma$-, $\omega$-, and $\rho$-mesons. Note that the mass $m_{\omega}$ of the $\omega$-meson
is fixed at indicated values in all functionals except NL3, PK1 and PK1R. The parameters 
are taken from Refs.\ \cite{NL1} (NL1), \cite{NL3} (NL3), \cite{NL3*} (NL3*), 
\cite{NLZ} (NL-Z), \cite{BRRMG.99} (NL-Z2), \cite{NLSH} (NL-SH),
\cite{NLRA1} (NL-RA1), \cite{PK1-PK1R-PKDD} (PK1,PK1R), \cite{TMA} (TMA), \cite{TM1} (TM1),
\cite{TW.99} (TW99), \cite{DD-ME1} (DD-ME1), \cite{DD-ME2} (DD-ME2), \cite{DD-MEdelta}
(DD-ME$\delta$), \cite{PK1-PK1R-PKDD} (PKDD). Note that we omitted mass-dependent terms for 
$g_{\omega}$ in the TMA parametrization which is a good approximation for heavy nuclei since 
$g_{\omega}=12.842+3.191A^{-0.4}$ \cite{TMA}. Vertical solid lines represent the mean values of 
the respective parameter over the set of indicated functionals. Red dashed and dotted lines 
show the $\pm 5$\% and $\pm 10$\% deviation bands with respect of these mean values.}
\label{Paramet}
\end{figure*}
 
  Above discussed features lead to the fact that the $m_{\sigma}$, 
$g_{\sigma}$ and $g_{\omega}$ parameters are well localized in the 
parameter hyperspace of all meson-exchange CEDFs (see Fig.\ \ref{Paramet}). 
Note that the absolute majority of these parameters are located 
within 5\% deviation band with respect to mean value. This similarity 
between the functionals becomes even more striking if we consider 
the ratios $\frac{g_i}{m_i}$ 
(see Figs.\ \ref{Paramet}c and f). In reality, many physical 
observables  depend on such ratios in the CDFT framework. For 
example, the vector and scalar fields of the CDFT are proportional 
to $\left( \frac{g_{\omega}}{m_{\omega}} \right)^2$ and 
$\left( \frac{g_{\sigma}}{m_{\sigma}} \right)^2$ in the lowest order \cite{Rei.89}.
Another example is the equation of the state of nuclear matter which depends
on the ratios $\frac{g_i}{m_i}$ \cite{TW.99}. Effective meson-nucleon 
coupling in nuclear matter is also determined by such ratios \cite{DD-ME2}.

  This is a unique feature of the CDFT not present in the non-relativistic DFTs. 
The later ones are characterized by a substantial spread of all parameters in 
optimum parametrizations (see discussion in Appendix \ref{param_spread}).

  On the contrary, the impact of the terms which  define isovector 
dependence, such as the $\rho$-meson, and density-dependent terms 
(such as $g_2$ and $g_3$) on total binding energies and charge 
radii is substantially smaller (see Fig.\ \ref{Paramet-range}a, b, d and e). 
For example, to get comparable changes in binding energy of $^{208}$Pb, 
the changes in the $g_2$, $g_3$ and especially $g_{\rho}$ parameters 
have to be substantially larger than those for the $m_{\sigma}$, $g_{\sigma}$ 
and $g_{\rho}$ parameters (Table \ref{table-factor}\footnote{The results
presented in Table \ref{table-factor} illustrate extreme sensitivity
of the results to precise value of the parameters. This is especially
true for the parameters related to the $\sigma$ and $\omega$ mesons. This
is a reason why in Table \ref{table-forces} all the parameters are 
given with six significant digits.}).  In general, the 
$g_\rho$ parameter has significantly larger impact on neutron skin than 
other parameters (see Fig.\ \ref{Paramet-range}c). However, in the 
vicinity of the optimum functional its impact is comparable with the 
ones of other parameters (see Fig.\ \ref{Paramet-range}f). Note 
also that the potential range of the variations of the $g_{\rho}$, $g_2$ 
and $g_3$ parameters is substantially larger as compared with the range 
of variations of the $g_{\sigma}$, $m_{\sigma}$ and $g_{\omega}$ parameters 
(see  Fig.\ \ref{Paramet-range}).

\begin{figure}
\centering
\includegraphics[width=6.0cm]{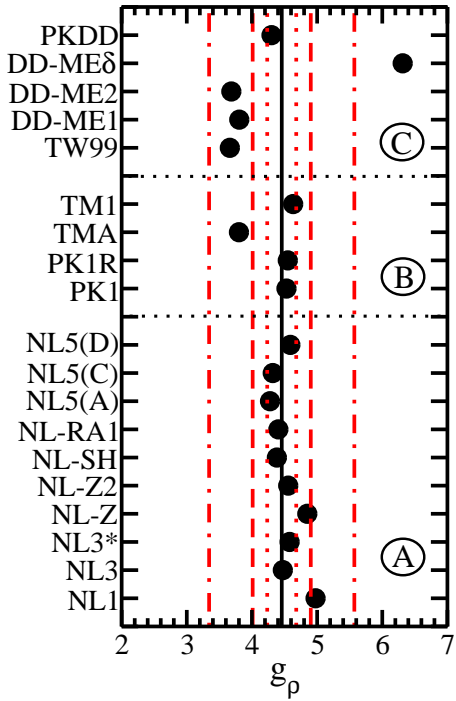}
\caption{(Color online) The same as Fig.\ \ref{Paramet} but 
for the $g_{\rho}$ parameter. Red dotted, dashed and dash-dotted 
lines show the $\pm 5$\%, $\pm 10$\% and $\pm 25$\% deviation 
bands with respect of mean values.}
\label{Paramet-grho}
\end{figure}

\begin{figure*}
\centering
\includegraphics[width=12.0cm]{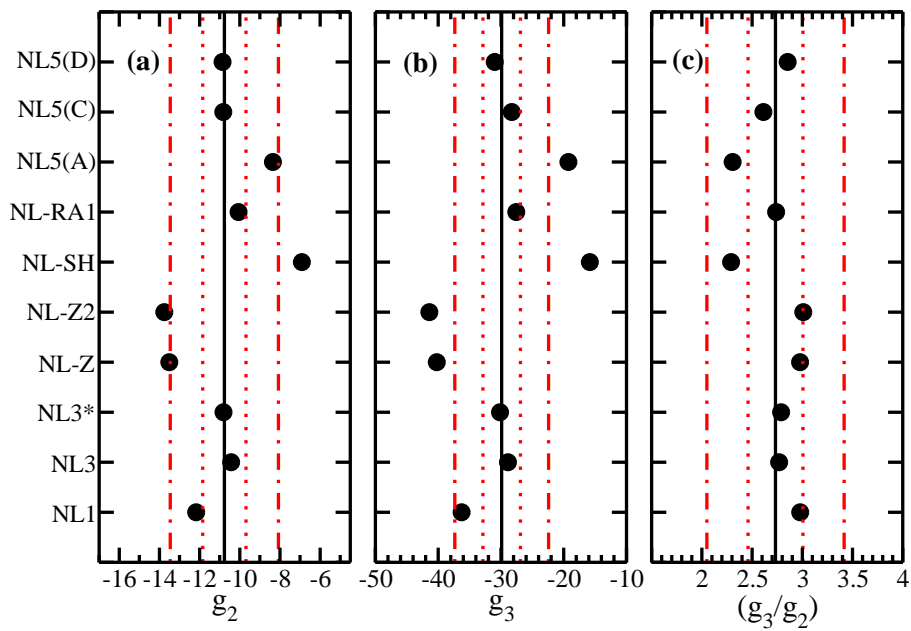}
\caption{(Color online) The same as Fig.\ \ref{Paramet} but for the 
$g_2$ and $g_3$ parameters. Red dotted and dash-dotted lines show the 
$\pm 10$\% and $\pm 25$\% deviation bands with respect of the mean 
values. Note that only the functionals which belong to group A are 
considered here.}
\label{Paramet-g2-g3}
\end{figure*}

 Fig.\ \ref{Paramet-grho} shows that the level of localization
of the $g_{\rho}$ parameter in the parameter hyperspace is lower
as compared with the parameters of the $\sigma$ and $\omega$ 
mesons. The largest deviations from the mean $g_{\rho}$ value, 
defined over the set of considered functionals, are seen for the 
models with explicit density dependencies (Group C functionals).
However, even for group A functionals, the deviations from mean 
values reach 10\%. The level of localization is even lower for 
the $g_2$ and $g_3$ parameters for which the deviations from the
mean values (defined for the set of considered functionals) 
could reach and even exceed 25\% limit (see Fig.\ \ref{Paramet-g2-g3}). 
However, it is interesting that for the considered functionals the 
$g_3/g_2$ ratio is very close to 2.75 (see Fig.\ \ref{Paramet-g2-g3}c). 
Only two functionals, namely, NLSH and NL5(A), exceed 10\% deviation 
band from the mean value for the $g_3/g_2$ ratio. Considering that 
these two parameters define the density dependence of the non-linear
meson coupling model, this consistency of the $g_3/g_2$ ratio over 
the studied functionals suggests hidden parametric correlations 
between the $g_2$ and $g_3$ parameters.

\begin{table}[ht]
\begin{center}
\caption{The factor $g$ by which one has to multiply the
indicated parameter of the NL5(C) functional in order to 
get 10 MeV increase in binding energy of $^{208}$Pb. Note that 
other parameters of the functional are kept at their original
values.
\label{table-factor}
}
\begin{tabular}{|c|c|} \hline 
     Parameter           &   Factor $g$     \\ \hline         
   $m_{\sigma}$            &     0.999801       \\ 
   $g_{\sigma}$            &     1.000170     \\ 
   $g_{\omega}$            &     0.999796     \\ 
   $g_{\rho}$             &      0.992981       \\ 
   $g_{2}$                &     1.002189      \\ 
   $g_{3}$                &     0.997833       \\  \hline
\end{tabular}
\end{center}
\end{table}

\begin{figure*}[htb]
\includegraphics[angle=-90,width=8.1cm]{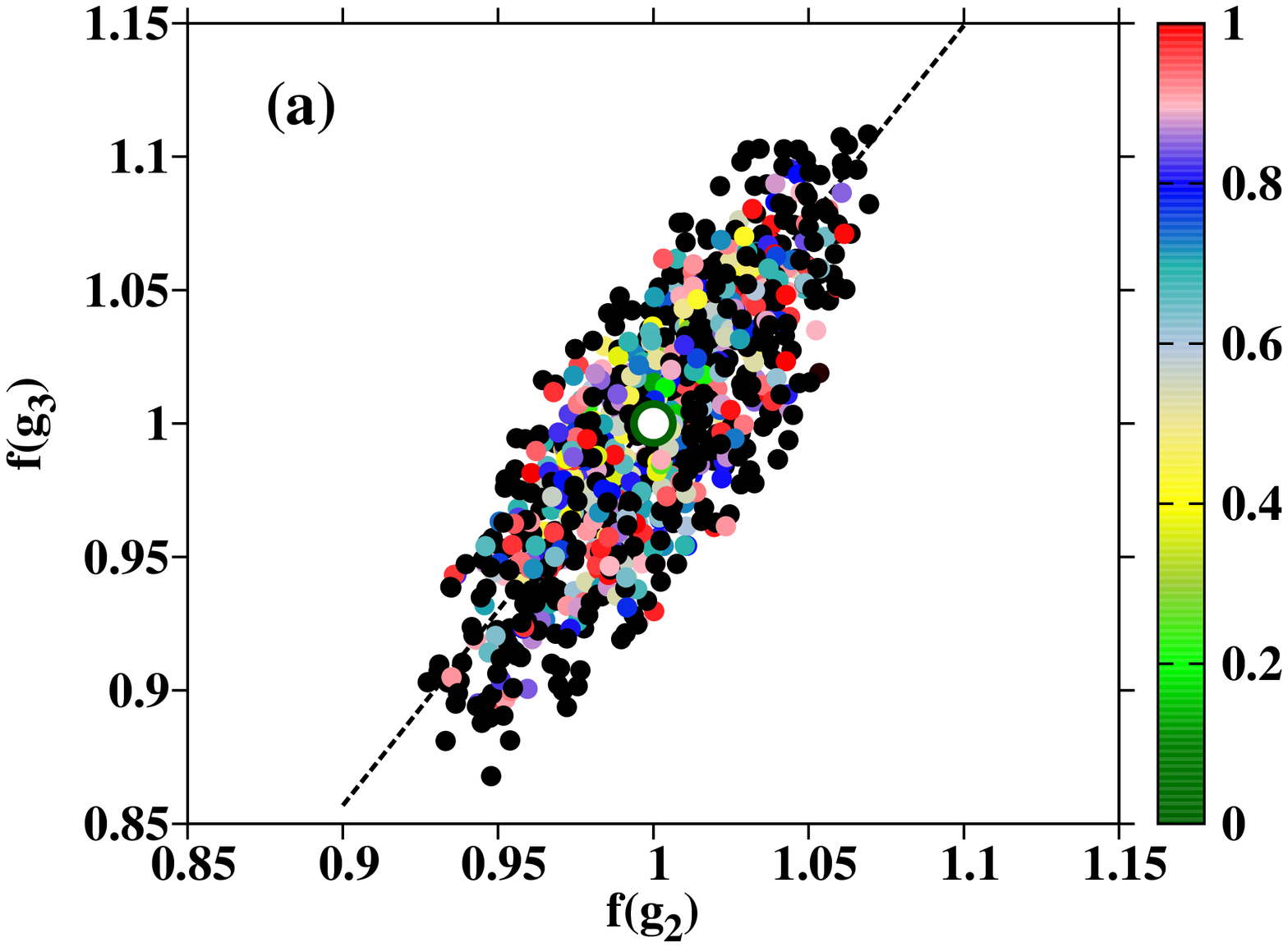}
\includegraphics[angle=-90,width=8.5cm]{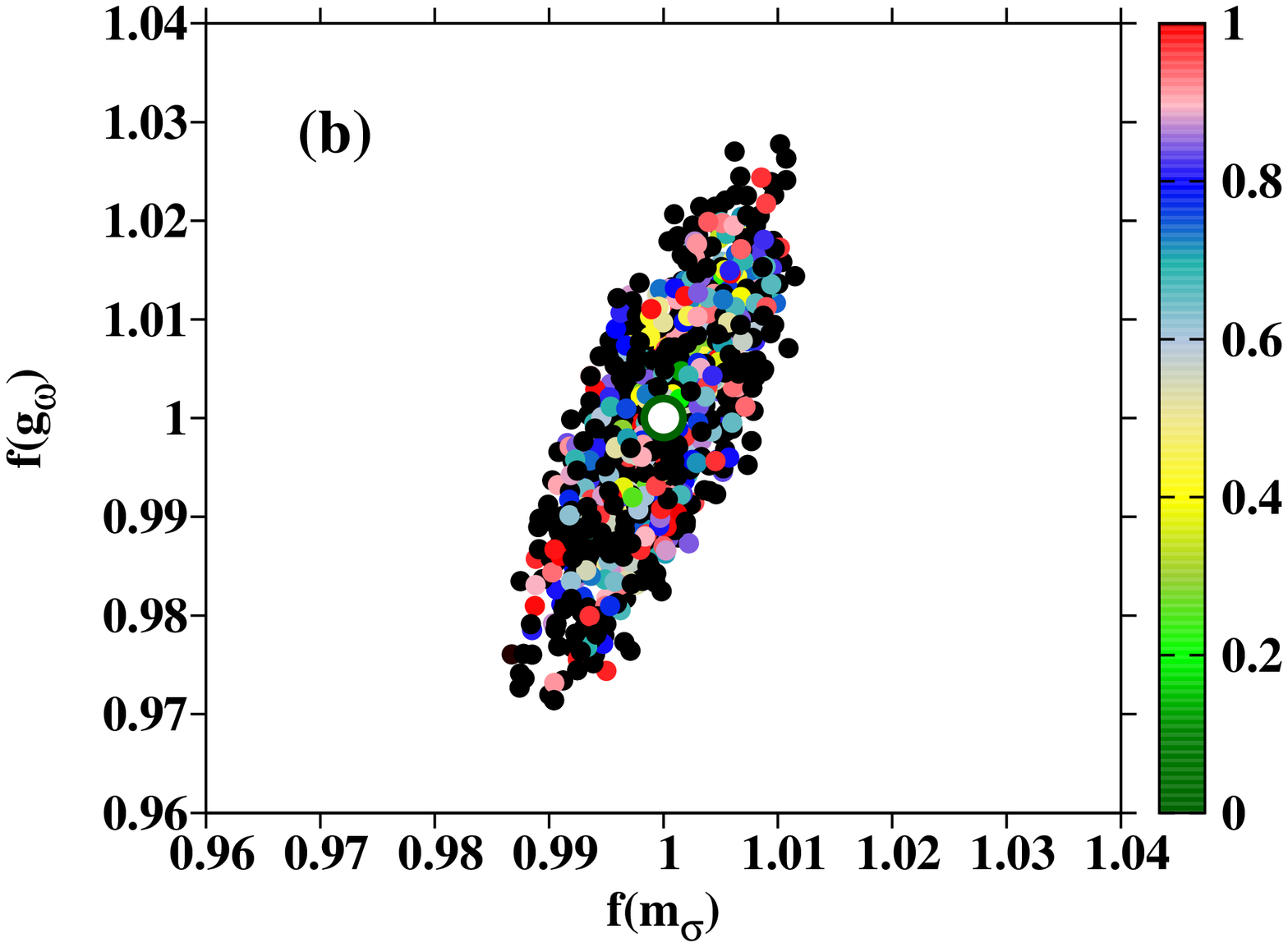}
\includegraphics[angle=-90,width=8.5cm]{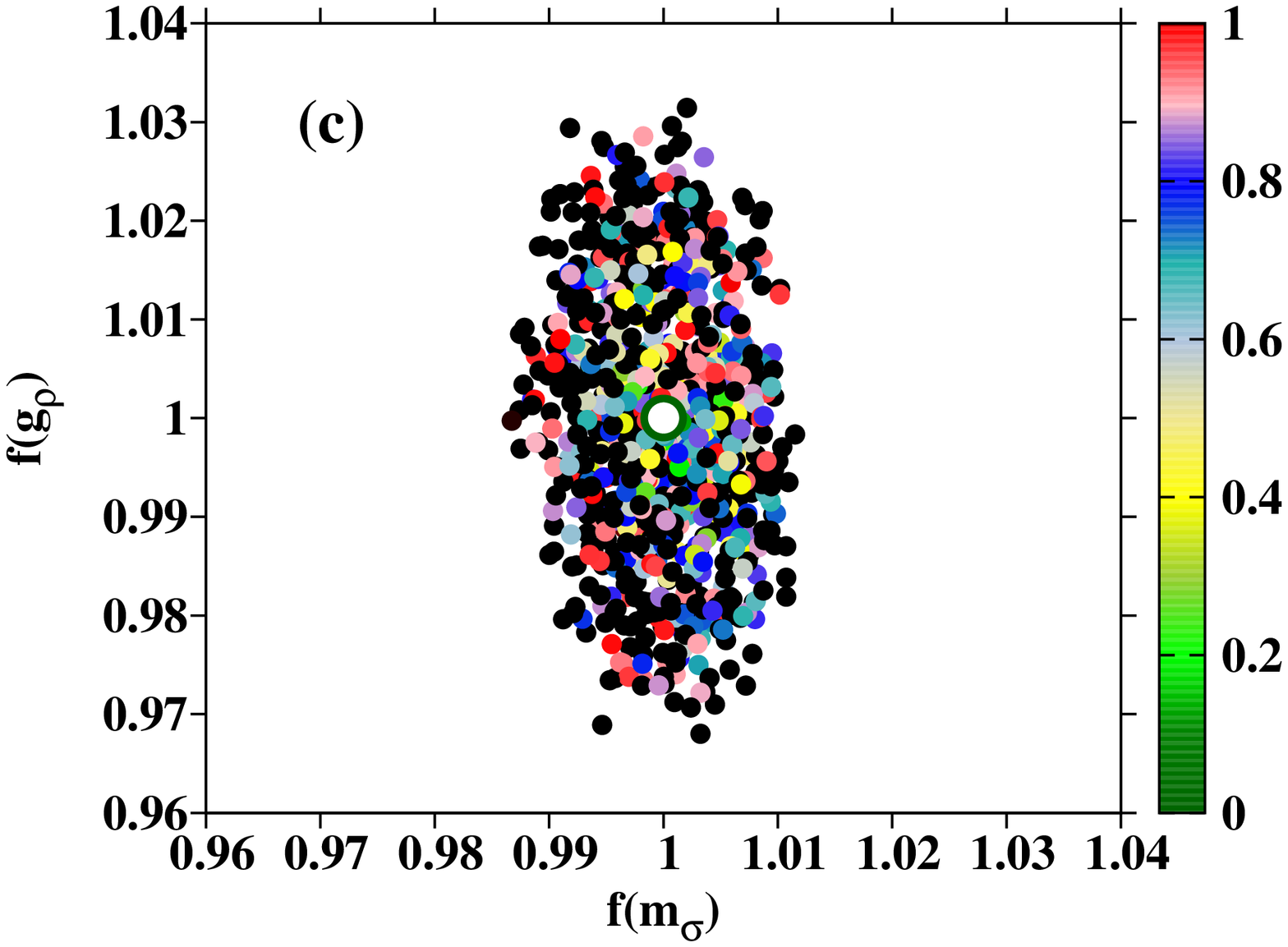}
\includegraphics[angle=-90,width=8.5cm]{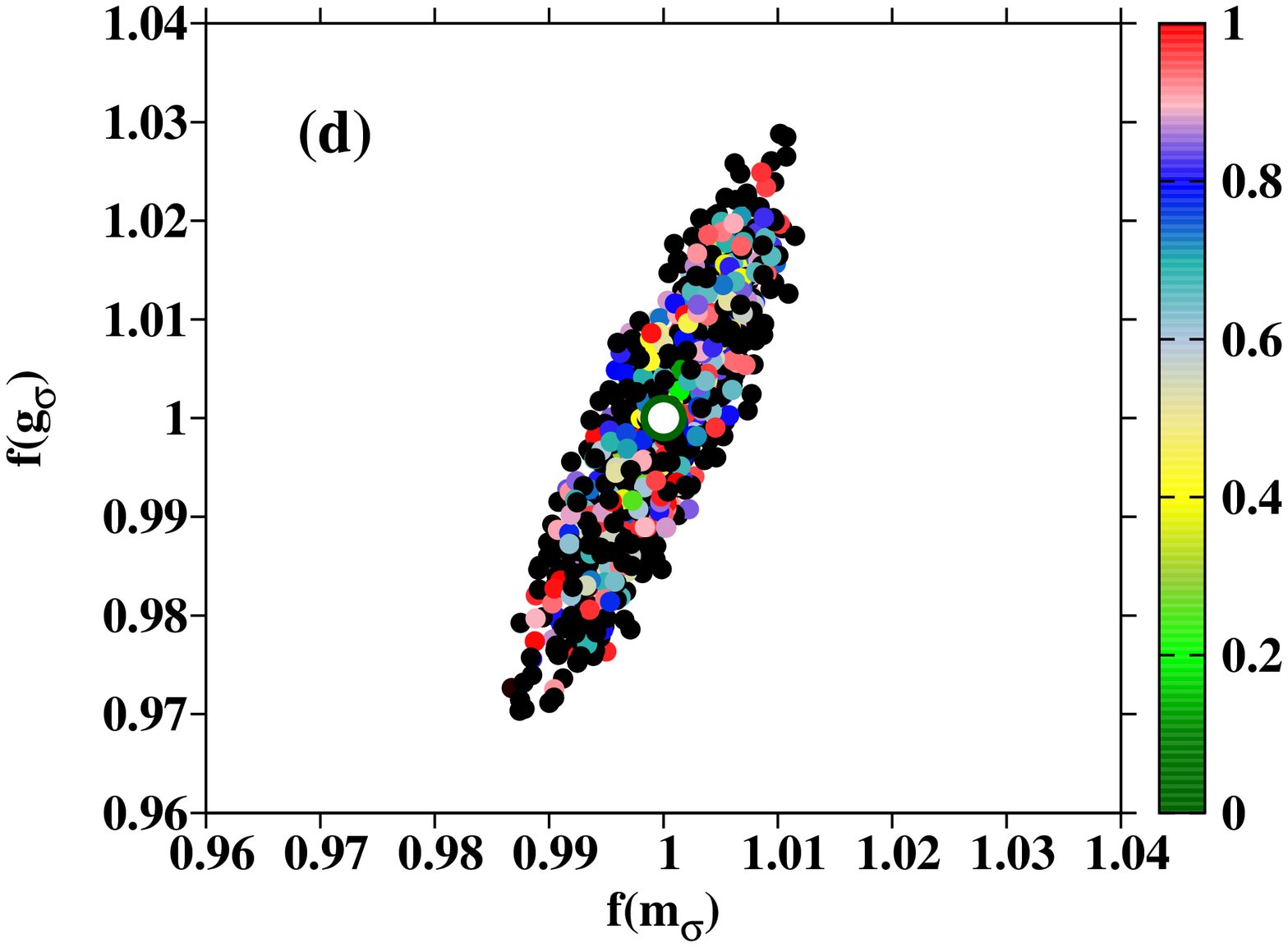}
\caption{ (Color online) Two-dimensional projections of the distribution 
of the functional variations in the 6-dimensional parameter hyperspace. 
The colors indicate the $a$ value of the $\chi^2_{norm}({\bf p})$ of the 
functional variation where the latter is expressed as $\chi^2_{norm}({\bf p}) 
= \chi^2_{norm}({\bf p}_0) + a$. The colormap is used for the functional
variations with $a\leq 1.0$; there are 500 such variations. Note that it
was found in the studies that further increase of the number of functional
variations does not change the results for statistical errors.
In addition, black circles are used for the functionals with $1.0 < a\leq 1.25$.
The optimum functional is shown by large open symbol. The dashed line in 
panel (a) shows the parametric correlations between $g_2$ and $g_3$ 
parameters defined by Eq.\ (\ref{g2g3-dep}).
}
\label{NL5-ksi-distrib}
\end{figure*}

\begin{figure*}[htb]
\includegraphics[angle=-90,width=8.1cm]{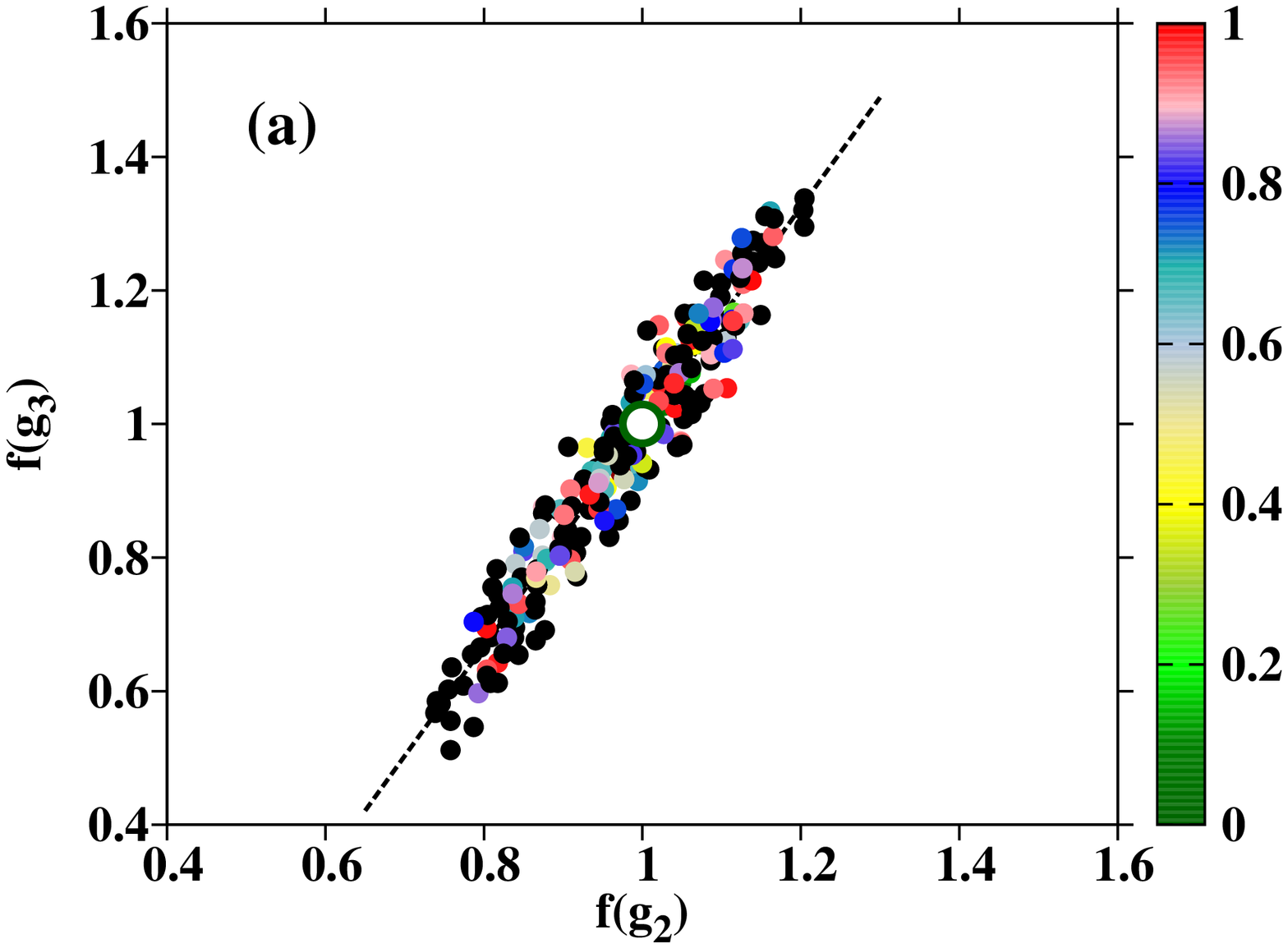}
\includegraphics[angle=-90,width=8.5cm]{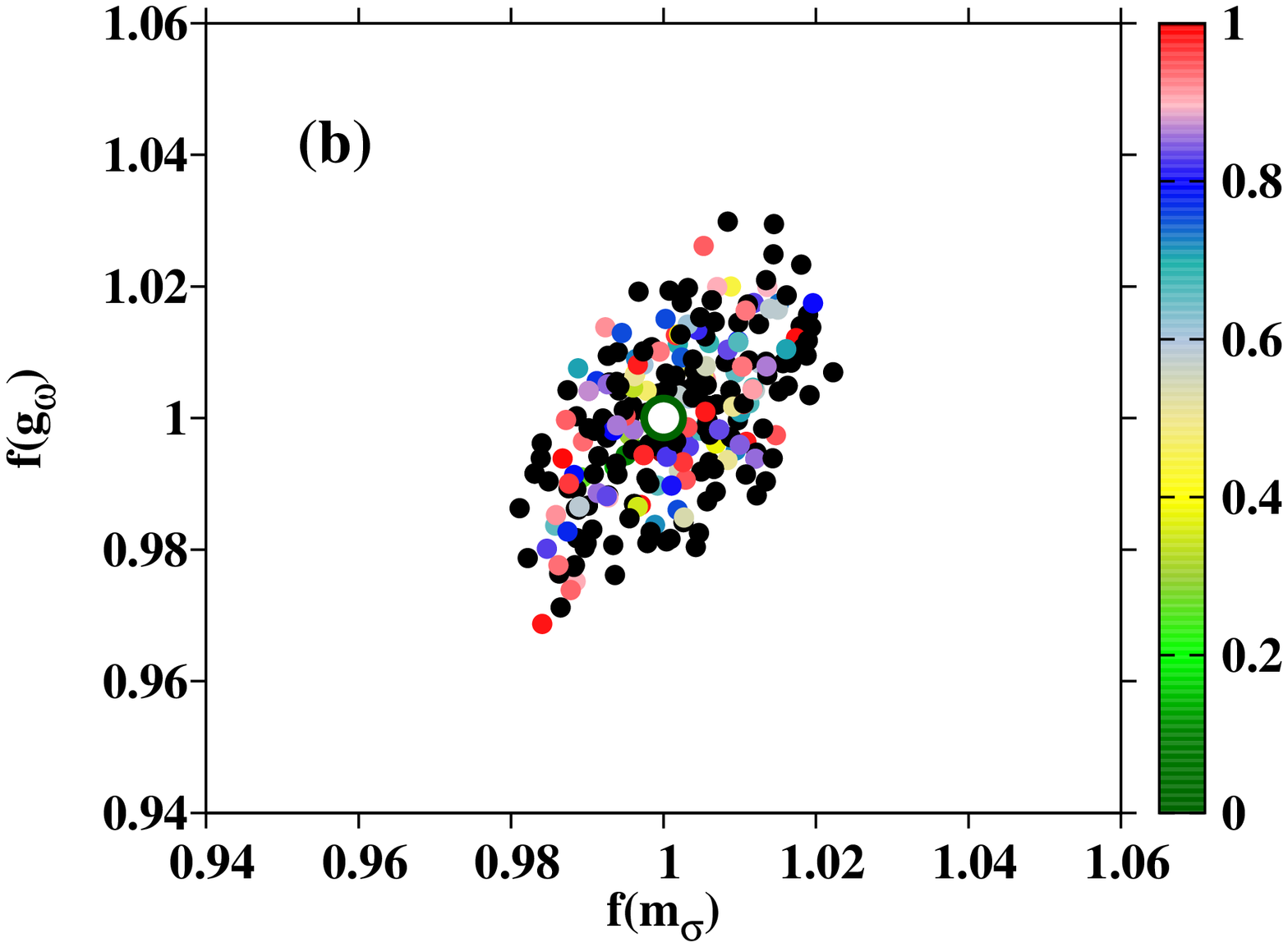}
\includegraphics[angle=-90,width=8.5cm]{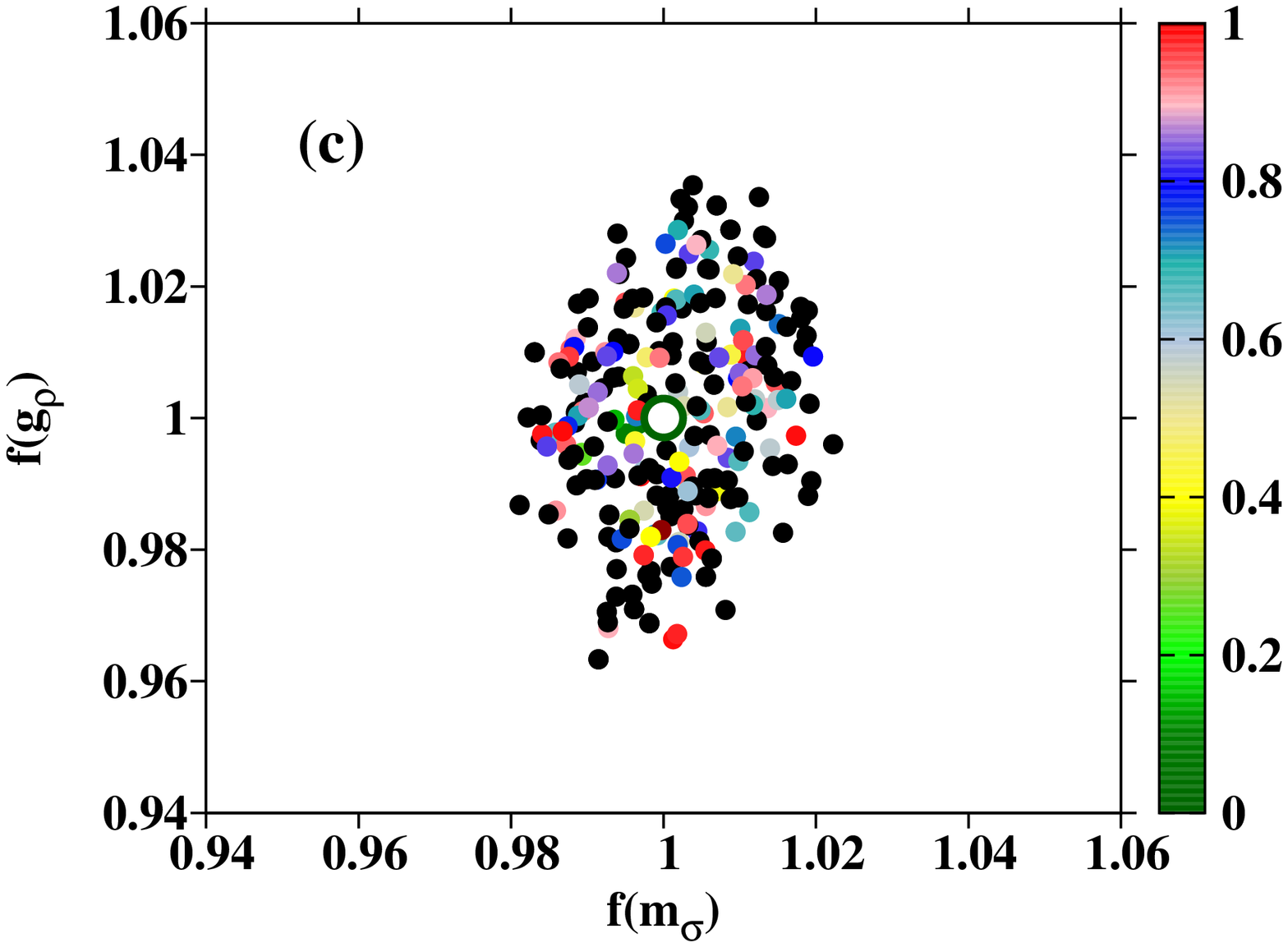}
\includegraphics[angle=-90,width=8.5cm]{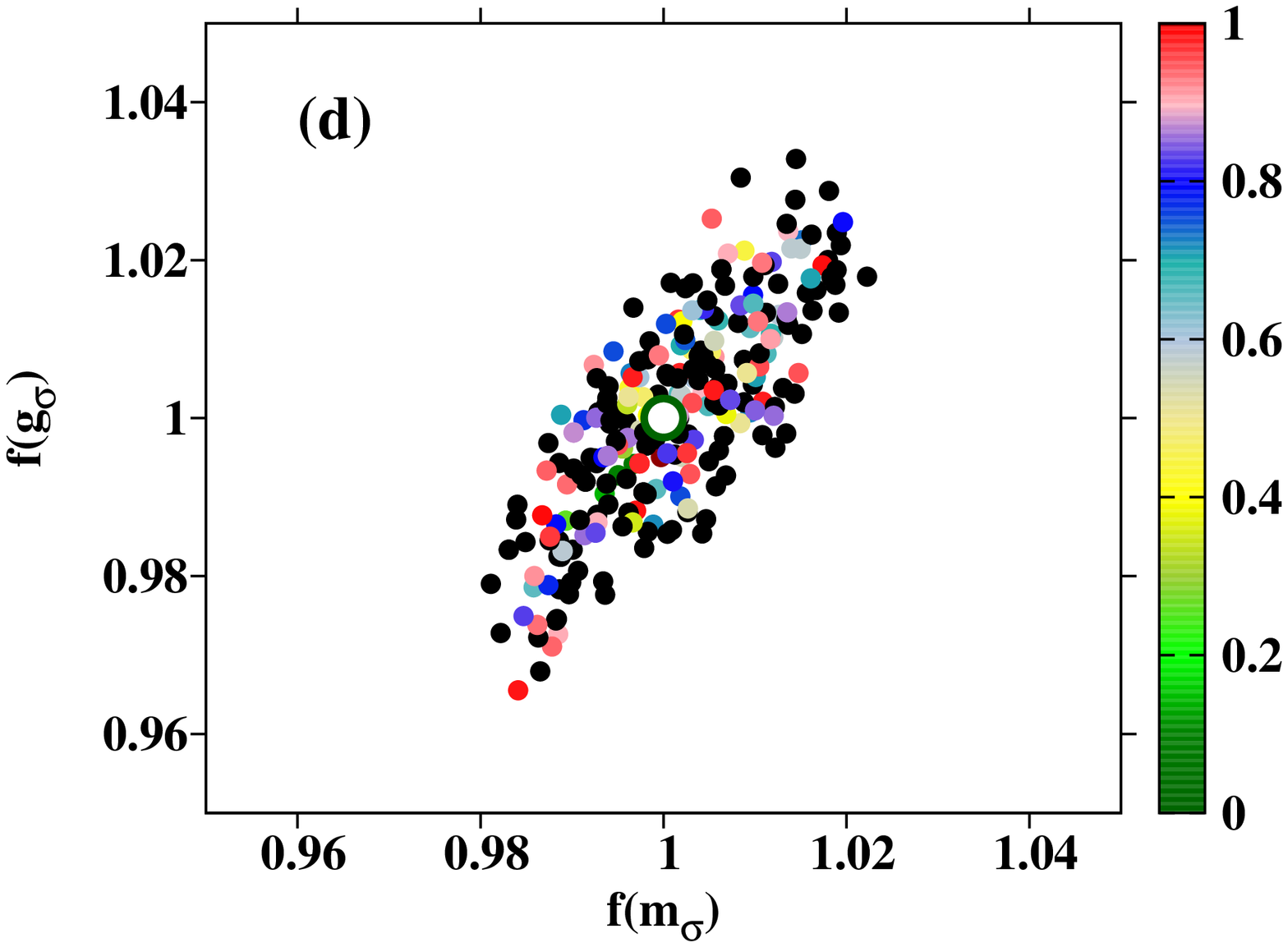}
\caption{(Color online) The same as Fig.\ \ref{NL5-ksi-distrib} 
but for the NL5(A) functional. Considering the increase of the size 
of the parameter hyperspace and extremely time-consuming nature of
numerical calculations only 150 functional variations satisfying 
the condition of Eq.\ (\ref{cond}) have been collected. This leads 
to some reduction of the accuracy of the calculations of statistical 
errors. However, the analysis of the NL5(C) results for which 500 
acceptable functional variations have been collected allows to 
conclude that statistical errors obtained with the 150 functional 
variations differ from those calculated with 500 variations only by 
approximately 5\%.}
\label{NL5(A)-ksi-distrib}
\end{figure*}

  The 2-dimensional distributions of the parameters of the 
acceptable functional variations for the NL5(C) CEDF are presented 
in Fig.\ \ref{NL5-ksi-distrib}. The parameters vary with respect of 
the central value of the distribution (which are typically given by 
the parameters of optimum functional) by at most 1.5\% for $m_\sigma$, 
3\% for $g_\sigma$, 3\% for $g_{\omega}$, 3\% for $g_{\rho}$, 7\% for 
$g_2$ and 10\% for $g_3$. However, these ranges of the parameter variations 
are dependent on the details of the fitting protocol. For example, in the NL5(A) 
functional  these ranges of the parameter variations are substantially 
larger for the $g_2$ and $g_3$ parameters for which they are around 
20\% and 40\%, respectively (see Fig.\ \ref{NL5(A)-ksi-distrib}a). 
These larger ranges for the $g_2$ and $g_3$ parameters in the acceptable 
NL5(A) functionals as compared with the NL5(C) ones are the consequence of the 
4-fold increase of adopted error for $K_0$ (from 2.5\% up to 10\% 
[see Table \ref{table-inp}]).

The analysis of these distributions can also provide the information on 
the presence of non-linear effects which are related either to non-linear 
dependencies of the observables on the coupling constants or complicated 
structure of the $\chi^2$ hypersurface exibiting several separated local 
minima \cite{BMR.04}. The position of the optimum functional in the plots 
of Figs.\ \ref{NL5-ksi-distrib} and \ref{NL5(A)-ksi-distrib} corresponds 
to the crossing point of the $f(par_i)=1.0$ and $f(par_j)=1.0$ lines. In 
absolute majority of the cases the distributions shown in Figs.\ 
\ref{NL5-ksi-distrib} and \ref{NL5(A)-ksi-distrib} have ellipsoid-like 
shapes with central point of the distribution at this crossing point. 
This means that non-linear effects are not affecting these distributions. 
The only exception is the case of the $f(g_3)-f(g_2)$ distribution for 
the NL5(A) functional in which the optimum functional is located off
the center of the distribution (see Fig.\ \ref{NL5(A)-ksi-distrib}a).

   In addition, the correlations between the parameters of the functional
can be easily defined from  these distributions.  Strong correlations 
between the $m_\sigma$ and $g_\sigma$ parameters are clearly visible in 
Figs.\ \ref{NL5-ksi-distrib}d and \ref{NL5(A)-ksi-distrib}d.
The same is true for the pair of the $m_\sigma$ and $g_\omega$ parameters
(see Figs.\ \ref{NL5-ksi-distrib}b and \ref{NL5(A)-ksi-distrib}b).
This is not surprising since these parameters enter the definition of the 
nucleonic potential via the sum of attractive scalar potential 
$S$ (which depends on $m_\sigma$ and $g_\sigma$) and repulsive vector potential 
$V$ (which depends on $m_\omega$ and $g_\sigma$) \cite{VALR.05}. 
Note that $m_\omega$ is fixed at $m_\omega=782.6$ MeV (see Table 
\ref{table-forces}). 
It is interesting to mention that these strong correlations 
between the $m_\sigma$ and $g_\omega$ parameters are
clearly visible in Figs.\ \ref{Paramet-range}a, b, d, and e 
where the modifications of the factors $f(g_{\omega})$ and 
$f(m_{\sigma})$ lead to almost the same changes in binding 
energies and charge radii.

  In contrast, the $g_{\rho}$ parameter does not correlate with 
other parameters of the functional. This is illustrated in Figs.\ 
\ref{NL5-ksi-distrib}c and  \ref{NL5(A)-ksi-distrib}c where the 
$f(g_{\rho})-f(m_{\sigma})$ distributions are plotted. The fact of 
the absence of the correlations are easy to understand since the 
$g_{\rho}$ parameter defines the isovector properties of the 
functionals while the $m_{\sigma}$ parameter the attractive scalar 
potential $S$.

 The parametric correlations are especially pronounced for the $g_2$ 
and $g_3$ parameters which define the density dependence of the 
functional via a non-linear meson coupling (see Eq. (\ref{g2g3-eq})).
Figs.\ \ref{NL5-ksi-distrib}a and \ref{NL5(A)-ksi-distrib}a clearly 
show that these two parameters are not independent and that the 
following linear dependence
\begin{equation}
f(g_3) = a f(g_2)  + b
\label{g2g3-dep}
\end{equation}
exists. The parameters $a$ and $b$, defined from $f(g_3)-f(g_2)$ 
distributions shown in Figs.\ \ref{NL5-ksi-distrib}a and 
\ref{NL5(A)-ksi-distrib}a, have the following values 
$a= 1.461276$ and $b = -0.458276$ for NL5(C) and $a= 1.643 $  
and $b = -0.64693 $ for NL5(A). These parametric correlations 
are more pronounced for the NL5(A) functional for which the 
$f(g_3)-f(g_2)$ distribution is narrower and more elongated.

  The NL5() functionals depends on 6 parameters. The present analysis 
strongly suggests that its parameter dependence could be reduced to 5 
independent parameters, namely, $m_\sigma$, $g_\sigma$, $g_\omega$, 
$g_\rho$, and $g_2$ if the parametric correlations given in Eq.\ 
(\ref{g2g3-dep}) are taken into account. However, this requires new 
refit of the functional with linear dependence of Eq.\ (\ref{g2g3-dep}) 
explicitly used in the fitting protocol. An alternative analysis in 
manifold boundary approximation method (Ref.\ \cite{NIV.17}) has shown 
that it is possible to reduce the dimension of the parameter hyperspace 
of the DD-PC1 CEDF from ten parameters to eight without sacrificing the 
quality of the reproduction of experimental and empirical data. Similar 
to the NL5() case, this reduction in Ref.\ \cite{NIV.17} takes place in 
the channel of the functional which defines its density dependence. In the 
context of the analysis of theoretical uncertainties there is one clear 
advantage of the reduction of the dimensionality of the parameter hyperspace 
via the removal of parametric correlations: such reduction leads to the 
decrease of statistical errors. This was illustrated in Ref.\ \cite{DD.17} 
on the example of the study of statistical errors in the single-particle 
energies and it is discussed for ground state observables in the present
manuscript in Sec.\ \ref{stat-errors-other}. 

\section{Statistical errors in the ground state observables of 
even-even nuclei.}
\label{sect-GS}

\begin{figure*}[htb]
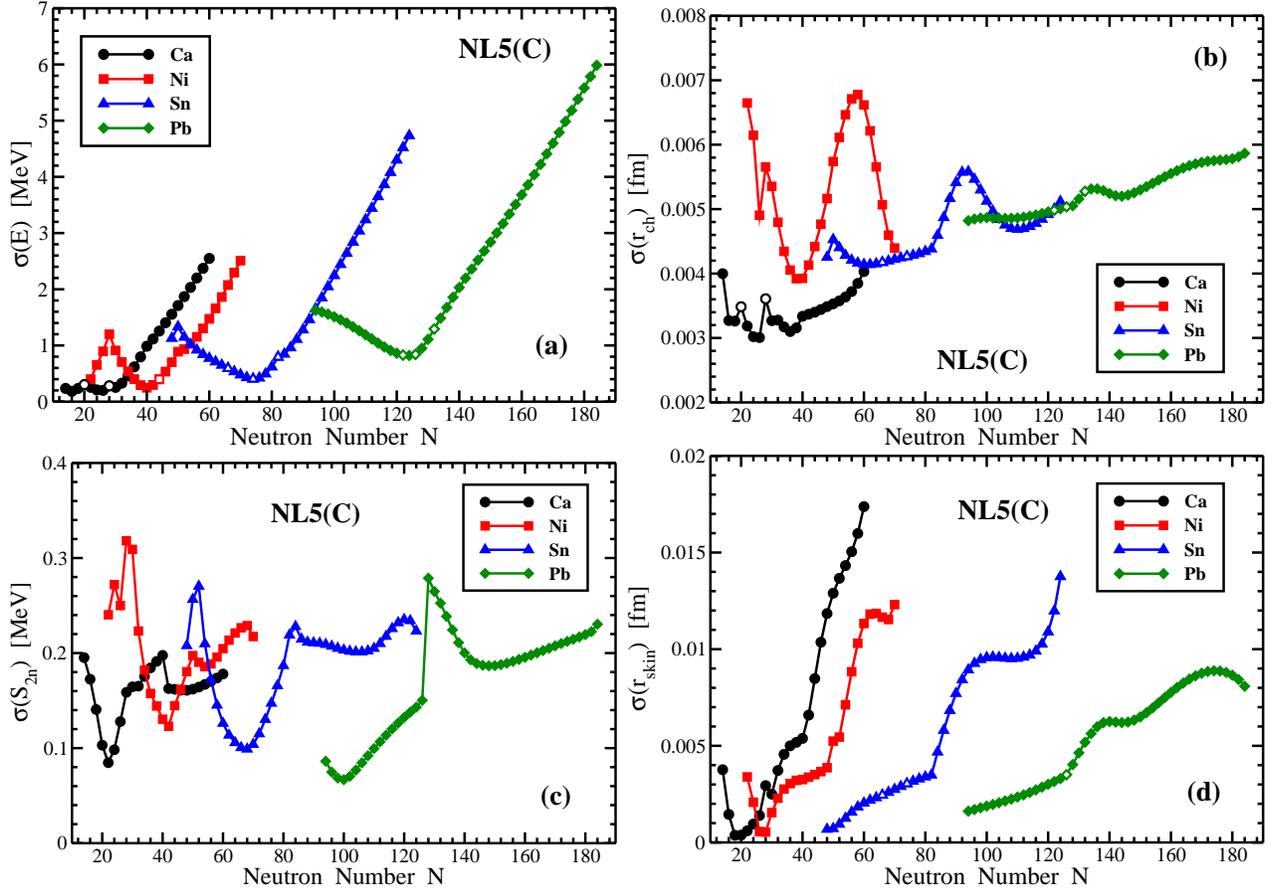

\includegraphics[angle=0,width=8.1cm]{fig-7-a.eps}
\includegraphics[angle=0,width=8.5cm]{fig-7-b.eps}
\includegraphics[angle=0,width=8.1cm]{fig-7-c.eps}
\includegraphics[angle=0,width=8.5cm]{fig-7-d.eps}
\caption{(Color online) The propagation of statistical errors
in binding energies [panel (a)], charge radii [panel (b)], 
two-neutron separation energies [panel (c)] and
neutron skins [panel (d)] with neutron number. The results are presented for 
the Ca ($Z=20$), Ni ($Z=28$), Sn ($Z=50$) and Pb ($Z=82$) isotopes 
between two-proton and two-neutron drip lines. Open symbols are 
used to indicate the nuclei whose experimental data of the type 
shown on the vertical axis of the panel has been used in the 
fitting protocol of the NL5(C) functional.}
\label{stat-errors}
\end{figure*}

\subsection{The case of the NL5(C) functional}
\label{stat-error-NL5(C)}

  In this section we will investigate statistical errors in the 
description of the ground state properties of spherical Ca, Ni, Sn 
and Pb even-even isotopes. Within the isotope chain the calculations 
cover all nuclei between the two-proton and two-neutron drip lines.
When possible the statistical errors obtained in the present study
will be compared with systematic uncertainties defined in Refs.\ 
\cite{AARR.14,AA.16}. In addition, they will be compared with
statistical errors obtained  in the Skyrme DFT study with UNEDF0
functional of Ref.\ \cite{GDKTT.13}.

  Statistical errors in binding energies obtained with the
CEDF NL5(C) and their propagation with the neutron number are shown 
in Fig.\ \ref{stat-errors}a. They are close to adopted errors of the
fitting protocol [0.1\% of binding energy (see Table \ref{table-inp})]
for the nuclei used in the fit. With increasing isospin the statistical 
errors in binding energies substantially increase  reaching 
$\sim 2.6$, $\sim 2.6$, $\sim 5.0$ and $\sim 6.0$ MeV at the 
two-neutron drip line in the Ca, Ni, Sn and Pb isotope chains, 
respectively.  However, they are significantly smaller at the 
neutron-drip line than those obtained in the Skyrme DFT studies of 
Ref.\ \cite{GDKTT.13}; by factors 4.6, 3.1, 3.4 and 2.3 for 
the Ca, Ni, Sn and Pb isotopes, respectively. Statistical errors 
in binding energies of these nuclei are by a factor 2-3 smaller 
than systematic uncertainties in the binding energies obtained
in Ref.\ \cite{AARR.14} (see Fig. 8 in this reference).
Note that the estimate of systematic uncertainties of Ref.\ 
\cite{AARR.14} are based only on four CEDFs, namely, NL3*, DD-PC1, 
DD-ME2 and DD-ME$\delta$. The investigation of Ref.\ \cite{AARR.15} 
suggests that the addition of the PC-PK1 functional could lead to 
a substantial increase of systematic uncertainties in binding 
energies. This is at least a case for the Yb ($Z=70$) isotopes for 
which they increase by a factor of 2.1 when the PC-PK1 results are 
added (see Fig. 3 in Ref.\ \cite{AA.16}).

  Statistical errors in charge radii $r_{ch}$ are presented in Fig.\ 
\ref{stat-errors}b. They are in the vicinity of 0.1\% of the calculated $r_{ch}$ 
values shown in Fig. 23 of Ref.\ \cite{AARR.14}. For the nuclei used in 
the fitting protocol, statistical errors are below the adopted errors 
of 0.2\% for $r_{ch}$. Calculated statistical errors are below 25\% of 
the rms deviations $\Delta r_{ch}^{rms}$ between calculated and experimental 
charge radii, which are typical for the state-of-the-art CEDFs and which 
are shown in Table VI of Ref.\ \cite{AARR.14}.  Systematic uncertainties 
in the predictions of charge radii of the Ca and Ni isotopes obtained 
from the set of the four functionals (see Fig. 24 of Ref.\ \cite{AARR.14}) 
are substantially larger (on average, by an approximate factor of 8 and
10, respectively) than relevant statistical errors. This difference goes 
down with the increase of proton number. For example, the situation in the 
Pb isotopes depends on the neutron number $N$. Statistical errors are 
only somewhat smaller than systematic uncertainties in the Pb nuclei with 
$N\sim 110$ and $N\sim 126$. On the other hand, they are smaller than 
statistical uncertainties by a factor of approximately 10 for the nuclei 
with $N\sim 102$. On average, for the Pb nuclei the statistical errors in 
charge radii are by a factor of approximately 4 smaller than relevant  
systematic uncertainties. Similar situation is observed also in the Sn 
isotopes, but the average difference between statistical errors and 
statistical uncertainties in charge radii is of the order of 7.

  Contrary to Skyrme DFT calculations with the UNEDF0 functional (see
Fig. 4b in Ref.\ \cite{GDKTT.13}), statistical errors in charge radii 
calculated with NL5(C) (see Fig.\ \ref{stat-errors}b in the present 
paper) do not show significant increase with neutron number. In reality, 
the $\sigma (r_{ch})$ values obtained with NL5(C) for the Sn and Pb isotopes 
show very modest increase of approximately 20\% on going from two-proton to 
two-neutron drip line. Note that the $\sigma (r_{ch})$ values show some 
fluctuations as a function of neutron number which are due to underlying 
shell structure; they become especially pronounced in the Ni isotopes. 
While statistical errors for charge radii of the Ca, Ni and Sn isotopes 
are comparable for Skyrme UNEDF0 and CDFT NL5(C) calculations for the 
nuclei near two-proton drip line, the situation changes drastically 
with the increase of neutron number so that for 
the Ca, Ni, Sn and Pb nuclei at the two-neutron drip line the 
$\sigma (r_{ch})$ values obtained in Skyrme calculations are by factor 
of 17-33 larger than those obtained in CDFT calculations with CEDF 
NL5(C).

\begin{figure*}[ht]
\includegraphics[angle=0,width=16.5cm]{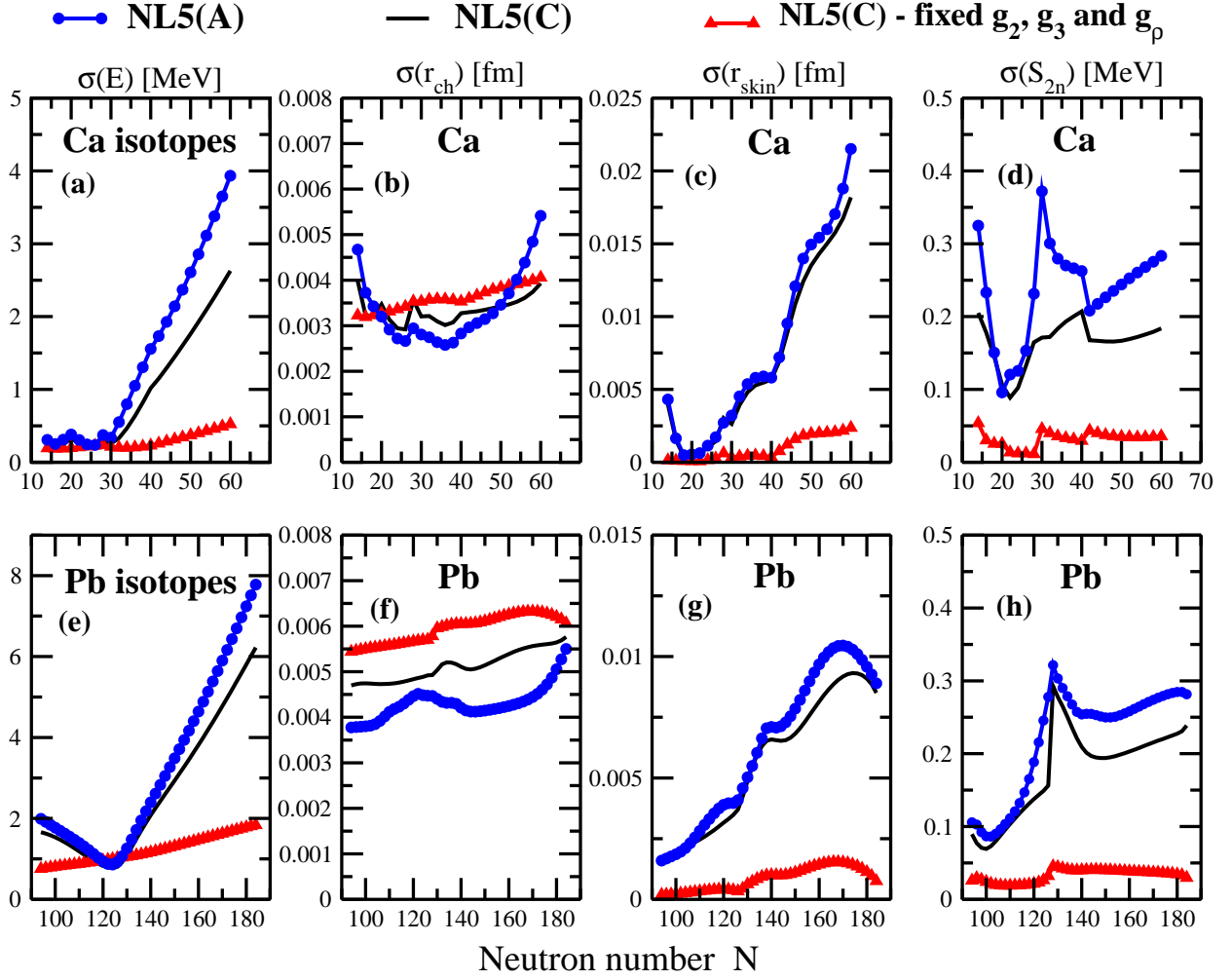}
\caption{(Color online) The propagation of statistical errors 
for the binding energies $(\sigma (E))$, charge radii $(\sigma (r_{ch}))$, 
neutron skins $(\sigma (r_{skin}))$ and two-neutron separation energies 
$(\sigma (S_{2n}))$ in the Ca and Pb isotope chains. The results of the
calculations with the NL5(C) and NL5(A) functionals are shown by solid
black line and blue solid line with solid circles, respectively. In
addition, red triangle symbols show statistical errors for the NL5(C) 
functional under the condition that the $g_{\rho}$, $g_2$ and $g_3$ 
parameters are fixed at the values corresponding to optimum NL5(C) 
functional.}
\label{SE-fixed_par}
\end{figure*}

  Statistical errors in two-neutron separation energies are displayed in
Fig.\ \ref{stat-errors}c. They are typically in the range of 0.1 - 0.3 MeV
and do not show a clear tendency of the increase on approaching two-neutron
drip line. These statistical errors show substantial fluctuations as a function
of neutron number with the changes in the slope of $\sigma(S_{2n})$ typically taking 
place in the vicinity of the shell ($N=20, 28, 50, 82$ and 126) and subshell 
($N=40$) closures. The calculated $\sigma(S_{2n})$ values are typically by a
factor of 3-4 smaller than the rms-deviations between theory and experiment 
for the state-of-the-art CEDF (see Table III in Ref.\ \cite{AARR.14}). It
is interesting to compare our results with the ones obtained in Skyrme 
DFT calculations of Ref.\ \cite{GDKTT.13}. While the $\sigma(S_{2n})$ values
are comparable for both models in the vicinity of the $\beta$-stability line,
they increase drastically with increasing neutron number in the Skyrme DFT 
calculations approaching $\sim 1.4$, $\sim 1.4$, $\sim 0.8$ and $\sim 0.75$ 
MeV for Ca, Ni, Sn and Pb nuclei at the two-neutron drip line, respectively. 
This trend is contrary to the one seen in the CDFT results.

  Statistical errors in the neutron skin thickness $r_{skin}$ are shown in 
Fig.\ \ref{stat-errors}d. They are close to zero near the $N=Z$ line but increase 
with increasing neutron number. This increase is rapid in the Ca, Ni and Sn 
isotopes but it is more moderate in the Pb isotopes. An interesting feature 
of the latter chain is the decrease of the $\sigma (r_{skin})$ values above 
$N\sim 170$ which is most likely due to underlying shell effects. The statistical 
errors in the neutron skin thickness are substantially larger in the Skyrme DFT 
calculations with the UNEDF0 and SV-min functionals (Ref.\ \cite{KENBGO.13}) than 
in the present RHB calculations with CEDF NL5(C). For example, for $^{208}$Pb the 
$\sigma (r_{skin})$ values are 0.058 fm, 0.037 fm and 0.0035 fm in the calculations 
with the UNEDF0, SV-min and NL5(C) functionals, respectively. In the neutron-rich Ca 
isotopes near the two-neutron drip line the $\sigma (r_{skin})$ values obtained in 
non-relativistic calculations are by a factor of approximately 7 larger than 
those obtained in the relativistic ones. The statistical errors in the neutron skin 
thickness shown in Fig.\ \ref{stat-errors}d are substantially smaller than systematic 
uncertainties shown in Fig.\ 25 of Ref.\ \cite{AARR.14}. In the vicinity of the 
two-neutron drip line, the latter ones reach 0.15 fm, 0.2 fm, 0.25 fm and 0.25 fm 
in the neutron rich Ca, Ni, Sn and Pb nuclei, respectively.

\subsection{The impact of the details of the fitting protocol
and of soft parameters on statistical errors.}
\label{stat-errors-other}

  It is important to understand how the details of the 
fitting protocol could affect the statistical errors in physical 
observables of interest. Considering the ingredients entering 
into fitting protocol and the uncertainties in the definition of
empirical/experimental values of physical observables and
adopted errors, the complete answer to this question requires 
enormous amount of numerical calculations which are beyond the 
scope of the present investigation. However, in order to get a 
crude estimate of potential changes in statistical errors 
additional calculations have been performed for the two cases
discussed below.

  In the first case, we analyse statistical errors obtained
with the NL5(A) functional. This functional differs from the
NL5(C) one by an increased adopted error for $K_0$ and the presence
of neutron skin of $^{90}$Zr in the fitting protocol (see
Table \ref{table-inp}). As a consequence of these features, 
the  acceptable NL5(A) functionals show larger range of variations 
for the $g_2$ and $g_3$ parameters as compared with the NL5(C) ones
(compare Figs.\ 
\ref{NL5-ksi-distrib}a and  \ref{NL5(A)-ksi-distrib}a). This 
leads to some increase in statistical errors for binding 
energies (especially in neutron-rich nuclei in which the 
$\sigma(E)$ values are increased by $\sim 30\%$  as compared 
with the NL5(C) results) and two-neutron separation energies 
(see Figs.\ \ref{SE-fixed_par}a, e, d and h). On the contrary, 
as compared with the NL5(C) results statistical errors in 
neutron skin are only slightly increased (see Fig.\ 
\ref{SE-fixed_par}c and g) and statistical errors in charge 
radii even decrease in the Pb isotopes and for some Ca 
isotopes (see Fig.\ \ref{SE-fixed_par}b and f).

  Despite these changes statistical errors for physical 
observables of interest obtained with the NL5(A) functional 
still remain substantially smaller than those obtained in 
Skyrme DFT calculations (compare results presented in
this subsection with the discussion in Sec.\ 
\ref{stat-error-NL5(C)}).

  In the second case, we use the NL5(C) functional but 
fix the $g_2$, $g_3$ and $g_{\rho}$ parameters at the values 
of the optimal functional during the Monte-Carlo procedure. 
As shown in Secs.\ \ref{fitting-prot} and \ref{sect-param}, 
the $g_{\sigma}$, $m_{\sigma}$ and $g_{\omega}$ parameters, which 
are allowed to change during Monte-Carlo procedure, are well 
localized in the parameter hyperspace and they vary in a very 
narrow range with respect of the optimum functional. 
Fig.\ \ref{SE-fixed_par} compares the results for the Ca and 
Pb nuclei presented  in Fig.\ \ref{stat-errors} with statistical 
errors obtained in such calculations. The freezing of the 
$g_2$, $g_3$ and $g_{\rho}$ parameters in the functional leads 
to a substantial decrease of statistical errors in binding 
energies of neutron-rich nuclei so that in two-neutron drip 
line nuclei they are only by a factor of approximately two larger 
than those for the nuclei used in the fitting protocol 
(Fig.\ \ref{SE-fixed_par}a and e). Note that the $\sigma(E)$ 
values obtained is such calculations are typically significanly 
smaller (especially in very neutron-rich nuclei) than those 
obtained in full NL5(C) calculations shown by solid black line. 
The only exception is the region of neutron numbers in which the 
experimental data on binding energies was used in the fitting 
protocol. Statistical errors in two-neutron separation energies 
become very small ($\sigma (S_{2n}) < 0.05$ MeV) in the 
calculations with fixed $g_2$, $g_3$ and $g_{\rho}$ parameters 
(see Fig.\ \ref{SE-fixed_par}d and h). They are also 
significantly smaller than those obtained in full NL5(C) 
calculations. Simultaneous freezing of the $g_2$, $g_3$ and 
$g_{\rho}$ parameters leads to a substantial decrease of statistical 
errors in neutron skin so that in the most of the nuclei they are 
very close to zero (see Fig.\ \ref{SE-fixed_par}c and g). On the 
contrary, the impact of freezing of the $g_2$, $g_3$ and $g_{\rho}$ 
parameters on statistical errors in charge radii is very limited 
(see Fig.\ \ref{SE-fixed_par}b and f)). It is interesting that it 
leads to slight increase of the $\sigma(r_{ch})$ as compared with 
those obtained in full NL5(C) calculations.

  This analysis clearly illustrates the importance of the discrimination 
of the impact of the {\it stiff} and {\it soft} parameters of CEDFs
on statistical errors. Such a separation of the parameters into
two types was discussed in Ref.\ \cite{NIV.17}. Even when the 
parameters of nuclear EDFs are adjusted to experimental/empirical/pseudo
data, their predictions are sensitive to only a few combinations of
parameters ({\it stiff} parameter combinations) and exhibit an exponential
decrease of sensitivity to variations of the remaining {\it soft} parameters 
that are only approximately constrained by data. In non-linear meson 
coupling models, the stiff parameters are represented by $m_{\omega}$, 
$g_{\omega}$ and $g_{\sigma}$. Their contribution to statistical errors is 
rather small and mostly independent of neutron number. On the contrary, 
the combination of soft parameters $g_2$, $g_3$ and $g_{\rho}$ leads to a
significant increase of statistical errors for binding energies
and neutron skins on approaching two-neutron drip line. They also 
lead to an increase of statistical errors in two-neutron separation
energies. Thus, one can conclude that the presence of soft parameters 
in the CEDFs is the major source of statistical errors.

\section{Statistical errors in the single-particle energies}
\label{sect-sp}

\begin{table}[ht]
\begin{center}
\caption{
Neutron and proton single-particle states in the $^{208}$Pb 
nucleus. Columns 2 and 5 show the mean energies $\bar{e}_i$ [in 
MeV] of these states obtained in the calculations with CEDF NL5(C). 
Columns 3 and 6 show their standard deviations $\sigma(e_i)$ [in 
MeV]. The positions of the proton and neutron Fermi levels are 
indicated. 
\label{table-sp-208Pb}
}
\begin{tabular}{|c|c|c|c|c|c|} \hline 
\multicolumn{3}{|c|}{Neutron} & \multicolumn{3}{c|}{Proton} \\ \hline
  Orbital     &   $\bar{e}_{\nu}$  & $\sigma(e_{\nu})$ &
 Orbital    & $\bar{e}_{\pi}$   & $\sigma(e_{\pi})$ \\ \hline
     1       &     2   &   3   &  4  &  5         &    6      \\ \hline
  1$s_{1/2}$ & -58.257 & 0.424 & 1$s_{1/2}$ & -47.100 & 0.401 \\
  1$p_{3/2}$ & -52.142 & 0.376 & 1$p_{3/2}$ & -41.573 & 0.360 \\
  1$p_{1/2}$ & -51.581 & 0.363 & 1$p_{1/2}$ & -40.933 & 0.345 \\
  1$d_{5/2}$ & -44.809 & 0.316 & 1$d_{5/2}$ & -34.691 & 0.307 \\
  1$d_{3/2}$ & -43.497 & 0.282 & 1$d_{3/2}$ & -33.249 & 0.270 \\
  2$s_{1/2}$ & -40.566 & 0.241 & 2$s_{1/2}$ & -29.725 & 0.221 \\
  1$f_{7/2}$ & -36.614 & 0.252 & 1$f_{7/2}$ & -26.864 & 0.251 \\
  1$f_{5/2}$ & -34.228 & 0.189 & 1$f_{5/2}$ & -24.320 & 0.187 \\
  2$p_{3/2}$ & -30.515 & 0.149 & 2$p_{3/2}$ & -20.045 & 0.150 \\
  2$p_{1/2}$ & -29.526 & 0.128 & 2$p_{1/2}$ & -19.054 & 0.136 \\
  1$g_{9/2}$ & -27.862 & 0.191 & 1$g_{9/2}$ & -18.402 & 0.199 \\
  1$g_{7/2}$ & -24.177 & 0.112 & 1$g_{7/2}$ & -14.556 & 0.128 \\
  2$d_{5/2}$ & -20.708 & 0.098 & 2$d_{5/2}$ & -10.465 & 0.122 \\
  2$d_{3/2}$ & -19.127 & 0.091 & 1$h_{11/2}$ &  -9.544 & 0.159 \\
 1$h_{11/2}$ & -18.809 & 0.141 & 2$d_{3/2}$ &  -8.903 & 0.121 \\
  3$s_{1/2}$ & -18.301 & 0.095 & 3$s_{1/2}$ &  -7.729 & 0.121 \\
             &         &      & \multicolumn{3}{c|}{Proton Fermi level} \\
  1$h_{9/2}$ & -13.774 & 0.108 & 1$h_{9/2}$ &  -4.367 & 0.134 \\
  2$f_{7/2}$ & -11.320 & 0.085 & 2$f_{7/2}$ &  -1.107 & 0.120 \\
 1$i_{13/2}$ &  -9.661 & 0.104 & 1$i_{13/2}$ &  -0.481 & 0.133 \\
  2$f_{5/2}$ &  -9.390 & 0.098 & 2$f_{5/2}$ &   0.802 & 0.131 \\
  3$p_{3/2}$ &  -8.591 & 0.087 & 3$p_{3/2}$ &   2.133 & 0.121 \\
  3$p_{1/2}$ &  -7.895 & 0.090 & 3$p_{1/2}$ &   2.828 & 0.125 \\
  \multicolumn{3}{|c|}{Neutron Fermi level}    &            &         &       \\ 
 1$i_{11/2}$ &  -3.515 & 0.148 &            &         &       \\
  2$g_{9/2}$ &  -2.688 & 0.080 &            &         &       \\
  3$d_{5/2}$ &  -0.873 & 0.062 &            &         &       \\
  2$g_{7/2}$ &  -0.855 & 0.090 &            &         &       \\
  4$s_{1/2}$ &  -0.599 & 0.050 &            &         &       \\
 1$j_{15/2}$ &  -0.597 & 0.081 &            &         &       \\
  3$d_{3/2}$ &  -0.281 & 0.061 &            &         &       \\
  4$p_{3/2}$ &   2.557 & 0.018 &            &         &       \\
  4$p_{1/2}$ &   2.669 & 0.019 &            &         &       \\ \hline
\end{tabular}
\end{center}
\end{table}

\begin{table}[ht]
\begin{center}
\caption{
The same as Table \ref{table-sp-208Pb} but for the $^{266}$Pb 
nucleus. The states in the energy range from $\sim -42$ MeV up to 
$\sim -20$ MeV are omitted in order to simplify the table. 
\label{table-sp-pb266}
}
\begin{tabular}{|c|c|c|c|c|c|} \hline 
\multicolumn{3}{|c|}{Neutron} & \multicolumn{3}{c|}{Proton} \\ \hline
  Orbital     &   $\bar{e}_{\nu}$  & $\sigma(e_{\nu})$ & Orbital    & $\bar{e}_{\pi}$   & $\sigma(e_{\pi})$ \\ \hline
     1       &     2   &   3   &  4  &  5         &    6      \\ \hline
  1$s_{1/2}$ & -53.045 & 0.364 & 1$s_{1/2}$ & -49.945 & 0.323 \\ 
  1$p_{3/2}$ & -47.922 & 0.333 & 1$p_{3/2}$ & -45.498 & 0.304 \\
  1$p_{1/2}$ & -47.567 & 0.326 & 1$p_{1/2}$ & -45.070 & 0.297 \\
     -----   &  -----  & ----- &   -----    &  ------ & ----- \\
  2$d_{3/2}$ & -20.427 & 0.153 &1$h_{11/2}$ & -18.396 & 0.199 \\
  3$s_{1/2}$ & -19.619 & 0.153 & 2$d_{3/2}$ & -18.075 & 0.176 \\
 1$h_{11/2}$ & -19.595 & 0.188 & 3$s_{1/2}$ & -16.727 & 0.169 \\
             &         &       &  \multicolumn{3}{c|}{Proton Fermi level} \\
  1$h_{9/2}$ & -16.052 & 0.159 & 1$h_{9/2}$ & -14.726 & 0.186 \\
  2$f_{7/2}$ & -13.384 & 0.142 & 2$f_{7/2}$ & -10.867 & 0.172 \\
  2$f_{5/2}$ & -11.882 & 0.138 &1$i_{13/2}$ & -10.415 & 0.184 \\
 1$i_{13/2}$ & -11.638 & 0.163 & 2$f_{5/2}$ &  -9.372 & 0.182 \\
  3$p_{3/2}$ & -11.057 & 0.130 & 3$p_{3/2}$ &  -7.655 & 0.171 \\
  3$p_{1/2}$ & -10.494 & 0.128 & 3$p_{1/2}$ &  -7.086 & 0.177 \\
 1$i_{11/2}$ &  -7.153 & 0.161 & 1$i_{11/2}$ &  -5.779 & 0.198 \\
  2$g_{9/2}$ &  -5.728 & 0.124 & 2$g_{9/2}$ &  -2.583 & 0.174 \\
  2$g_{7/2}$ &  -4.138 & 0.123 & 1$j_{15/2}$ &  -2.234 & 0.175 \\
  3$d_{5/2}$ &  -3.847 & 0.100 & 2$g_{7/2}$ &  -0.896 & 0.186 \\
 1$j_{15/2}$ &  -3.691 & 0.143 & 3$d_{5/2}$ &   0.851 & 0.168 \\
  4$s_{1/2}$ &  -3.342 & 0.088 & 3$d_{3/2}$ &   1.540 & 0.173 \\
  3$d_{3/2}$ &  -3.252 & 0.097 & 4$s_{1/2}$ &   2.130 & 0.162 \\
  \multicolumn{3}{|c|}{Neutron Fermi level} &   &         &       \\
 2$h_{11/2}$ &   1.038 & 0.096 & 1$j_{13/2}$ &   3.185 & 0.216 \\
  4$p_{3/2}$ &   1.292 & 0.034 &            &         &       \\
 1$j_{13/2}$ &   1.406 & 0.164 &            &         &       \\
  4$p_{1/2}$ &   1.460 & 0.033 &            &         &       \\
  3$f_{7/2}$ &   1.546 & 0.051 &            &         &       \\
  3$f_{5/2}$ &   1.998 & 0.045 &            &         &       \\
  2$h_{9/2}$ &   2.387 & 0.086 &            &         &       \\ \hline
\end{tabular}
\end{center}
\end{table}

   The energies of the single-particle states $e_i$ represent 
another quantity which affects many physical observables (see Refs.\ 
\cite{LA.11,AL.15,AS.11,AARR.15,AANR.15}). In this section we 
investigate statistical errors in the description of the energies 
of the single-particle states $e_i$; they are quantified by the 
standard deviations $\sigma(e_i)$. We focus on three nuclei, namely, 
$^{208}$Pb, $^{266}$Pb and $^{304}$120. The first one is well known doubly 
magic nucleus which serves as a testing ground in many theoretical 
studies. Second nucleus is the last bound neutron-rich Pb isotope 
in absolute majority of theoretical studies (see Fig.\ 14 in Ref.\ 
\cite{AARR.14}); it is characterized by large $N=184$ shell 
gap (see Fig.\ 6a in Ref.\ \cite{AARR.15}). Third nucleus is the 
$Z=120, N=184$  superheavy nucleus which is considered as doubly magic 
in a number of studies (see, for example, Ref.\ \cite{BRRMG.99}); 
this conclusion is however model dependent (see Refs.\ 
\cite{BRRMG.99,BNR.01,A250,AANR.15} and references quoted therein). 
The comparison of the results obtained for these three nuclei will 
allow to assess the propagation of statistical errors on going from 
the valley of beta-stability towards the extremes of neutron number 
and charge.

\begin{table}[ht]
\begin{center}
\caption{The same as Table \ref{table-sp-208Pb} but for the $^{304}$120
nucleus. Neutron states in the energy range from $\sim -50$ MeV up to 
$\sim -25$ MeV and proton states in the energy range from $\sim -35$ 
MeV up to $\sim -12$ MeV are omitted in order to simplify the table. 
\label{table-sp-z120-n184}
}
\begin{tabular}{|c|c|c|c|c|c|} \hline 
\multicolumn{3}{|c|}{Neutron} & \multicolumn{3}{c|}{Proton} \\ \hline
  Orbital     &   $\bar{e}_{\nu}$    & $\sigma(e_{\nu})$ & Orbital  & $\bar{e}_{\pi}$  & $\sigma(e_{\pi})$ \\ \hline
     1       &     2   &   3   & 4   &    5       &    6    \\ \hline
  1$s_{1/2}$ & -57.748 & 0.386 & 1$s_{1/2}$ & -41.480 & 0.367 \\
  1$p_{3/2}$ & -53.698 & 0.362 & 1$p_{3/2}$ & -37.545 & 0.349 \\
  1$p_{1/2}$ & -53.459 & 0.359 & 1$p_{1/2}$ & -37.246 & 0.343 \\
     -----   & -----   & -----  &   -----    & ------  & ----- \\
  3$s_{1/2}$ & -24.421 & 0.107 & 3$s_{1/2}$ &  -9.349 & 0.134 \\
  1$h_{9/2}$ & -23.315 & 0.105 & 1$h_{9/2}$ &  -8.169 & 0.126 \\
  2$f_{7/2}$ & -19.452 & 0.098 & 1$i_{13/2}$ &  -4.150 & 0.153 \\
 1$i_{13/2}$ & -18.893 & 0.132 & 2$f_{7/2}$ &  -3.903 & 0.123 \\
  2$f_{5/2}$ & -17.953 & 0.092 & 2$f_{5/2}$ &  -2.418 & 0.121 \\
             &         &       & \multicolumn{3}{c|}{Proton Fermi level} \\
  3$p_{3/2}$ & -15.293 & 0.092 & 3$p_{3/2}$ &  -0.302 & 0.132 \\
  3$p_{1/2}$ & -14.947 & 0.093 & 3$p_{1/2}$ &   0.017 & 0.134 \\
 1$i_{11/2}$ & -14.288 & 0.103 & 1$i_{11/2}$ &   0.612 & 0.130 \\
  2$g_{9/2}$ & -11.194 & 0.086 &1$j_{15/2}$ &   3.758 & 0.133 \\
 1$j_{15/2}$ & -10.809 & 0.103 & 2$g_{9/2}$ &   4.230 & 0.119 \\
  2$g_{7/2}$ &  -9.206 & 0.099 &            &         &       \\
  3$d_{5/2}$ &  -7.336 & 0.085 &            &         &       \\
  3$d_{3/2}$ &  -7.014 & 0.085 &            &         &       \\
  4$s_{1/2}$ &  -6.372 & 0.077 &            &         &       \\
   \multicolumn{3}{|c|}{Neutron Fermi level} &  &         &       \\
 1$j_{13/2}$ &  -5.195 & 0.136 &            &         &       \\
 2$h_{11/2}$ &  -3.278 & 0.084 &            &         &       \\
 1$k_{17/2}$ &  -2.669 & 0.084 &            &         &       \\
  2$h_{9/2}$ &  -1.227 & 0.100 &            &         &       \\
  3$f_{7/2}$ &  -0.626 & 0.066 &            &         &       \\
  3$f_{5/2}$ &  -0.188 & 0.067 &            &         &       \\
  4$p_{3/2}$ &   0.080 & 0.051 &            &         &       \\
  4$p_{1/2}$ &   0.259 & 0.052 &            &         &       \\
  5$s_{1/2}$ &   2.353 & 0.016 &            &         &       \\
  4$d_{5/2}$ &   2.822 & 0.017 &            &         &       \\ \hline
\end{tabular}
\end{center}
\end{table}

  Tables \ref{table-sp-208Pb}, \ref{table-sp-pb266} and \ref{table-sp-z120-n184}
show the calculated mean energies $\bar{e}_i$ of the single-particle states and 
related standard deviations $\sigma(e_i)$ obtained in these nuclei. The general 
trend, which is clearly seen in these tables, is the decrease of statistical errors 
on going from the bottom of nucleonic potential towards continuum. The states at 
the bottom  of potential are characterized by $\sigma(e_i)\sim 0.35$ MeV both for 
proton and neutron subsystems for all nuclei under consideration. In $^{208}$Pb, 
the neutron and proton states are characterized by $\sigma(e_i)\sim 0.1$ MeV in 
the vicinity of the respective Fermi levels which have similar energies. 
The addition of neutrons leading to $^{266}$Pb moves the proton Fermi level to 
lower energies (deeper into the potential) and neutron Fermi level closer to 
continuum limit. As a consequence, the $\sigma(e_i)$ values for the proton 
states in the vicinity of the proton Fermi level increase to $\sim 0.15$ MeV 
(see Table \ref{table-sp-pb266}). On the contrary, with  the exception of the 
high-$j$ $\nu 1 j_{15/2}$ and $\nu 1 j_{13/2}$ states for which $\sigma(e_i)\sim 0.15$ 
MeV, the $\sigma(e_i)$ values for the neutron states in the vicinity of the 
neutron Fermi level is less than 0.1 MeV (see Table \ref{table-sp-pb266}).
Because the neutron and proton Fermi levels are located at $\sim -6.4$ MeV 
and $\sim -2.4$ MeV, respectively, in the $^{184}$120 nucleus [which is not far 
away from their values in $^{208}$Pb], the $\sigma(e_i)$ values for the 
single-particle states in this nucleus are comparable with the ones in $^{208}$Pb 
(compare tables \ref{table-sp-208Pb} and \ref{table-sp-z120-n184}).

\begin{table}[ht]
\begin{center}
\caption{Relative energies ${\Delta e}_i(m,j) = e_i(m) - e_i(j)$
of the pairs of neutron ($i=\nu$) and  
proton ($i=\pi$) single-particle states in the $^{208}$Pb nucleus. The pairs of
neighboring states $m$ and $j$ shown in the columns 1 and 4 are defined 
from the single-particle spectra obtained with the  NL5(C) CEDF. The columns 2 
and 5  show their mean relative energies $\overline{\Delta e}_i(m,j)$ 
[in MeV] while the columns 3 and 6 show standard deviations $\sigma(\Delta e_i(m,j))$ 
[in MeV].  All the pairs of the states up to the label  ``proton/neutron Fermi level''
have both members located below respective Fermi levels. Next line after this 
statement indicates the pair of the states one of the members of which is located 
below the large shell gap (either proton $Z=82$ or neutron $N=126$ one) and another 
above this gap. Subsequent lines show the pairs of the states both members of 
which are located above respective Fermi levels (above proton $Z=82$ or neutron 
$N=126$ shell gaps).
\label{table-rel-sp-pb208}
}
\begin{tabular}{|c|c|c|c|c|c|} \hline 
\multicolumn{3}{|c|}{Neutron} & \multicolumn{3}{c|}{Proton} \\ \hline
 Orbital         &           &         &  Orbital           &           &         \\        
 pairs $(m,j)$   &   $\overline{\Delta e}_{\nu}$    & $\sigma(\Delta e_{\nu})$ & pairs $(m,j)$  & 
$\overline{\Delta e}_{\pi}$  & $\sigma(\Delta e_{\pi})$ \\ \hline
     1                     &     2 &   3   &    4                     &    5  &   6   \\ \hline
  1$s_{1/2}$ - 1$p_{3/2}$  & 6.115 & 0.054 & 1$s_{1/2}$ - 1$p_{3/2}$  & 5.527 & 0.049 \\
  1$p_{3/2}$ - 1$p_{1/2}$  & 0.560 & 0.013 & 1$p_{3/2}$ - 1$p_{1/2}$  & 0.640 & 0.016 \\
  1$p_{1/2}$ - 1$d_{5/2}$  & 6.772 & 0.053 & 1$p_{1/2}$ - 1$d_{5/2}$  & 6.242 & 0.045 \\
  1$d_{5/2}$ - 1$d_{3/2}$  & 1.313 & 0.037 & 1$d_{5/2}$ - 1$d_{3/2}$  & 1.442 & 0.042 \\
  1$d_{3/2}$ - 2$s_{1/2}$  & 2.931 & 0.047 & 1$d_{3/2}$ - 2$s_{1/2}$  & 3.524 & 0.058 \\
  2$s_{1/2}$ - 1$f_{7/2}$  & 3.951 & 0.033 & 2$s_{1/2}$ - 1$f_{7/2}$  & 2.860 & 0.049 \\
  1$f_{7/2}$ - 1$f_{5/2}$  & 2.386 & 0.073 & 1$f_{7/2}$ - 1$f_{5/2}$  & 2.545 & 0.079 \\
  1$f_{5/2}$ - 2$p_{3/2}$  & 3.713 & 0.055 & 1$f_{5/2}$ - 2$p_{3/2}$  & 4.274 & 0.063 \\
  2$p_{3/2}$ - 2$p_{1/2}$  & 0.989 & 0.028 & 2$p_{3/2}$ - 2$p_{1/2}$  & 0.991 & 0.028 \\
  2$p_{1/2}$ - 1$g_{9/2}$  & 1.664 & 0.090 & 2$p_{1/2}$ - 1$g_{9/2}$  & 0.652 & 0.106 \\
  1$g_{9/2}$ - 1$g_{7/2}$  & 3.685 & 0.116 & 1$g_{9/2}$ - 1$g_{7/2}$  & 3.846 & 0.121 \\
  1$g_{7/2}$ - 2$d_{5/2}$  & 3.469 & 0.051 & 1$g_{7/2}$ - 2$d_{5/2}$  & 4.092 & 0.058 \\
  2$d_{5/2}$ - 2$d_{3/2}$  & 1.581 & 0.039 & 2$d_{5/2}$ - 1$h_{11/2}$ & 0.920 & 0.098 \\
 2$d_{3/2}$ - 1$h_{11/2}$  & 0.318 & 0.122 & 1$h_{11/2}$ - 2$d_{3/2}$ & 0.641 & 0.133 \\
 1$h_{11/2}$ - 3$s_{1/2}$  & 0.507 & 0.113 & 2$d_{3/2}$ - 3$s_{1/2}$  & 1.174 & 0.025 \\
              &            &       &   \multicolumn{3}{c|}{below proton Fermi level} \\
  3$s_{1/2}$ - 1$h_{9/2}$  & 4.527 & 0.094 & 3$s_{1/2}$ - 1$h_{9/2}$  & 3.362 & 0.095 \\
  1$h_{9/2}$ - 2$f_{7/2}$  & 2.455 & 0.078 & 1$h_{9/2}$ - 2$f_{7/2}$  & 3.260 & 0.081 \\
  2$f_{7/2}$ - 1$i_{13/2}$ & 1.659 & 0.094 & 2$f_{7/2}$ - 1$i_{13/2}$ & 0.626 & 0.103 \\
  1$i_{13/2}$ - 2$f_{5/2}$ & 0.271 & 0.128 & 1$i_{13/2}$ - 2$f_{5/2}$ & 1.283 & 0.136 \\
  2$f_{5/2}$ - 3$p_{3/2}$  & 0.799 & 0.033 & 2$f_{5/2}$ - 3$p_{3/2}$  & 1.331 & 0.031 \\
  3$p_{3/2}$ - 3$p_{1/2}$  & 0.696 & 0.017 & 3$p_{3/2}$ - 3$p_{1/2}$  & 0.695 & 0.016 \\
    \multicolumn{3}{|c|}{below neutron Fermi level}  &                &       &       \\
  3$p_{1/2}$ - 1$i_{11/2}$ & 4.380 & 0.109 &                          &       &       \\ 
  1$i_{11/2}$ - 2$g_{9/2}$ & 0.827 & 0.118 &                          &       &       \\
  2$g_{9/2}$ - 3$d_{5/2}$  & 1.816 & 0.022 &                          &       &       \\
  3$d_{5/2}$ - 2$g_{7/2}$  & 0.018 & 0.036 &                          &       &       \\
  2$g_{7/2}$ - 4$s_{1/2}$  & 0.256 & 0.044 &                          &       &       \\
  4$s_{1/2}$ - 1$j_{15/2}$ & 0.003 & 0.084 &                          &       &       \\
  1$j_{15/2}$ - 3$d_{3/2}$ & 0.316 & 0.090 &                          &       &       \\
  3$d_{3/2}$ - 4$p_{3/2}$  & 2.837 & 0.044 &                          &       &       \\
  4$p_{3/2}$ - 4$p_{1/2}$  & 0.113 & 0.003 &                          &       &       \\ \hline
\end{tabular}
\end{center}
\end{table}

  The detailed analysis of the 
results of the calculations shows that the
freedom to rebalance the depths of the proton and neutron potentials is 
a major source of these statistical errors in the energies of the 
single-particle states. Indeed, it was observed that when proton potential 
becomes deeper as compared with the one in the optimum functional, the 
neutron potential  becomes less deep as compared with the one in
optimum functional. This leads to more/less bound proton/neutron single-particle 
states and allows to keep total energy of the system close to the one in 
the optimum functional. The opposite situation with deeper neutron and less 
deep proton potentials takes place with similar frequency.

  In general, statistical errors in the absolute energies of the 
single-particle states could affect model predictions for the position of 
the two-neutron drip line. Indeed, as discussed in Ref.\ \cite{AARR.15} 
its position sensitively depends on the positions (in absolute energy) 
and the distribution of the single-particle states (and especially high-$j$ 
intruder and extruder ones) located around the continuum limit. However, in 
the nuclei around $^{266}$Pb the standard deviations $\sigma(e_i)$ for such 
neutron single-particle states are safely below 0.1 MeV (see Table \ref{table-sp-pb266}); 
the only exception is the $\nu 1 j_{13/2}$ orbital for which $\sigma(e_i)=0.164$ 
MeV. Thus, it is reasonable to expect that the impact of statistical errors 
in the energies of the single-particle states on the position of two-neutron 
drip line will be rather modest. Moreover, these statistical errors 
are substantially smaller than systematic uncertainties in the predictions of 
the energies of single-particle states which for many orbitals exceed 1 MeV 
in nuclei near two-neutron drip line (see Figs. 11c and 6c in Ref.\  
\cite{AARR.15}).  These facts suggest that the theoretical uncertainties in 
the prediction of the position of two-neutron drip line are dominated
by systematic ones.

  While the accuracy of the prediction of the position of the neutron
drip line is sensitive to calculated absolute energies of the single-particle 
states, the accuracy of the reproduction of the single-particle spectra 
depends mostly on the predictions of the relative energies of the 
single-particle states. Tables \ref{table-rel-sp-pb208}, \ref{table-rel-sp-pb266} 
and \ref{table-rel-sp-z120-n184} show the mean relative energies 
$\overline{\Delta e}_i$ of the pairs of neighboring single-particle states (as 
defined in the NL5(C) functional) and related  standard deviations 
$\sigma(\Delta e_i)$. One can see that in all nuclei the $\sigma(\Delta e_i)$
values are substantially smaller than the $\overline{\Delta e}_i$ values. They are also 
much smaller than the deviations between theory and experiment for one-(quasi)-particle 
configurations  in spherical \cite{LA.11,A.15-jpg,AL.15} and deformed \cite{AS.11,DABRS.15} 
nuclei. Here we assume that statistical errors in the description of the energies of 
deformed  single-particle states are similar to spherical ones which is a reasonable 
assumption considering that deformed states emerge from spherical ones. Thus, one can conclude
that systematic uncertainties in the energies of the single-particle states 
are more important than statistical ones for the predictions of the single-particle 
spectra.

\begin{table}[ht]
\begin{center}
\caption{The same as Table \ref{table-rel-sp-pb208} but for the $^{266}$Pb 
nucleus. The states in the energy range from $\sim -42$ MeV up to $\sim -20$ 
MeV are omitted in order to simplify the table.
\label{table-rel-sp-pb266}
}
\begin{tabular}{|c|c|c|c|c|c|} \hline 
\multicolumn{3}{|c|}{Neutron} & \multicolumn{3}{c|}{Proton} \\ \hline
 Orbital         &           &         &  Orbital           &           &         \\        
 pairs $(m,j)$   &   $\overline{\Delta e}_{\nu}$    & $\sigma(\Delta e_{\nu})$ & pairs $(m,j)$  & 
$\overline{\Delta e}_{\pi}$  & $\sigma(\Delta e_{\pi})$ \\ \hline
     1                     &   2   &   3   &    4                     &    5  &   6   \\ \hline  
  1$s_{1/2}$ - 1$p_{3/2}$  & 5.123 & 0.037 & 1$s_{1/2}$ - 1$p_{3/2}$  & 4.447 & 0.031 \\
  1$p_{3/2}$ - 1$p_{1/2}$  & 0.355 & 0.008 & 1$p_{3/2}$ - 1$p_{1/2}$  & 0.428 & 0.010 \\
            -----          & ----- & ----- &   -----                  & ------& ----- \\
  2$d_{5/2}$ - 2$d_{3/2}$  & 1.143 & 0.029 &2$d_{5/2}$ - 1$h_{11/2}$  & 0.811 & 0.071 \\
  2$d_{3/2}$ - 3$s_{1/2}$  & 0.807 & 0.016 &1$h_{11/2}$ - 2$d_{3/2}$  & 0.322 & 0.096 \\
 3$s_{1/2}$ - 1$h_{11/2}$  & 0.024 & 0.077 & 2$d_{3/2}$ - 3$s_{1/2}$  & 1.348 & 0.019 \\
                           &       &       &   \multicolumn{3}{c|}{below proton Fermi level} \\
 1$h_{11/2}$ - 1$h_{9/2}$  & 3.543 & 0.101 & 3$s_{1/2}$ - 1$h_{9/2}$  & 2.001 & 0.063 \\
  1$h_{9/2}$ - 2$f_{7/2}$  & 2.669 & 0.055 & 1$h_{9/2}$ - 2$f_{7/2}$  & 3.859 & 0.061 \\
  2$f_{7/2}$ - 2$f_{5/2}$  & 1.502 & 0.034 &2$f_{7/2}$ - 1$i_{13/2}$  & 0.451 & 0.079 \\
 2$f_{5/2}$ - 1$i_{13/2}$  & 0.244 & 0.094 &1$i_{13/2}$ - 2$f_{5/2}$  & 1.043 & 0.108 \\
 1$i_{13/2}$ - 3$p_{3/2}$  & 0.581 & 0.088 & 2$f_{5/2}$ - 3$p_{3/2}$  & 1.717 & 0.025 \\
  3$p_{3/2}$ - 3$p_{1/2}$  & 0.562 & 0.014 & 3$p_{3/2}$ - 3$p_{1/2}$  & 0.570 & 0.014 \\ 
 3$p_{1/2}$ - 1$i_{11/2}$  & 3.342 & 0.075 &3$p_{1/2}$ - 1$i_{11/2}$  & 1.307 & 0.074 \\
 1$i_{11/2}$ - 2$g_{9/2}$  & 1.425 & 0.084 &1$i_{11/2}$ - 2$g_{9/2}$  & 3.196 & 0.082 \\
  2$g_{9/2}$ - 2$g_{7/2}$  & 1.590 & 0.031 &2$g_{9/2}$ - 1$j_{15/2}$  & 0.349 & 0.079 \\
  2$g_{7/2}$ - 3$d_{5/2}$  & 0.291 & 0.035 &1$j_{15/2}$ - 2$g_{7/2}$  & 1.338 & 0.105 \\
 3$d_{5/2}$ - 1$j_{15/2}$  & 0.156 & 0.085 & 2$g_{7/2}$ - 3$d_{5/2}$  & 1.748 & 0.034 \\
 1$j_{15/2}$ - 4$s_{1/2}$  & 0.349 & 0.093 & 3$d_{5/2}$ - 3$d_{3/2}$  & 0.688 & 0.014 \\
  4$s_{1/2}$ - 3$d_{3/2}$  & 0.090 & 0.014 & 3$d_{3/2}$ - 4$s_{1/2}$  & 0.590 & 0.018 \\
    \multicolumn{3}{|c|}{below neutron Fermi level}  &                &       &       \\
 3$d_{3/2}$ - 2$h_{11/2}$  & 4.290 & 0.015 &4$s_{1/2}$ - 1$j_{13/2}$  & 1.056 & 0.109 \\
 2$h_{11/2}$ - 4$p_{3/2}$  & 0.255 & 0.064 &                          &       &       \\
 4$p_{3/2}$ - 1$j_{13/2}$  & 0.113 & 0.138 &                          &       &       \\
 1$j_{13/2}$ - 4$p_{1/2}$  & 0.054 & 0.137 &                          &       &       \\
  4$p_{1/2}$ - 3$f_{7/2}$  & 0.086 & 0.022 &                          &       &       \\
  3$f_{7/2}$ - 3$f_{5/2}$  & 0.452 & 0.011 &                          &       &       \\
  3$f_{5/2}$ - 2$h_{9/2}$  & 0.389 & 0.041 &                          &       &       \\ \hline
\end{tabular}
\end{center}
\end{table}

 The underlying single-particle structure is responsible for the differences 
in the predictions of the ground state deformations of superheavy nuclei near 
the $Z=120$ and $N=184$ lines (Ref.\ \cite{AANR.15}). These nuclei could be 
either spherical or oblate dependent on employed CEDF. Thus, it is important 
to estimate statistical errors in the predictions of the $Z=120$ and $N=184$ spherical
shell gaps formed between the $\pi 2 f_{5/2} - \pi 3 p_{3/2}$ and $\nu 4 s_{1/2} - \nu 1j_{13/2}$
pairs of the states (see Figs. 1b and 1d in Ref.\ \cite{AANR.15}). These errors are
very small ($\sigma(\Delta e_{\pi}) = 0.030$ MeV) for the $Z=120$ shell gap which is
characterized by the mean size of 2.116 MeV (see Table \ref{table-rel-sp-z120-n184}).
They are bigger for the $N=184$ shell gap ($\sigma(\Delta e_{\nu}) = 0.102$ MeV)
the mean size of which is equal to 1.177 MeV (Table \ref{table-rel-sp-z120-n184}).
These statistical errors are substantially smaller than the systematic uncertainties
in the shell gap sizes (see Fig. 2a in Ref.\ \cite{AANR.15}) which, as a result, are 
almost fully responsible for the differences in the predictions of the ground state 
properties of the superheavy nuclei under discussion.

\begin{table}[ht]
\begin{center}
\caption{The same as Table \ref{table-rel-sp-pb208} but for the 
$^{304}$120 nucleus. Neutron states in the energy range from $\sim -50$ 
MeV up to $\sim -25$ MeV and proton states in the energy range from 
$\sim -35$ MeV up to $\sim -12$ MeV are omitted in order to simplify 
the table. 
\label{table-rel-sp-z120-n184}
}
\begin{tabular}{|c|c|c|c|c|c|} \hline 
\multicolumn{3}{|c|}{Neutron} & \multicolumn{3}{c|}{Proton} \\ \hline
 Orbital         &           &         &  Orbital           &           &         \\        
 pairs $(m,j)$   &   $\overline{\Delta e}_{\nu}$    & $\sigma(\Delta e_{\nu})$ & pairs $(m,j)$  & 
$\overline{\Delta e}_{\pi}$  & $\sigma(\Delta e_{\pi})$ \\ \hline
     1                     &   2   &   3   &    4                      &    5  &   6   \\ \hline
  1$s_{1/2}$ - 1$p_{3/2}$   & 4.050 & 0.025 & 1$s_{1/2}$ - 1$p_{3/2}$   & 3.935 & 0.020 \\
  1$p_{3/2}$ - 1$p_{1/2}$   & 0.239 & 0.004 & 1$p_{3/2}$ - 1$p_{1/2}$   & 0.299 & 0.006 \\
            -----           & ----- & ----- &               -----       & ----- & ----- \\
  3$s_{1/2}$ - 1$h_{9/2}$   & 1.106 & 0.026 & 3$s_{1/2}$ - 1$h_{9/2}$   & 1.180 & 0.032 \\
  1$h_{9/2}$ - 2$f_{7/2}$   & 3.863 & 0.044 &1$h_{9/2}$ - 1$i_{13/2}$   & 4.019 & 0.070 \\
 2$f_{7/2}$ - 1$i_{13/2}$   & 0.559 & 0.072 &1$i_{13/2}$ - 2$f_{7/2}$   & 0.247 & 0.082 \\
 1$i_{13/2}$ - 2$f_{5/2}$   & 0.941 & 0.105 & 2$f_{7/2}$ - 2$f_{5/2}$   & 1.486 & 0.035 \\
                            &       &       &   \multicolumn{3}{c|}{below proton Fermi level} \\
  2$f_{5/2}$ - 3$p_{3/2}$   & 2.660 & 0.022 & 2$f_{5/2}$ - 3$p_{3/2}$   & 2.116 & 0.030 \\
  3$p_{3/2}$ - 3$p_{1/2}$   & 0.346 & 0.006 & 3$p_{3/2}$ - 3$p_{1/2}$   & 0.318 & 0.006 \\
 3$p_{1/2}$ - 1$i_{11/2}$   & 0.659 & 0.058 & 3$p_{1/2}$ -1$i_{11/2}$   & 0.595 & 0.058 \\
 1$i_{11/2}$ - 2$g_{9/2}$   & 3.094 & 0.075 &1$i_{11/2}$ -1$j_{15/2}$   & 3.146 & 0.109 \\
 2$g_{9/2}$ - 1$j_{15/2}$   & 0.385 & 0.083 &1$j_{15/2}$ - 2$g_{9/2}$   & 0.473 & 0.090 \\
 1$j_{15/2}$ - 2$g_{7/2}$   & 1.603 & 0.123 & 2$g_{9/2}$ - 2$g_{7/2}$   & 1.946 & 0.045 \\
  2$g_{7/2}$ - 3$d_{5/2}$   & 1.870 & 0.020 &                           &       &       \\
  3$d_{5/2}$ - 3$d_{3/2}$   & 0.322 & 0.005 &                           &       &       \\
  3$d_{3/2}$ - 4$s_{1/2}$   & 0.642 & 0.011 &                           &       &       \\
    \multicolumn{3}{|c|}{below neutron Fermi level}  &                  &       &       \\
  4$s_{1/2}$ - 1$j_{13/2}$  & 1.177 & 0.102 &                           &       &       \\
 1$j_{13/2}$ - 2$h_{11/2}$  & 1.917 & 0.107 &                           &       &       \\
 2$h_{11/2}$ - 1$k_{17/2}$  & 0.609 & 0.082 &                           &       &       \\
 1$k_{17/2}$ - 2$h_{9/2}$   & 1.442 & 0.113 &                           &       &       \\
  2$h_{9/2}$ - 3$f_{7/2}$   & 0.600 & 0.041 &                           &       &       \\
  3$f_{7/2}$ - 3$f_{5/2}$   & 0.438 & 0.004 &                           &       &       \\
  3$f_{5/2}$ - 4$p_{3/2}$   & 0.269 & 0.020 &                           &       &       \\
  4$p_{3/2}$ - 4$p_{1/2}$   & 0.179 & 0.003 &                           &       &       \\
  4$p_{1/2}$ - 5$s_{1/2}$   & 2.094 & 0.037 &                           &       &       \\
  5$s_{1/2}$ - 4$d_{5/2}$   & 0.468 & 0.003 &                           &       &       \\
  4$d_{5/2}$ - 4$d_{3/2}$   & 0.039 & 0.001 &                           &       &       \\ \hline
\end{tabular}
\end{center}
\end{table}

  The results presented in Tables \ref{table-rel-sp-pb208}, \ref{table-rel-sp-pb266}, 
and \ref{table-rel-sp-z120-n184} provide also the information on statistical errors in
the description of spin-orbit splittings. Indeed, these tables contain the pairs of the 
orbitals which form spin-orbit doublets such as $p_{3/2}-p_{1/2}$, $d_{5/2} - d_{3/2}$,
$f_{7/2} - f_{5/2}$, $g_{9/2} - g_{7/2}$  and $h_{11/2} - h_{9/2}$. For the majority of the
spin-orbit doublets standard deviations $\sigma(\Delta e_i(m,j))$ are of the order
of 2.7\% of their mean splitting energies ${\Delta e}_i(m,j)$. Indeed, for 14 spin-orbit
doublets of $^{208}$Pb seen in Table \ref{table-rel-sp-pb208}, the ratio 
$\sigma(\Delta e_i(m,j))/{\Delta e}_i(m,j)$ is located in the range from 0.023 up to 
0.031. In $^{266}$Pb and $^{304}$120 nuclei, the standard deviations 
$\sigma(\Delta e_i(m,j))$ are of the order of 2.4\% and 2.0\% of their 
mean splitting energies ${\Delta e}_i(m,j)$, respectively (see Tables 
\ref{table-rel-sp-pb266} and \ref{table-rel-sp-z120-n184}). Thus, statistical 
errors (as compared with those seen in $^{208}$Pb) in the description of 
spin-orbit splittings do not increase on going towards the extremes of 
neutron number or charge.

\begin{figure*}[htb]
\includegraphics[angle=-90,width=8.5cm]{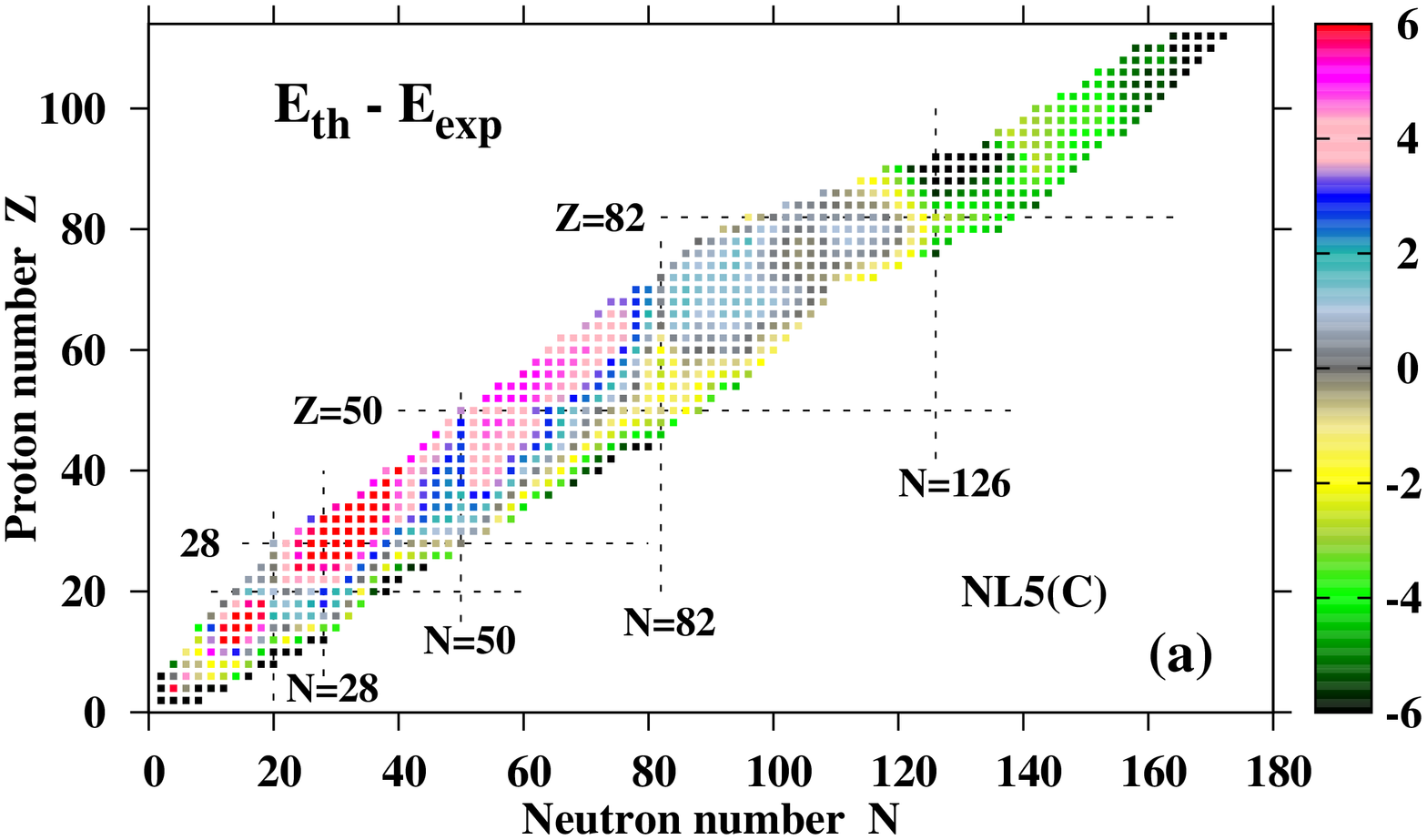}
\includegraphics[angle=-90,width=8.5cm]{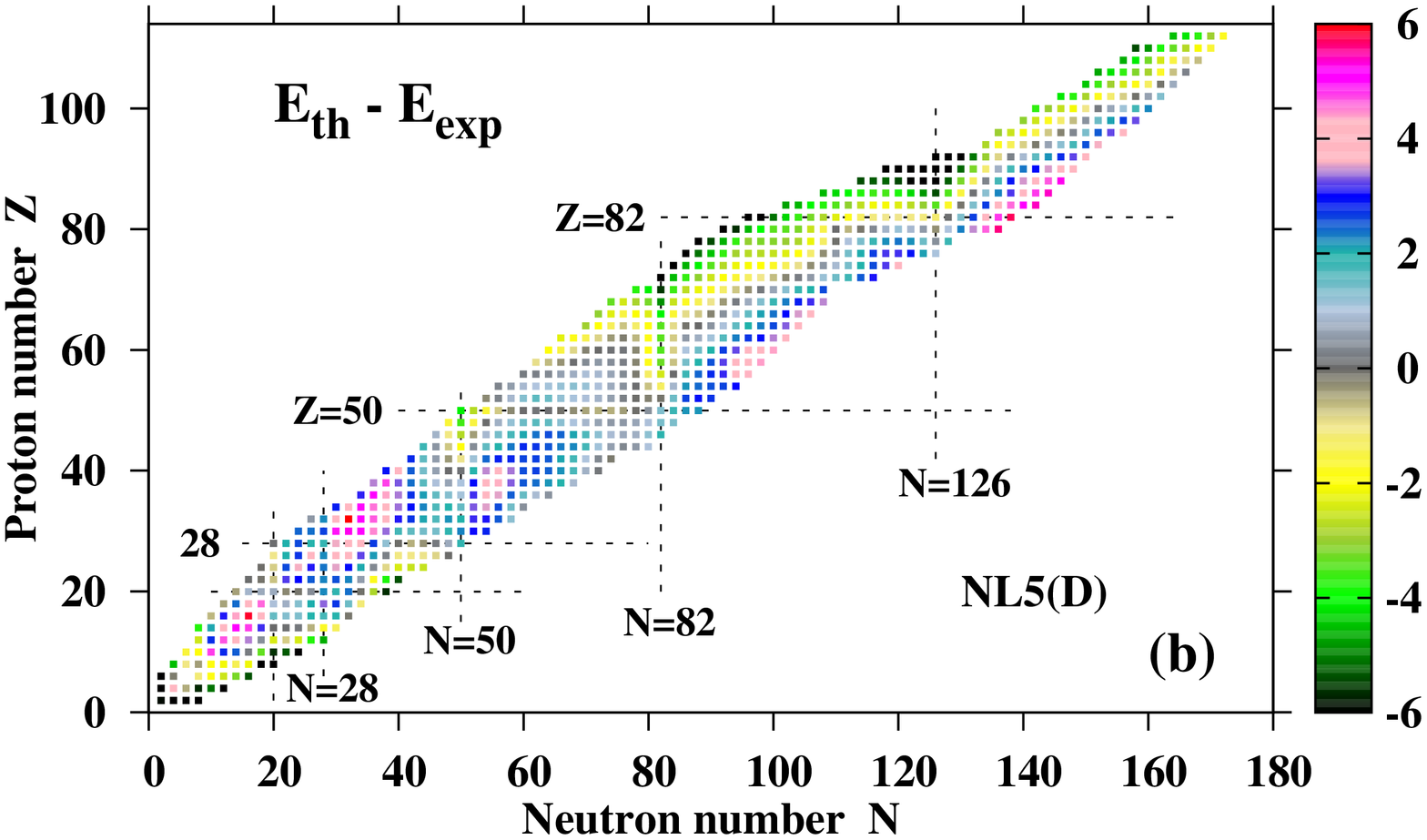}
\includegraphics[angle=-90,width=8.5cm]{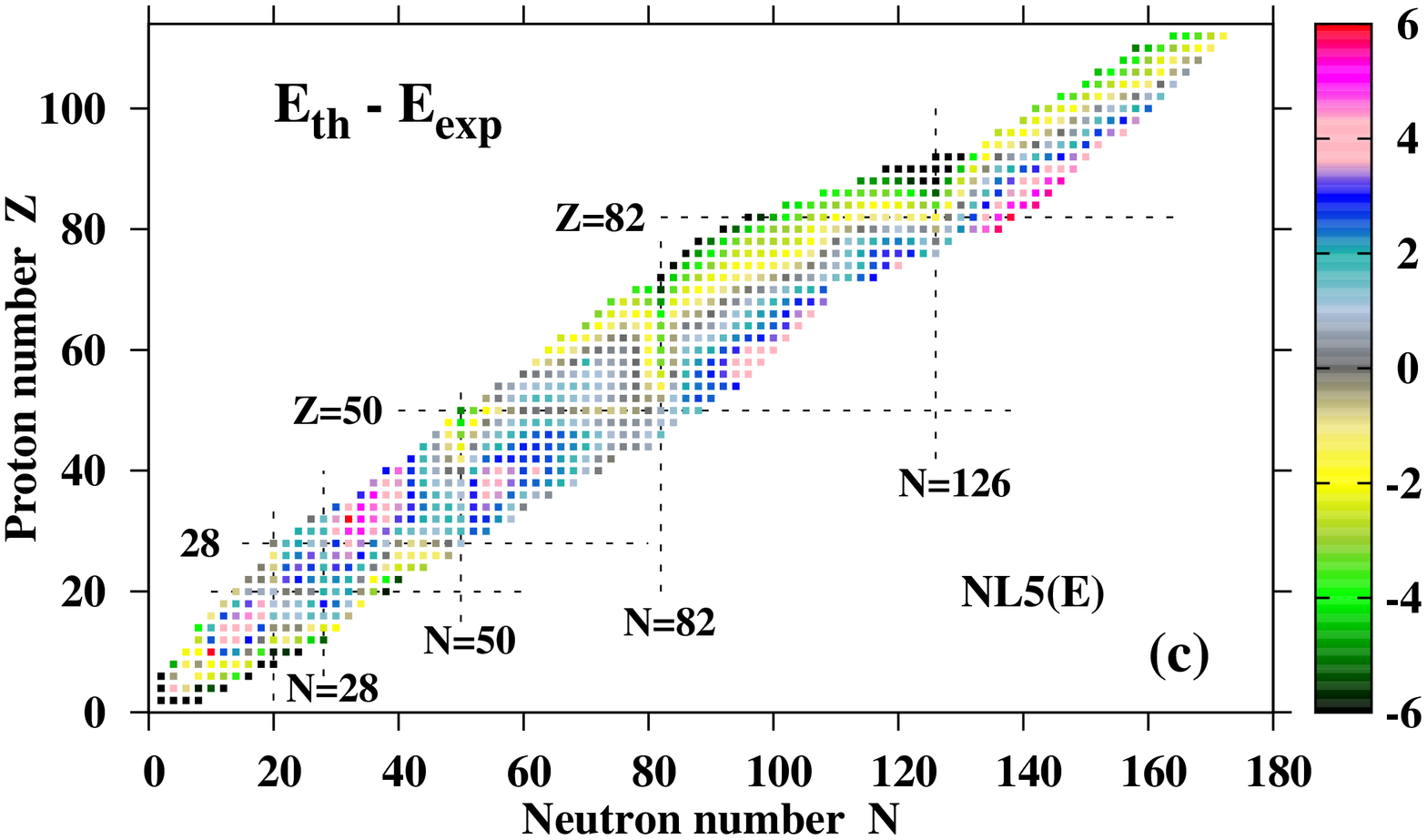}
\includegraphics[angle=-90,width=8.5cm]{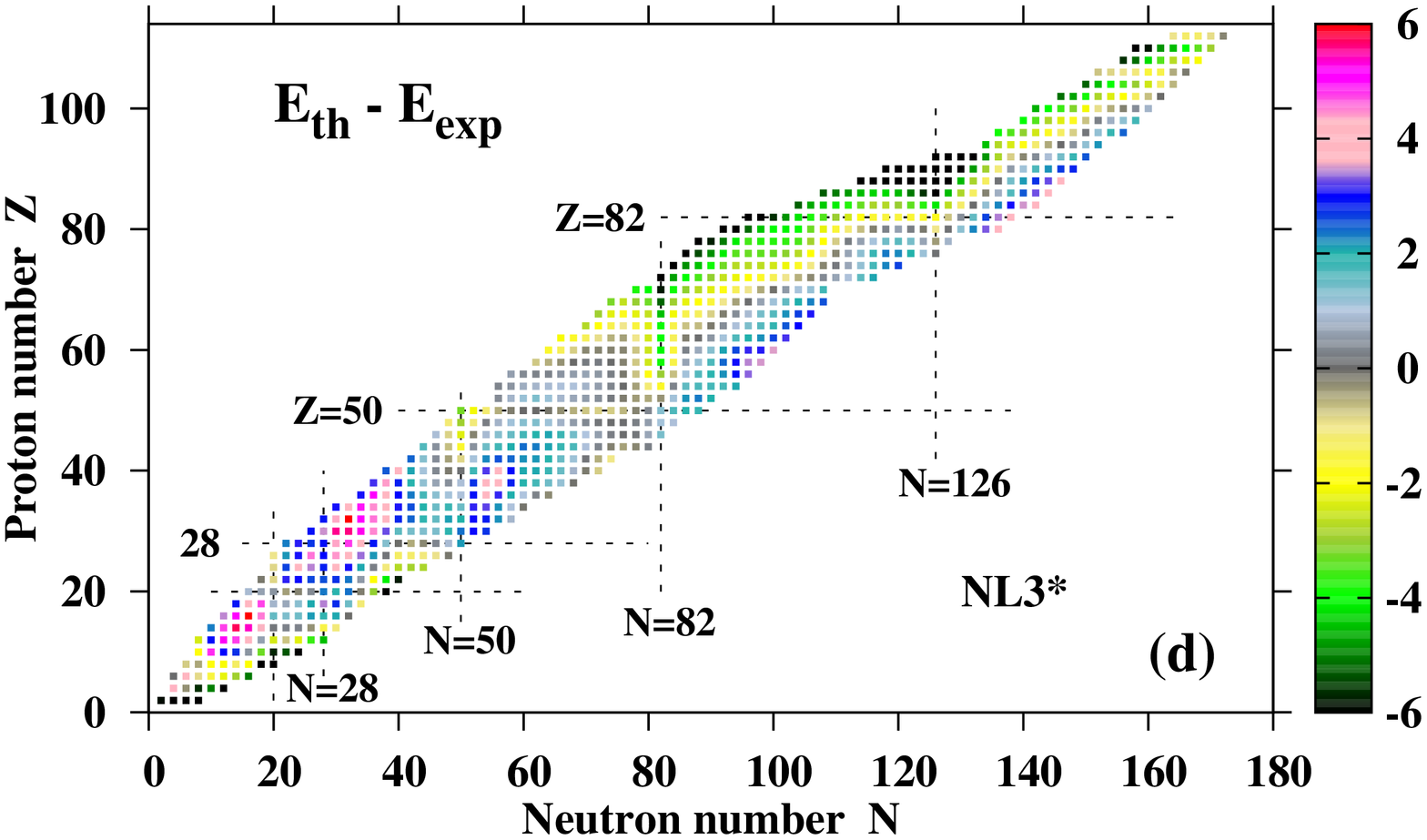}
\caption{(Color online) The differences $E_{th}-E_{exp}$ between 
calculated and experimental binding energies for the indicated 
CEDFs. The experimental data are taken from Ref.\ \cite{AME2012} and 830 
even-even nuclei, for which measured and estimated masses are 
available, are included. If $E_{th}-E_{exp} < 0$, the nucleus is more 
bound in the calculations than in experiment. 
\label{Global-masses}
}
\end{figure*}

  Statistical errors in the description of the single-particle energies
have also been analysed in lighter $^{48}$Ca and $^{132}$Sn nuclei. They 
show the same general features as those discussed above in $^{208,266}$Pb 
and $^{304}$120. As compared with $^{208}$Pb, statistical errors in the 
energies of the single-particle states located in the vicinity of the 
Fermi levels are similar/somewhat smaller in $^{132}$Sn/$^{48}$Ca.

  It is interesting to compare our CDFT results with those obtained in
Skyrme DFT framework with the UNEDF0 functional and presented in Table I 
of Ref.\ \cite{GDKTT.13}. For the neutron/proton states of $^{208}$Pb 
shown in this table the statistical errors obtained in the Skyrme DFT 
calculations are on average by a factor of 2.05/1.46 larger than those 
obtained in our CDFT calculations (compare Table \ref{table-sp-208Pb} in 
the present manuscript with Table I of Ref.\ \cite{GDKTT.13}).
 In addition, there is one principal difference between the CDFT and 
Skyrme DFT results. The standard deviations $\sigma(\Delta e_i(m,j))$ 
for the spin-orbit splittings are very small in the CDFT calculations; 
they are typically on the level of 2-3\% of total size of spin-orbit 
splitting. On the contrary, they are substantially larger (both in 
relative and absolute senses) in the Skyrme DFT calculations (see Table 
I in Ref.\ \cite{GDKTT.13}).

\section{CONCLUSIONS}
\label{concl}

  Statistical errors in ground state observables and 
single-particle properties of spherical even-even nuclei 
and their propagation to the limits of nuclear landscape 
have been investigated in covariant density functional 
theory for the first time.  The main results can be 
summarized as follows:

\begin{itemize}

\item
  Statistical errors in  binding energies, charge radii, 
neutron skins and two-neutron separation energies have been
studied for the Ca, Ni, Sn and Pb nuclei located between 
two-proton and two-neutron drip lines. While statistical
errors for binding energies and neutron skins drastically
increase on approaching two-neutron drip line, such a trend
does not exist for statistical errors in charge radii and
two-neutron separation energies. The latter  is contrary to 
the trends seen in Skyrme density functional theory. Statistical 
errors obtained in the CDFT calculations are substantially 
smaller than related systematic uncertainties.

\item
  The absolute energies of the single-particle states in the 
vicinity of the Fermi level are characterized by low 
statistical errors ($\sigma(e_i)\sim 0.1$ MeV). This is also
true for relative energies of the single-particle states. 
These statistical errors are substantially smaller than 
systematic uncertainties in the predictions of the absolute
and relative energies of the single-particle states. Thus, they 
are not expected to modify in a substantial way the predictions 
of a given CEDF. This is true both for known nuclei and for 
nuclear extremes such as the vicinity of neutron-drip line and 
the region of superheavy elements.

\item
  The statistical errors in the predictions of spin-orbit splittings
are rather small. For the spin-orbit doublets in studied nuclei, the 
standard deviations $\sigma(\Delta e_i(m,j))$ are of the order of 2.4\% of 
their mean splitting energies $\overline{\Delta e}_i(m,j)$. These errors 
are quite robust and they do not increase on going towards the extremes 
of  neutron number or charge.

\item
  Statistical errors in the description of physical observables related 
to the ground state and single-particle degrees of freedom are substantially 
lower in CDFT as compared with Skyrme DFT. A special feature of CDFT 
due to which the parameters of the $\omega$ and $\sigma$ mesons, 
defining the basis features of the nucleus such as a nucleonic
potential, are well localized in very narrow range of the parameter 
hyperspace, is responsible for that. Note that fixing the $g_{\rho}$, 
$g_2$ and $g_3$ parameters of the model leads to drastic reduction of 
statistical errors as compared with the case when all parameters of 
the non-linear functional are permitted to vary in Monte-Carlo 
procedure.

\item
  The present investigation reveals strong correlations between a 
number of the  parameters defining the non-linear CEDFs. Note that 
these correlations are dependent on the details of fitting protocol. 
They are especially pronounced for the $g_2$ and $g_3$ parameters 
responsible for the density dependence of the model. This suggests 
that these parameters are not independent. Thus, the accounting of 
these parametric correlations will allow in future to reduce the 
number of free parameters of non-linear meson coupling models 
from six to five.

\end{itemize}

   Considering the structure of non-linear meson coupling models and typical
features of existing non-linear CEDFs and their fitting protocols,
it is reasonable to expect that different non-linear functionals 
will provide comparable statistical errors for the physical observables 
of interest. This was illustrated by comparison of  the results for 
the NL5(C) and NL5(A) functionals.

Note that obtained statistical errors represent in a sense their upper
limit since the fitting protocol includes only limited set of nuclei and
empirical data. It is expected that the increase of the size of the dataset 
in the fitting protocol will lead to further reduction of statistical errors \cite{DNR.14}.

  There are clearly many physical observables which are left 
outside the present study. However, based on the present results 
we can evaluate statistical errors of some of them. One of the 
examples is the energies of the single-particle states in deformed 
nuclei.   The wavefunctions of the deformed single-particle states 
are the mixtures of the contributions coming from different spherical 
single-particle states. Thus, the statistical errors for the energies 
of deformed single-particle states are expected to be of similar 
magnitude as those for spherical states. However, the fluctuations 
in their magnitudes are expected to be smaller as compared with 
spherical states because of the above mentioned mixing. For the same 
reasons, the statistical errors in charge radii and neutron 
skins of deformed nuclei are expected to be comparable with 
spherical ones.

\begin{figure*}[htb]
\includegraphics[angle=-90,width=8.5cm]{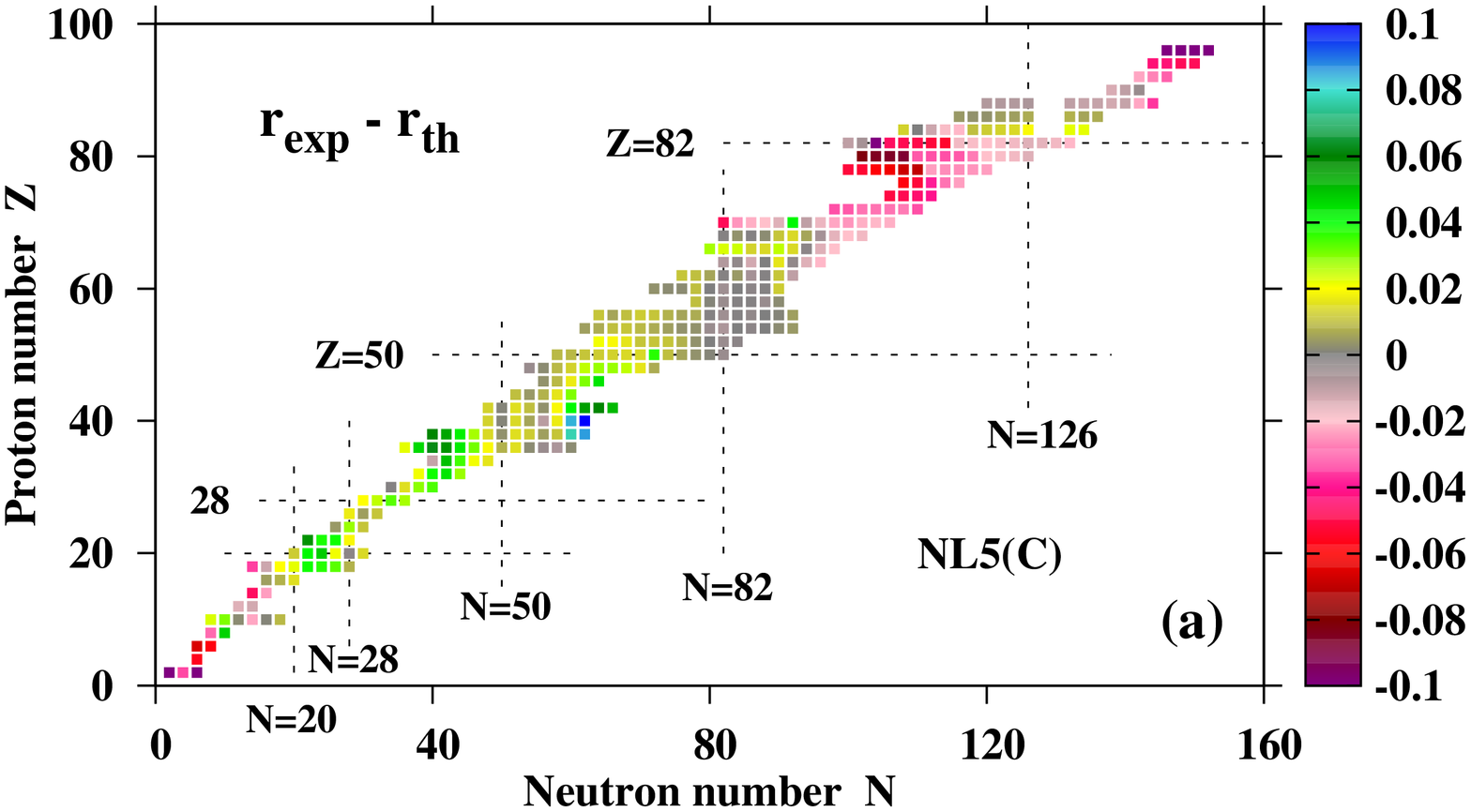}
\includegraphics[angle=-90,width=8.5cm]{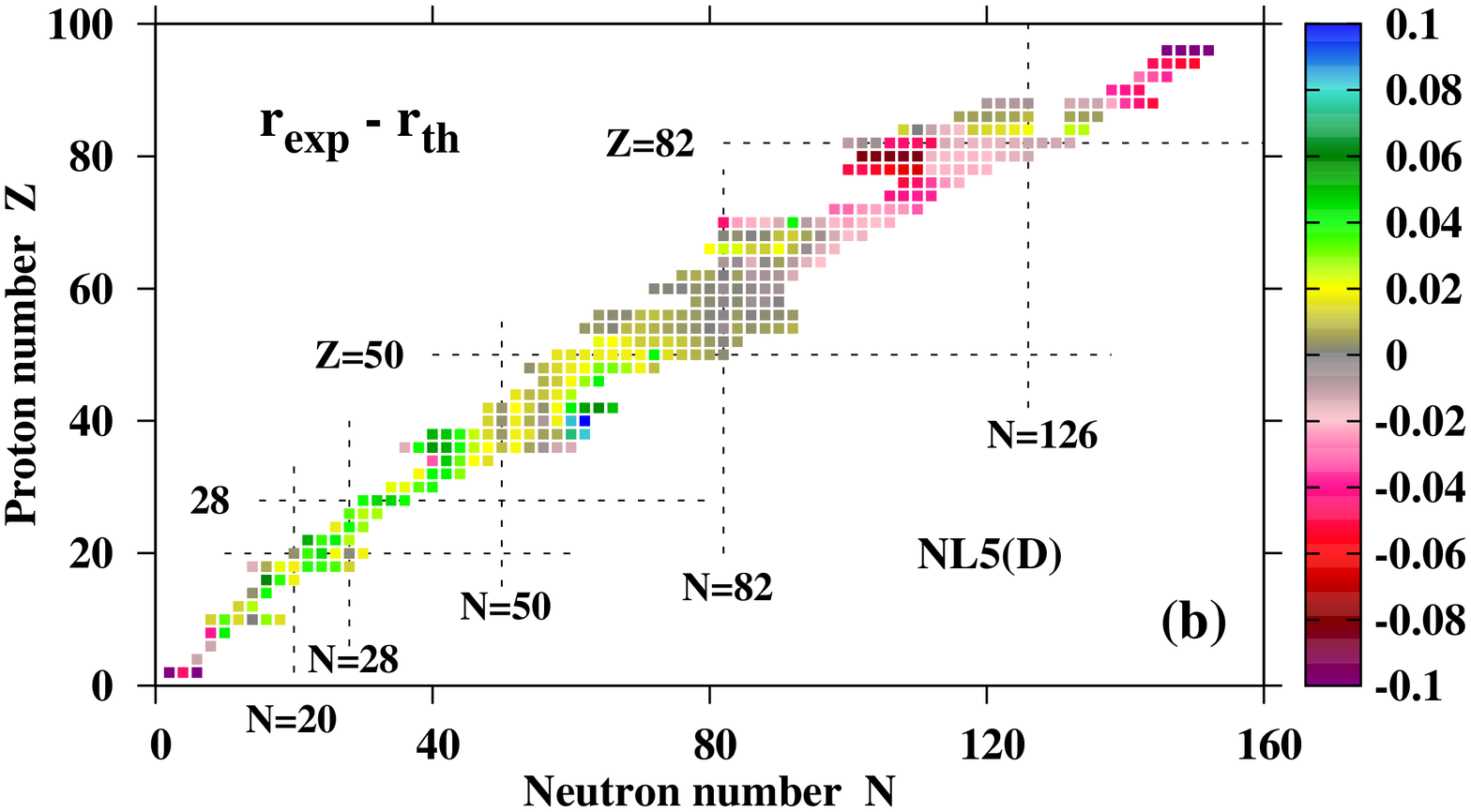}
\includegraphics[angle=-90,width=8.5cm]{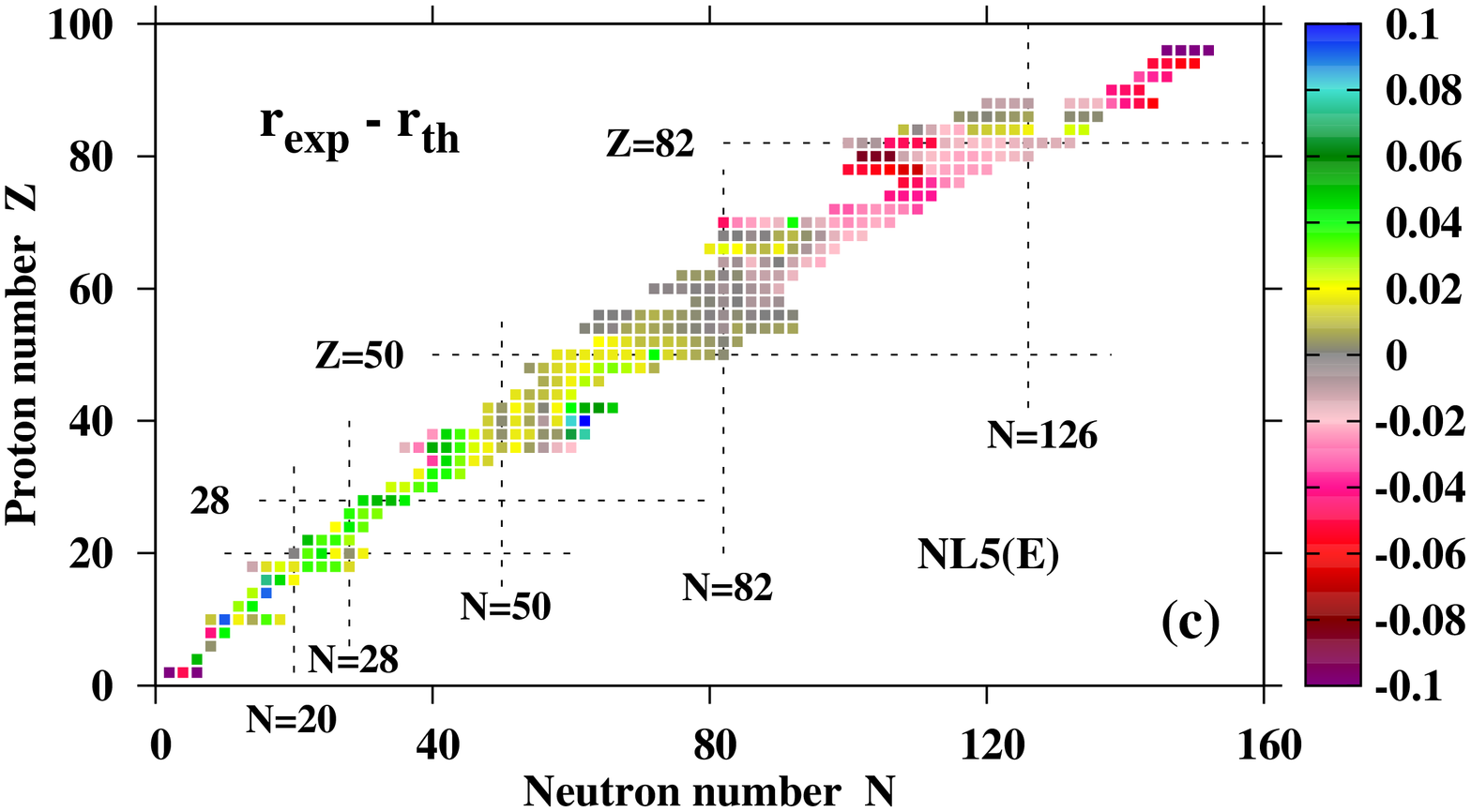}
\includegraphics[angle=-90,width=8.5cm]{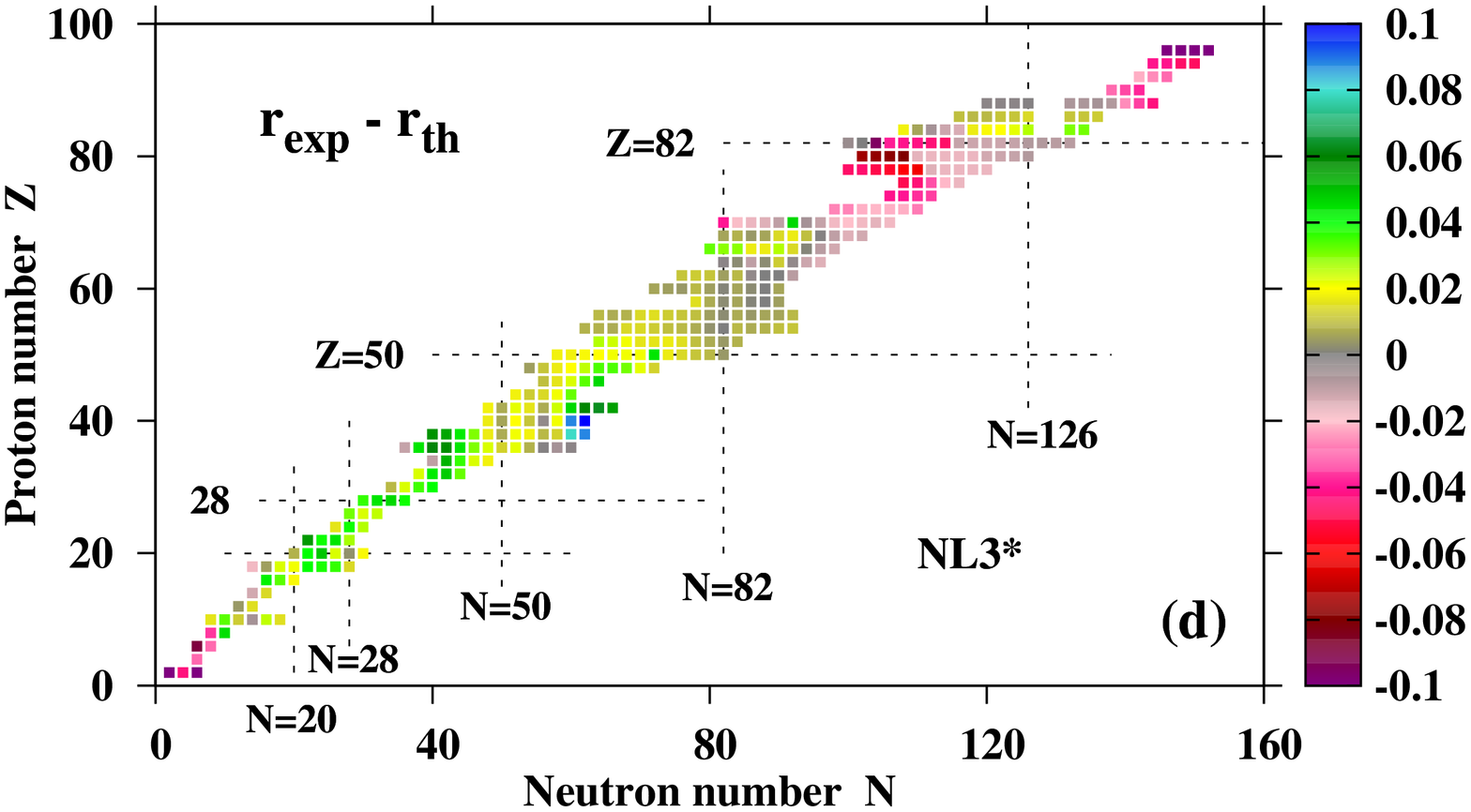}
\caption{(Color online)  The difference between measured and calculated 
charge radii $r_{ch}$ for indicated functionals. The experimental data are 
taken from Ref.\  \cite{AM.13}.}
\label{Global-radii}
\end{figure*}

\begin{figure*}[htb]
\includegraphics[angle=-90,width=8.5cm]{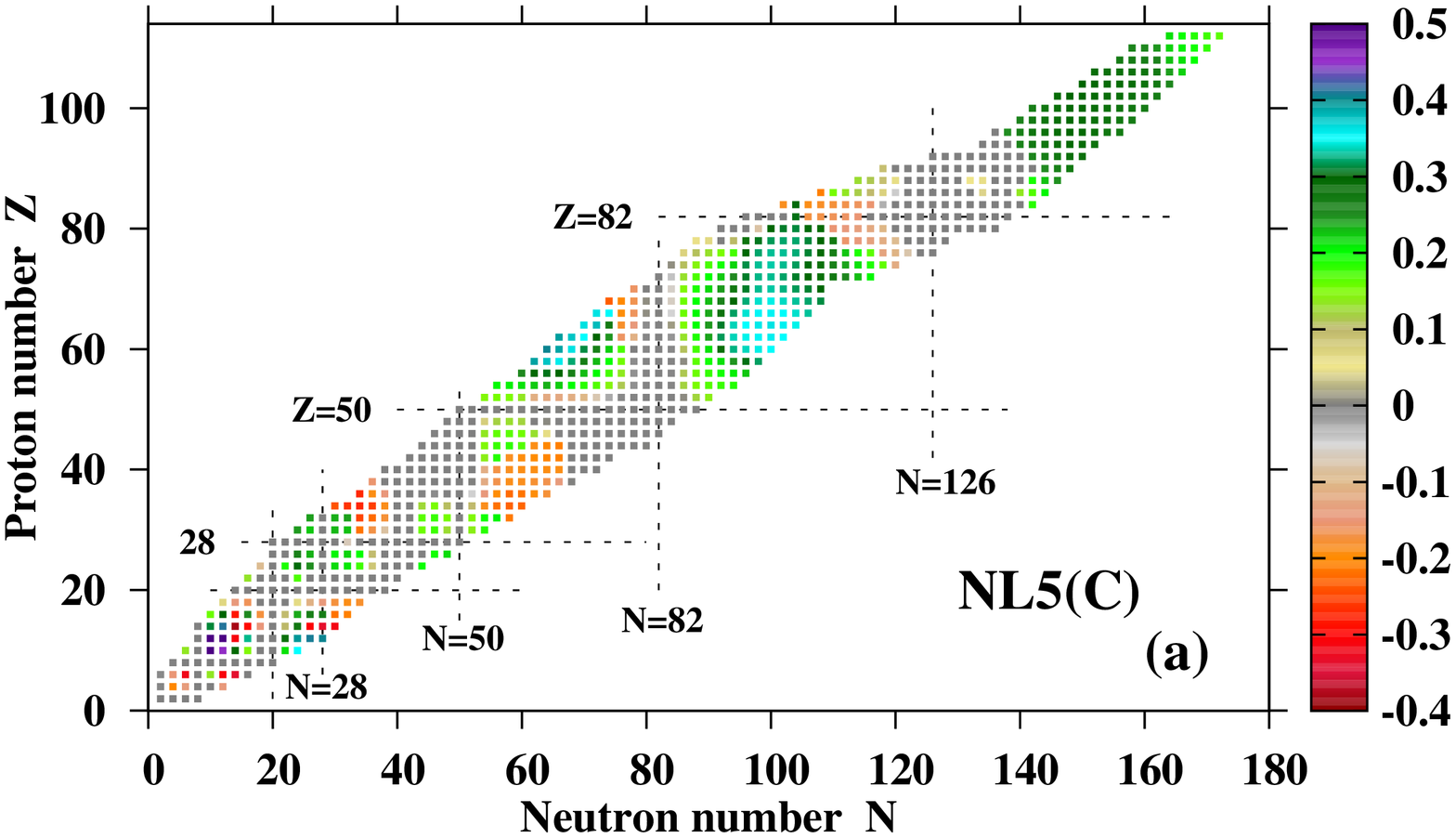}
\includegraphics[angle=-90,width=8.5cm]{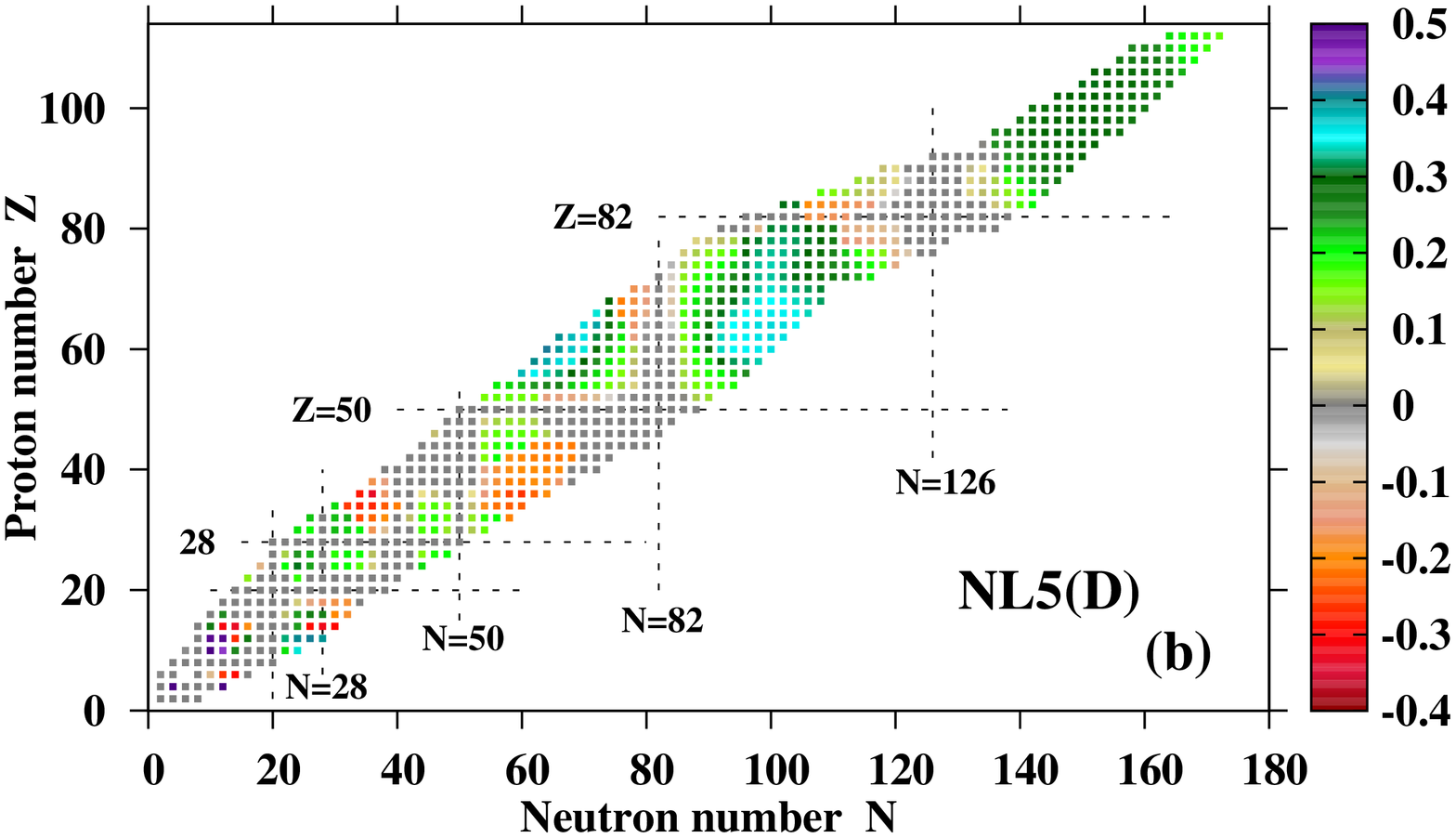}
\includegraphics[angle=-90,width=8.5cm]{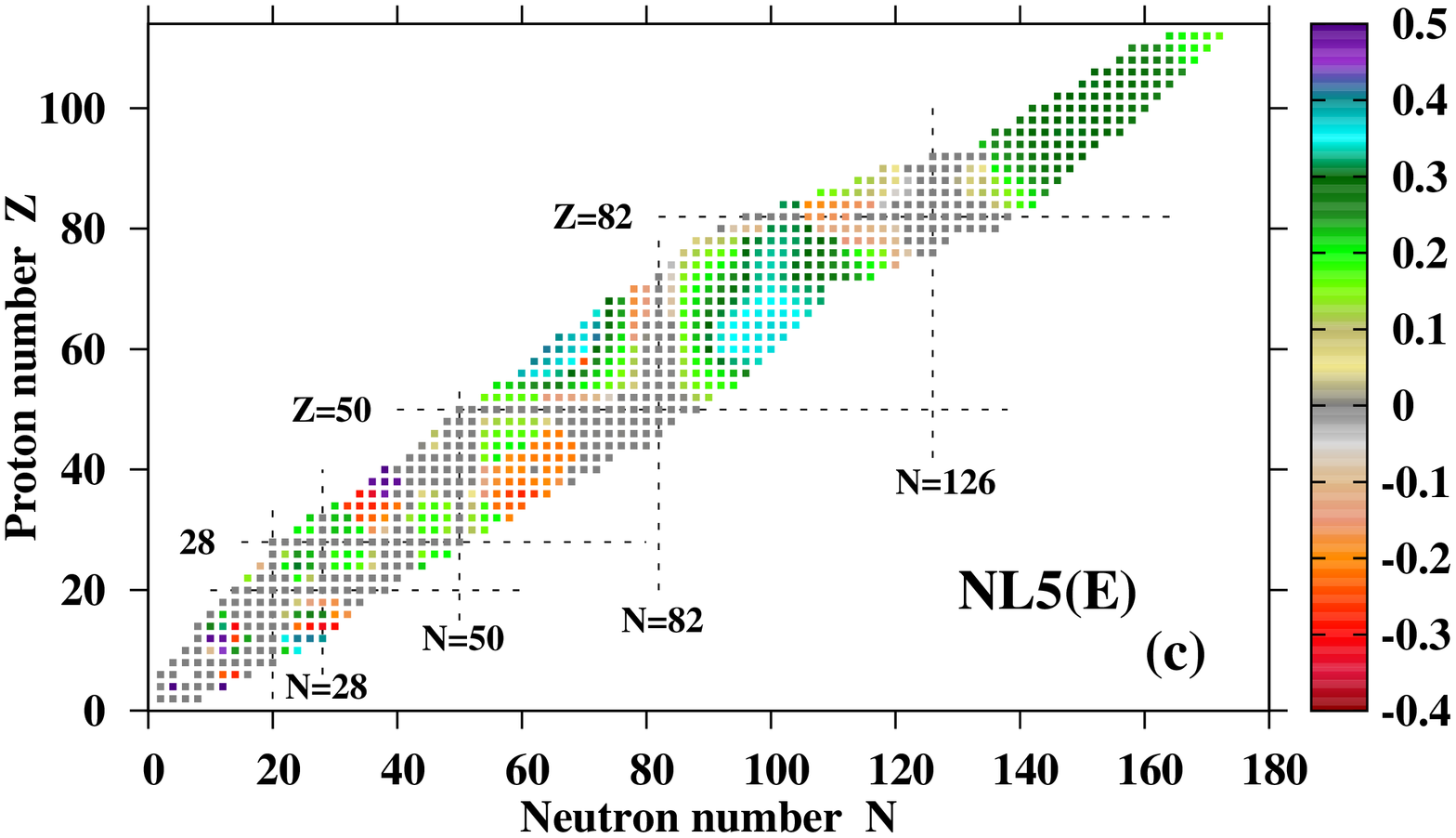}
\includegraphics[angle=-90,width=8.5cm]{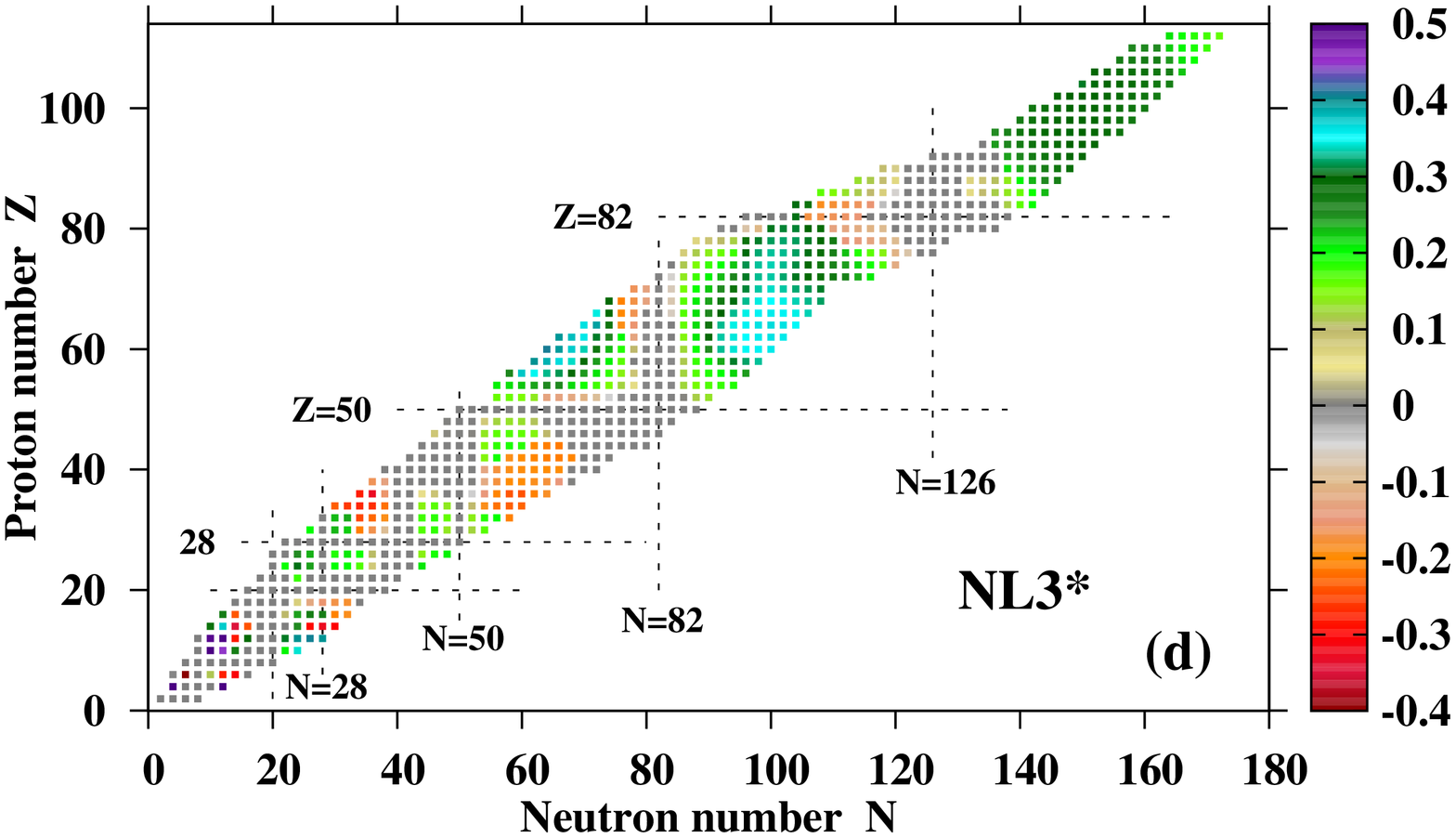}
\caption{(Color online) Charge quadrupole deformations $\beta_2$ of the
ground states in even-even nuclei obtained in the RHB calculations with 
indicated CEDFs.}
\label{Global-deformations}
\end{figure*}

\begin{table*}[h]
  \begin{center}
  \caption{ The rms deviations $\Delta{E_{rms}}$, $\Delta{(S_{2n})_{rms}}$ 
          ($\Delta{(S_{2p})_{rms}}$) and $\Delta{(r_{ch})_{rms}}$ 
           between calculated and experimental binding energies $E$,
           two-neutron (-proton) separation energies $S_{2n}$ $(S_{2p})$           
           and charge radii $r_{ch}$ for indicated CEDFs. For first three 
           observables,  they are given  with respect of 
           "measured" (640 nuclei) and "measured+estimated"  (830 nuclei)
           sets of experimental masses.  
           For the calculations of the $\Delta{(r_{ch})_{rms}}$ values, all 
           experimental data are used in column 6, while the data on radii 
           of He ($Z = 2$) and Cm ($Z = 96$) isotopes are excluded in column 7. 
           See text for the discussion of these cases.
\label{all-data}
}
  \begin{tabular}{|c|c|c|c|c|c|c|}
  \hline\hline
   CEDF & measured & \multicolumn{3}{|c|}{measured+estimated} & \multicolumn{2}{|c|}{charge radii}   
   \\  \hline
    & $\Delta{E_{rms}}$ [MeV] & $\Delta{E_{rms}}$ [MeV] &$\Delta{(S_{2n})_{rms}}$ [MeV] &$\Delta{(S_{2p})_{rms}}$ [MeV]& $\Delta{(r_{ch})_{rms}}$ [fm] & $\Delta{(r_{ch})_{rms}}$ [fm] \\\hline
       1           &     2    &  3     &   4   &  5    &    6     &     7     \\ \hline
    NL5(C)      & 3.41 & 3.71 & 1.37 & 1.54 & 0.040 & 0.0284 \\
    NL5(D)      & 2.83 & 2.90 & 1.22 & 1.29 & 0.041 & 0.0277 \\
    NL5(E)      & 2.73 & 2.81 & 1.23 & 1.29 & 0.042 & 0.0288 \\
    NL3*        & 2.96 & 3.00 & 1.23 & 1.29 & 0.041 & 0.0283 \\
  \hline
  \end{tabular}
  \end{center}
\end{table*}

\begin{table*}[htb]
\begin{center}
\caption{The values of the parameters of different sets for the finite range Gogny forces
(see equations in the appendix of Ref.\ \cite{D1S} or Eq. (3) in Ref.\ \cite{GHGP.09}
for the definition of the Gogny force). The $\mu_i$ $(i=1,2)$ parameters describe the range 
of the interaction. The strength of the Wigner, Bartlett, Heisenberg and Majorana terms is given 
by $W_i$, $B_i$, $H_i$ and $M_i$, respectively. The parameters $t_0$, $x_0$ and $\alpha$ 
define the density dependence of the interaction, while $W_{LS}$ zero-range two-body 
spin-orbit interaction.}
\label{table-Gogny}
\begin{tabular}{| c| c| c| c| c|  c| c|} \hline 
\multicolumn{1}{|c|}{Parameters} &\multicolumn{2}{c|}{D1S\cite{D1S}} &\multicolumn{2}{c|}{D1\cite{D1}}&\multicolumn{2}{c|}{D1M\cite{GHGP.09}} \\ \hline
$\mu_i$ (fm)        &     0.7  &    1.2           &     0.7 &    1.2 &     0.5 &    1.0    \\ \hline
$W_i$  [MeV]       & -1720.30 &  103.64   &  -402.4 &  -21.30&-12797.57&  490.95    \\ \hline
$B_i$   [MeV]       & 1300.00 & -163.48    &   -100.0 &  -11.77& 14048.85& -752.27    \\ \hline
$H_i$   [MeV]       &-1813.53 &  162.81    &  -496.2 &   37.27&-15144.43&  675.12     \\ \hline
$M_i$   [MeV]       & 1397.60 & -223.93.   &  -23.56&  -68.81&11963.89 & -693.57      \\ \hline
\multicolumn{1}{|c|}{$t_0$ [MeV]} &\multicolumn{2}{c|}{1390.6} &\multicolumn{2}{c|}{1350.0}&\multicolumn{2}{c|}{1562.22 } \\ \hline
\multicolumn{1}{|c|}{$x_0$} &\multicolumn{2}{c|}{ 1} &\multicolumn{2}{c|}{1 }&\multicolumn{2}{c|}{1 } \\ \hline
\multicolumn{1}{|c|}{$\alpha$} &\multicolumn{2}{c|}{1/3} &\multicolumn{2}{c|}{1/3 }&\multicolumn{2}{c|}{1/3 } \\ \hline
\multicolumn{1}{|c|}{$W_{LS}$ [MeV] } &\multicolumn{2}{c|}{-130.0 } &\multicolumn{2}{c|}{115.0 }&\multicolumn{2}{c|}{115.56} \\ \hline
\end{tabular}
\end{center}
\end{table*}

\begin{table*}[htb]
\begin{center}
\caption{The values of the parameters for different Skyrme forces. When needed
the parameters are reduced  to the form of the Skyrme force employed for the BSk class 
of the functionals (see Eq.\ (1) in Ref.\ \cite{GCP.13}). The parameters $t_0$ and $x_0$ 
define the central term.  $t_1$, $x_1$, $t_2$, $x_2$ are the parameters of the non-local 
terms.  Density dependence is defined by  $t_3$, $x_3$ and $\gamma$ parameters.
The $W_0$ and $W_1$ parameters define the spin-orbit interaction.  The BSk28 and 
BSk29 functionals have many additional parameters as compared with other functionals; 
some of them ($t_4$,  $x_4$, $t_5$, $x_5$) are shown here but others (see 9 bottom lines 
of Table 1 in Ref.\  \cite{Gor.15}) are omitted for simplicity.
\label{table-Skyrme}
}
\begin{tabular}{| c| c| c| c| c| c| c|} \hline 
Parameters                &   UNEDF0 \cite{UNEDF0} &    UNEDF1 \cite{UNEDF1}    & SLy4 \cite{SLy4} & SKM* \cite{SkM*} & BSk28 \cite{Gor.15}& BSk29 \cite{Gor.15} \\ \hline\hline
$t_0[MeV fm^3]$          & -1883.68781034     & -2078.32802326              & -2488.91    & -2645.00   &  -3988.86  &  -3970.40     \\ \hline
$t_1[MeV fm^5]$           &   277.50021224     &   239.40081204              &   486.82    & 410.00     &   395.769  &    394.880    \\ \hline
$t_2[MeV fm^5]$           &   608.43090559     &  1575.11954190              &  -546.39    & -135.00    &            &               \\ \hline
$t_3[MeV fm^{3+3\alpha}]$   & 13901.94834463     & 14263.64624708              &  13777.0    & 15595.0    &   22774.4  &   22649.3     \\ \hline 
$t_4[MeV fm^{5+3\beta}]$    &                    &                             &             &            &   -100.000 &  -100.000     \\ \hline
$t_5[MeV fm^{5+3\gamma}]$   &                    &                             &             &            &   -150.000 &  -150.000     \\ \hline
$x_0$                     & 0.00974375         & 0.05375692                  &  0.834      & 0.09       &  0.928026  &  0.964850     \\ \hline
$x_1$                     & -1.77784395        & -5.07723238                 &  -0.344     & 0.00       &  0.0274980 & -0.0047741    \\ \hline
$x_2$                     & -1.67699035        & -1.36650561                 &  -1.000     & 0.00       &            &               \\ \hline
$t_2x_2[MeV fm^5]$        &                    &                             &             &            &  -1388.61  & -1388.95      \\ \hline
$x_3$                     & -0.38079041        & -0.16249117                 &  1.354      & 0.00       &   1.09482  & 1.14453       \\ \hline
$x_4$                     &                    &                             &             &            &   2.0000   & 2.00000       \\ \hline
$x_5$                     &                    &                             &             &            &  -11.0000  &-11.0000        \\ \hline
$W_0[MeV fm^5]$           & 33.9006 & 109.6845943    &  123.0      & 130.0      &   80.489   &  64.600         \\ \hline
$W_1[MeV fm^5]$           &                     &                            &             &            &  180.411   &     0           \\ \hline
$\alpha$                  &   0.32195599        & 0.27001801                 &   1/6       & 1/6        &   1/12     &    1/12          \\ \hline
$\beta $                  &                     &                            &             &            &   1/2      &    1/2           \\ \hline
$\gamma$                  &                     &                            &             &            &    1/12    &    1/12          \\ \hline
$\nu$                     &                     &                            &             &            &    1       &      0           \\ \hline
$y_\omega$                 &                     &                            &             &            &    0       &      2           \\ \hline
\end{tabular}
\end{center}
\end{table*}

  Another example is time-odd mean fields. They have an impact on 
a considerable number of physical observables in the systems with 
broken time-reversal symmetries \cite{VALR.05,AA.10}. However, 
their impact depends on the $g_{\omega}$ and $m_{\omega}$ parameters 
\cite{AA.10,TO-rot}, which according to our results vary very 
little (see Fig.\ \ref{Paramet}a, b, c, \ref{NL5-ksi-distrib}b and 
\ref{NL5(A)-ksi-distrib}b). Note that  $m_{\omega}$ is fixed in
many functionals.  These facts suggest that statistical errors in 
time-odd mean fields and the components of physical observables 
related to time-odd mean fields (such as the contribution to the 
moment of inertia due to time-odd mean fields \cite{TO-rot} or 
additional binding due to nuclear magnetism \cite{AA.10}) 
should be reasonably small. 

\section{ACKNOWLEDGMENTS}

This material is based upon work supported by the U.S. 
Department of Energy,  Office of Science, Office of 
Nuclear Physics under Award No. DE-SC0013037.

\appendix
\section{Global performance of the NL5(*) covariant 
                                  energy density functionals }
\label{Glob-perfom}                                  
 
  The global performance of the NL5(C), NL5(D) and NL5(E) functionals
in the description of ground state properties of even-even nuclei  is 
presented in Figs.\ \ref{Global-masses} and \ref{Global-radii} and summarized 
in Table \ref{all-data}. Experimental data on binding energies of 835 even-even 
nuclei is taken from Ref.\  \cite{AME2012}; note that  there are 640 measured 
and 195 estimated masses of even-even nuclei in the AME2012 mass
evaluation\footnote{For simplicity, we exclude 5 superheavy nuclei with
$Z=114, 116$ and 118 from comparison between theory and experiment
since the definition of the ground state (prolate normal deformed or
superdeformed) in these nuclei requires extensive and time consuming 
calculations 
in the axial RHB code with octupole deformation and triaxial RHB
code. However, the exclusion of these nuclei has very little effect on
rms deviations presented in the columns 2-5 of Table \ref{all-data}.}.
Experimental data on charge radii of 351 even-even nuclei is taken 
from Ref.\ \cite{AM.13}. The global performance of the NL5(*) functionals is also 
compared with the one of the NL3* functional studied in details in Refs.\ 
\cite{AARR.14,AANR.15,AA.16}.  Note that the NL3* is the state-of-the-art 
functional for the non-linear coupling meson exchange models (see Ref.\ 
\cite{AARR.14}) with well  documented record of successful applications to the 
ground state properties of even-even nuclei \cite{NL3*,AARR.14,AA.16},
octupole deformation in the ground states of even-even nuclei \cite{AAR.16},  giant 
resonances \cite{NL3*}, the energies and spectroscopic factors of dominant components 
of single-particle states in odd-mass nuclei \cite{LA.11,AL.15},  rotating nuclei 
\cite{NL3*,AO.13},  fission barriers \cite{AAR.10,AAR.12},  superheavy nuclei 
\cite{AANR.15} etc.

  The rms deviations $\Delta{E_{rms}} \sim 3.7$  MeV between calculated and experimental 
binding energies $E$ for the NL5(C) CEDF are very similar to those obtained with original 
NL3 functional (see Ref.\ \cite{LRR.99}). The NL3* functional \cite{NL3*,AARR.14} with 
$\Delta{E_{rms}} = 3.0$ MeV represents 
an improved version of this functional.  The NL5(D) and especially NL5(E) functionals 
provide further improvement of global description of masses as compared with the NL3*
one (see Table \ref{all-data}).  They produce comparable with NL3*  description of two-neutron 
($S_{2n}$) and two-proton ($S_{2n}$) separation energies (see Table \ref{all-data}).  With minor 
differences the distribution of the $E_{th} - E_{exp}$  
quantities in the $(Z,N)$ plane is similar for the NL5(D), NL5(E) and NL3* functionals
(Fig.\ \ref{Global-masses}). On the contrary, there are substantial differences between
the NL5(C) and NL3* functionals in that respect.
     
   All functionals give comparable rms deviations for charge radii
$\Delta(r_{ch})_{rms}$ (see Table \ref{all-data}). Note that the last
column in Table \ref{all-data} excludes experimental data on He 
($Z=2$) (3 data points) and Cm ($Z=96$) (4 data points) nuclei 
(see detailed motivation in Sect. X of Ref.\ \cite{AARR.14}). It is clear 
that DFTs cannot describe very light nuclei such as He. In addition, 
experimental charge radii of  the Cm ($Z = 96$) nuclei are lower than 
those of Pu ($Z = 94$) and U ($Z = 92$) \cite{AM.13}. Such feature 
goes against a general trend of the increase of charge radii with 
the increase of proton number for comparable deformations and could 
be described neither in CDFT (see Ref.\ \cite{AARR.14}) nor in 
non-relativistic DFT calculations with Gogny D1S functional  (see 
Supplemental Material to Ref.\ \cite{DGLGHPPB.10}).

\section{The examples of the spread of the parameters 
                                  in non-relativistic functionals}
\label{param_spread}                                  

     Tables \ref{table-Gogny} and \ref{table-Skyrme} illustrate the spread of model 
parameters  in the Gogny and Skyrme energy density functionals.  These are
state-of-the-art finite range Gogny functionals D1S \cite{D1S} and D1M \cite{GHGP.09} 
and state-of-the-art zero-range Skyrme functionals  UNEDF0 \cite{UNEDF0}, 
UNEDF1\cite{UNEDF1}, BSk28 \cite{Gor.15} and BSk29 \cite{Gor.15}. They are
compared with older Skyrme functionals SLy4 \cite{SLy4} and SkM* \cite{SkM*} and
first Gogny functional D1 \cite{D1}. Apart of D1, all other functionals are still in
extensive use. 

 D1 \cite{D1}  is the first Gogny functional. However, it was
found in Ref.\ \cite{D1S-a} that it does not reproduce fission barriers in actinides. Thus, the 
D1S functional was fitted in Ref.\ \cite{D1S-a}: fitting protocol of this functional 
includes experimental data on fission barriers in addition to the data used
in the fitting protocol of the D1 functional. One can see in Table \ref{table-Gogny}
that for the same radial ranges $\mu_i$ of the D1 and D1S forces,  the absolute 
values of the strength parameters $W_i$, $B_i$, $H_i$ and $M_i$ of the central 
force of D1S differ by almost an order of magnitude from those of D1.
In addition, the strength $W_{LS}$ of the spin-orbit interaction
has different signs in these two functionals. The D1S functional has been 
successfully  applied to the description of many nuclear phenomena \cite{PM.14}
and it is still widely used by many practitioners of the Gogny DFT. The D1M 
functional has been fitted globally to nuclear masses in Ref.\ \cite{GHGP.09}. Shorter ranges of 
interaction are used in it (see Table \ref{table-Gogny}); as a result, the strenghts
of the central force terms change (as compared with D1S) by approximately one 
order of magnitude for $i=1$ and by a factor of approximately 4 for $i=2$.
As compared with D1S, the $t_0$ strength is modified by more that 10\%
but the strength of spin-orbit interaction remains unchanged.

   Contrary to the Gogny functionals there is significant number of Skyrme energy
density functionals (SEDF); the compilation of Ref.\ \cite{Sk-nm} reveals 240 SEDFs
developed by 2012. However, one should recognize that only small subset of these functionals 
is used in the calculations of finite nuclei in a repetitive manner. Table \ref{table-Skyrme}
lists some of them. As compared with Gogny functionals shown in Table \ref{table-Skyrme}, 
the variations in the parameters of SEDF (see Table \ref{table-Skyrme}) are less drastic. 
However, even the parameters of recent classes of the  SEDF, namely UNEDF* 
\cite{UNEDF0,UNEDF1} and BSk* \cite{Gor.15}, differ substantially. This reflects the 
differences  in the form of SEDFs and  the details of the fitting protocols; see also 
Sect. 5.C. of Ref.\ \cite{BHP.03} for short review  of  fitting protocols employed in nuclear 
DFTs.

   These results clearly show the difference between relativistic and non-relativistic
DFTs. In CDFT, the parameters of  the $\sigma$- and $\omega$-mesons are well 
localized for all functionals  (see Fig.\ \ref{Paramet}) independent of the form of the density dependence, the 
properties of the $\rho$-meson or the details of the fitting protocol.  Basically starting 
from first successful  CEDF NL1 \cite{NL1}, fitted 33 years ago, the parameters of the $\sigma$- 
and $\omega$-mesons of all successful functionals developed over the years and 
quoted in the present manuscript are very close to each other (see Fig.\ \ref{Paramet}).  
This is due to the fact that nucleonic potential in CDFT is build as a sum of large attractive 
scalar potential (related to $\sigma$-meson)  and large repulsive vector potential (related to 
$\omega$-meson) which limits possible range of the variations of the parameters. On 
the contrary, as illustrated in Tables \ref{table-Gogny} and \ref{table-Skyrme} the level of the 
localization of the parameters is lower (and the spread of the parameters is higher) in 
non-relativistic functionals.

\bibliography{references23}
\end{document}